\documentclass{mystyle}  

\usepackage{rotating} 
\usepackage{amssymb,amsmath}   
\usepackage{amsfonts} 
\usepackage{bm}

\makeatletter
\newcommand{\rb}[1]{\raisebox{1.7ex}[1.7ex]{#1}}
\newcommand{\la}{\left \langle}
\newcommand{\ra}{\right \rangle}
\newcommand{\tx}{\tilde x}
\newcommand{\ty}{\tilde y}
\newcommand{\tz}{\tilde z}
\newcommand{\ttt}{\tilde t}

\def\@dottedtocline#1#2#3#4#5{\ifnum #1>\c@tocdepth \else
  \vskip \z@ \@plus.2\p@
  {\leftskip #2\relax \rightskip \@tocrmarg \parfillskip -\rightskip
    \parindent #2\relax\@afterindenttrue
   \interlinepenalty\@M
   \leavevmode
   \@tempdima #3\relax
   \advance\leftskip \@tempdima \null\hskip -\leftskip
    {#4}\nobreak
        \hfill \nobreak
           \hb@xt@\@pnumwidth{%
             \hfil\normalfont \normalcolor #5}\par}\fi}
\def\numberline#1{\hb@xt@\@tempdima{#1.\hfil}}
\makeatother

\begin{document}

\setcounter{chapter}{0}

\chapter{Central and Non-central HBT from AGS to RHIC}

\markboth{B. Tom\'a\v sik and U.A. Wiedemann}{HBT from AGS to RHIC}

\author{Boris Tom\'a\v sik and Urs Achim Wiedemann}

\address{Theory Division, CERN, CH-1211 Geneva 23, Switzerland}

\begin{abstract}
We review the status of particle interferometry in ultra-relativistic 
nucleus-nucleus collisions. The theoretical focus is on the 
model-independent space-time interpretation of HBT radius parameters
and its extension to the geometrical and dynamical asymmetries 
generated in finite impact parameter collisions. On the experimental 
side, we give a complete account of all presently available data 
for beam energies above 2 $A$GeV. We discuss what these data imply
for the dynamics and space-time extension of the collision region 
and for the condition under which particles decouple from this region.
\end{abstract}

\newpage

\tableofcontents  

\newpage
\section{Introduction}
\label{sec1}
One aspect in which a theory of high-energy nucleus-nucleus 
collisions has to go beyond conventional nuclear and high-energy physics
is in understanding the space-time extension of the collision region
and its collective dynamical evolution. On the microscopic
level, basic quantities governing the reaction dynamics
such as the probability that a particle
can escape the collision region without further interaction
(``freeze-out'') or that it participates in
three-body interactions, clearly depend on the space-time
extension of the collision region and the relative velocity
of the particle with respect to the medium. On a macroscopic
level, the study of the equation of state of hot and dense
matter rests on assessing the energy density and pressure 
attained in nucleus-nucleus collisions. The operational
definition of these quantities clearly involves the 
measurement of volumes and the determination of the degree
of collectivity of the expansion.

It has been long recognised in both 
astrophysics\cite{HanburyBrown:1954wr,HanburyBrown:1956pf} and 
particle physics\cite{Goldhaber:sf} 
that correlations between identical bosons give access
to the geometry of particle emitting sources. While statistical
requirements for particle correlation measurements are significant,
interferometric techniques are often the method of choice when 
conventional space-time measurement techniques (like paralaxe measurements
or scattering with external probes) are not feasible. In the context of 
nucleus-nucleus collisions, the use of identical 
two-particle ``Hanbury-Brown Twiss'' (HBT) correlations was first 
forcefully advocated in the 1970s in a series of pioneering works 
published independently in the East\cite{Kopylov:qw,Kopylov:qq,Shuryak:1972kq} 
and the West\cite{Cocconi:1974pr,Gyulassy:yi}.
Those were followed by the first detailed discussions of how to 
separate quantum interference effects from final state 
interactions~\cite{Gyulassy:xb,Lednicky:1981su}. By now,
particle interferometry provides the most direct and most
detailed space-time picture of the late stage of nucleus-nucleus
collisions.

2 $A$GeV, the lowest beam energy delivered by the Alternating Gradient
Synchrotron (AGS) at the Brookhaven National Laboratory (BNL),
is a natural choice for the lowest energy included in a review
on relativistic nucleus-nucleus collisions.
Non-relativistic collisions at lower beam energies 
(up to several 100 $A$MeV) create evaporative sources with long 
lifetimes ($\sim$ several 100 fm/$c$) which are dominated by proton 
and neutron emission. Strong final state effects dominate 
correlation measurements. A comprehensive review of the rather different
physics at these lower energies exists\cite{Boal:yh}. Above 2 $A$GeV,
pion abundances become important, source
lifetimes are much shorter ($\sim 10$ fm/c) and quantum interference
effects dominate over the strong (but not necessarily over the
electro-magnetic) final state interactions. Moreover, the increase
of pion multiplicity per event with increasing beam energy
helps to overcome the major obstacle of particle interferometry
at lower energies, namely the lack of statistics.

The increase in event multiplicity and the accumulation of increasing
event samples lead throughout the 1990s to a rapid refinement
of analysis tools. Early analyses often show
``one-dimensional'' two-particle correlators as functions 
of one relative momentum variable only. These were soon superseded by 
``three-dimensional'' representations. Then, the increasing statistics
allowed to resolve ``three-dimensional'' correlation functions with
respect to the average transverse pair momentum and rapidity. Only 
recently, the next step of these refinements was taken when the
first measurements of HBT with respect to the orientation of 
the reaction plane became available. Parallelling this experimental 
progress and sometimes anticipating it was a theoretical effort which 
related systematically these more and more differential interferometric
measurements to more and more detailed geometrical and
dynamical properties of the particle emitting source. By the end
of the CERN SPS heavy ion program in 2000, several 
reviews\cite{Wiedemann:1999qn,Heinz:1999rw,Csorgo:1999sj} gave
detailed summaries of these developments. 

We were asked to contribute to the present review volume only two
years later, at a time when  HBT measurements from RHIC
are published but no thorough theoretical analysis of these findings
is yet available. In this situation, we decided to emphasise in a 
topical review those two aspects of particle interferometry which
are likely to play an important role in the further 
discussion of RHIC data and which are not yet adequately represented 
in reviews. On the experimental side, our focus will be on a
comprehensive overview of all data from AGS, SPS and
RHIC. This is timely not only because several experiments at the
AGS and SPS just finalised their HBT analyses. It is also
of obvious use for a discussion of the energy-dependence of HBT
correlation measurements which plays an important role in the HBT 
studies at RHIC. On the theoretical side, our presentation reflects
the fact that the first measurements of azimuthally dependent HBT 
radius parameters in finite impact parameter collisions at AGS and
RHIC became available after 1999. That these novel observables may 
provide a complementary space-time picture to the large elliptic 
flow measured at RHIC gives an additional motivation for discussing
their theoretical basis in detail.

This review is organised as follows: In section \ref{sec2}, we
introduce the formalism for calculating identical two-particle
correlations from particle emission functions. In particular,
we discuss in detail the case of non-central collisions.
In section \ref{sec-models}, we briefly summarise models for the 
particle emitting source which have been used in calculating HBT 
correlations from ultra-relativistic heavy-ion collisions. 
Finally, in section~\ref{Sec3}, we give a complete account of the 
experimental situation and we discuss shortly the most important features
of the presently available data. Since most data are for central
collisions, readers primarily interested in the data summary may
jump from the introductory sections~\ref{sec2aa} and \ref{sec2b}
directly to section~\ref{Sec3}. The extension to non-central
collisions in section~\ref{sec2c} and the corresponding experimental
overview in section~\ref{ss-azdep} can be read independently. Also,
the short overview of model calculations in section~\ref{sec-models}
allows for an independent reading.

\section{The Gaussian source formalism}     
\label{sec2}

The two-particle momentum correlator between identical bosons
with momenta $p_1$ and $p_2$ is defined as the quotient of 
two-particle and one-particle spectra\footnote{ 
Throughout this review, we use italics for four-vectors 
and bold for three-vectors.},
  \begin{equation}
     C(\bm{q},\bm{K}) 
     = \frac{\frac{d^6N}{d\bm{p}_1^3\, d\bm{p}_2^3}}{%
                     \frac{d^3N}{d\bm{p}_1^3}\,  \frac{d^3N}{d\bm{p}_2^3}}\, .
       \label{eq1} 
  \end{equation}
Experimentally, this ratio is constructed for event samples,
normalising {\it actual pairs} from the same event by {\it mixed
pairs} from different events. Usually, the two-particle correlator 
is written in terms of the corresponding relative and average pair 
momenta 
\begin{eqnarray}
  q &=& p_1 - p_2\, , \label{eq3} \\
  K &=& \frac{1}{2}\left( p_1 + p_2\right)\, . \label{eq4}
\end{eqnarray}
If the particle spectra entering (\ref{eq1}) are corrected for
final state interactions, then the two-particle correlator
$C(\bm{q},\bm{K})$ is related to the emission function 
$S(x,K)$~\cite{Shuryak:1972kq,Gyulassy:yi,Pratt:su,Pratt:cc,Bertsch:1993nx,Chapman:xa}
which is the Wigner phase-space density of the particle emitting 
system
  \begin{eqnarray}
    C(\bm{q}, \bm{K}) 
       \approx 1 + 
     \frac{\left\vert \int d^4x\, S(x,K)\, e^{iq{\cdot}x}\right\vert^2 } 
      {\left\vert \int d^4x\, S(x,K)\right\vert^2 }\, . 
  \label{eq2} 
  \end{eqnarray}
For a derivation of (\ref{eq2}) in different formalisms
(coherent source formalism and Gaussian wave-packet formalism),
we refer to the review\cite{Wiedemann:1999qn}. Equation
(\ref{eq2}) is approximate since it is based on the
smoothness approximation $S(x,K-\textstyle{\frac{1}{2}}q) 
S(y,K+\textstyle{\frac{1}{2}}q)  \approx S(x,K) S(y,K)$
valid for small relative momenta. Moreover, the following
discussion of (\ref{eq2}) will regularly employ the on-shell 
approximation $K_0 \approx \sqrt{ \bm{K}^2 + m^2}$ which
neglects the off-shell components of $K$. Corrections to
these approximations were studied systematically\cite{Chapman:1994ax}
and were found to be small except for especially constructed
toy models whose phase-space volumes were very small
(i.e. comparable to the lower bound coming from the Heisenberg
uncertainty relation). For particle correlations in nucleus-nucleus
collisions, equation (\ref{eq2}) is a well-defined starting point.
The emission function $S(x,K)$ can be viewed as the
probability that a particle with momentum $K$ is emitted from 
the space-time point $x$ in the collision region. 

The aim of HBT two-particle 
interferometry is to extract from the experimentally measured 
correlator $C(\bm{q},\bm{K})$ as much information about 
$S(x,K)$ as possible. This includes i) information about the
geometrical extent of the collision region at the time
of last hadronic scattering (freeze-out), ii) information about
the freeze-out time and emission duration of the source, 
iii) information about the collective velocity of the expanding collision 
region in the directions parallel and orthogonal to the beam,
iv) information about the azimuthal dependence of the geometrical
and dynamical properties of the collision region in finite
impact parameter collisions. From these primary measurements, 
various derived quantities can be extracted, in particular
the particle phase-space density at freeze-out.

\subsection{Gaussian correlators in terms of space-time variances }     
\label{sec2aa}

Experimental data of two-particle correlations are usually
parametrised by a Gaussian ansatz in terms of HBT radius
parameters $R_{ij}(\bm{K})$ and the $\lambda$-intercept
parameter
  \begin{eqnarray}
     C(\bm{q},\bm{K}) &=&  1 + \lambda(\bm{K})\, \exp\left[ 
      -\sum_{ij} R_{ij}^2(\bm{K})\, \bm{q}_i\, \bm{q}_j\right]\, .
  \label{eq5}  
  \end{eqnarray}
Here, the indices $i$, $j$ run over three of the four components 
of $q$. The fourth component of $q$ in (\ref{eq3}) is fixed by 
the requirement that the final state particles are on-shell
\begin{eqnarray}
  q\cdot K = \frac{1}{2}\left(p_1^2 - p_2^2\right) = 0\, 
  \Longrightarrow\,  q_0 = \frac{\bm{K}\cdot \bm{q}}{K_0}\, .
  \label{eq6}
\end{eqnarray}
Different choices of the three independent components of 
$q = (q_0,\bm{q})$ correspond to different Gaussian parametrisations. 

The connection between the HBT radius parameters and the space-time 
structure of the source is based on a Gaussian approximation to the 
emission function
 \begin{eqnarray}
   S(x,K) &\approx& S(\bar x(K),K)
          \exp\left[ - \frac{1}{2} \tilde x^\mu(K)\,
            B_{\mu\nu}(K)\,\tilde x^\nu(K)\right]\, .
 \label{eq7}
 \end{eqnarray}
The space-time coordinates $\tilde{x}_{\mu}$ in (\ref{eq7})
are defined relative to the ``effective source centre'' $\bar x(K)$
for bosons emitted with momentum
$\bm{K}$\cite{Chapman:yv,Herrmann:1994rr}  
 \begin{equation}
  \tilde x^\mu (K) = x^\mu - \bar x^\mu(K)\, , \qquad
  \bar x^\mu(K) = \langle x^\mu \rangle(K) \, ,
  \label{eq8} 
 \end{equation}
where $\langle \dots \rangle$ denotes an average with the 
emission function $S(x,K)$:
  \begin{equation}
  \langle f\rangle(K) = 
  \frac{\int d^4x\, f(x)\, S(x,K)}{\int d^4x\, S(x,K)}\, .
  \label{eq9}
  \end{equation}
The choice
 \begin{equation}
  (B^{-1})_{\mu\nu}(K)
  = \langle \tilde x_\mu \tilde x_\nu \rangle(K)
  \label{eq10}
 \end{equation}
ensures that the Gaussian ansatz (\ref{eq7}) has the same
rms widths in space-time as the full emission function.
Inserting (\ref{eq7}) into the basic relation (\ref{eq2}),
the correlator takes a Gaussian form, 
 \begin{equation}
   C(\bm{q},\bm{K}) = 1 + \exp\left[ - q_\mu q_\nu 
   \langle \tilde x^\mu \tilde x^\nu \rangle (\bm{K}) \right]\, .
  \label{eq11} 
 \end{equation}
Since the correlator depends only
on the relative distances $\tilde{x}^{\mu}$ with respect to the
source centre, no information can be obtained about the 
position $\bar{x}(\bm{K})$ of the centre of the emission.

\begin{figure}[th]\epsfxsize=10cm 
\centerline{\epsfbox{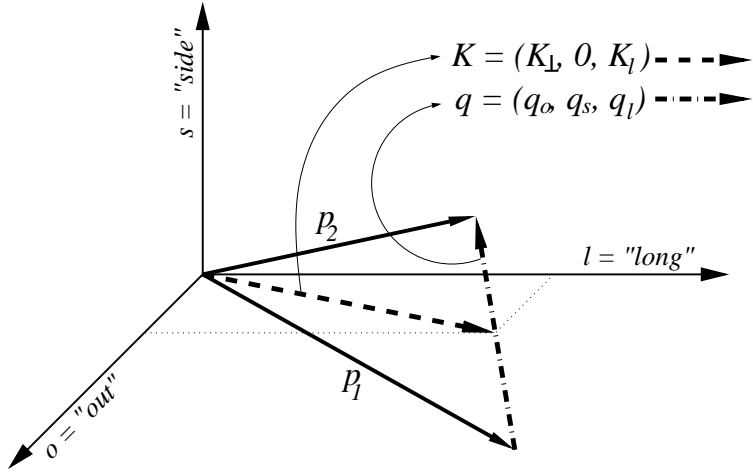}}
\caption{The out-side-longitudinal {\it osl} coordinate system. 
}\label{fig1}
\end{figure}
%

Finally, we comment on $\lambda(\bm{K})$ in eq.~(\ref{eq5})
which parametrises the intercept of the correlation function for 
$q = 0$. This $\lambda$-parameter is unity
for a chaotic source and it is smaller than 1 for a source with
partially coherent particle emission. In practice, however, there
are many other reasons why deviations from $C(q=0)=2$ may be
observed. For example, contributions to pion 
production\cite{Wiedemann:1996ig} from long-lived resonances lead 
to an extended source component which is reflected in a narrow peak of 
the correlation function. If this peak is narrower than the experimental 
resolution in relative momentum, a reduced intercept is observed. Also,
misidentified particles in the sample of identical particle pairs 
reduce the correlation strength $\lambda({\bf K})$.
In addition to this, many technical details 
of the analysis -- such as Coulomb correction or finite momentum resolution
-- can influence the value of $\lambda$.

\subsection{Central (azimuthally symmetric) collisions}
\label{sec2b}
\subsubsection{The Cartesian Bertsch-Pratt parametrisation}
\label{sec2b1}
In the {\it out-side-longitudinal} ({\it osl}) coordinate system,
defined in Fig.~\ref{fig1}, the longitudinal ({\it long}) direction 
points along the beam axis. In the transverse plane, the
{\it out} direction is chosen parallel to the transverse component
of the pair momentum ${K}_{\perp}$, the remaining Cartesian
component denotes the {\it side} direction. In this way,
the out-axis is rotated independently for each particle pair. 
The azimuthal symmetry
of the particle source in central collisions then translates 
in the {\it osl} coordinate system into a reflection symmetry with 
respect to the {\it side}-direction:
  \begin{eqnarray}
    S(x,y,z,t;K_{\perp},K_l) 
    = S(x,-y,z,t;K_{\perp},K_l)\, .
    \label{eq13}
  \end{eqnarray}
This $\tilde{y} \to -\tilde{y}$ symmetry translates to a
$q_s \to -q_s$ symmetry of the two-particle correlation
function (\ref{eq11}). The Cartesian Bertsch-Pratt 
parametrisation\cite{Pratt:su,Bertsch:1988db,Chapman:yv} 
exploits this reflection symmetry to write the two-particle
correlator as
\begin{eqnarray}
 &&\hspace{-1cm}
    C(\bm{q},\bm{K}) = \exp\lbrack
        -R_o^2(\bm{K})\, q_o^2 - R_s^2(\bm{K})\, q_s^2
       - R_l^2(\bm{K})\, q_l^2 - 2\, R_{ol}^2(\bm{K})\, q_o q_l \rbrack\, .
\end{eqnarray}
This parametrisation is based on the three 
Cartesian spatial components $q_o$ ({\it out}), $q_s$ ({\it side}),
$q_l$ ({\it long}) of the relative momentum $q$. The temporal 
component is eliminated via the mass-shell constraint (\ref{eq6})
  \begin{eqnarray}
    q^0 = \bm{\beta} \cdot \bm{q}\, ,\qquad
    \mbox{\boldmath${\beta}$\unboldmath} = 
    \left(\beta_\perp, 0,\beta_l\right)\, ,
    \label{eq12}
  \end{eqnarray}
where $\mbox{\boldmath${\beta}$\unboldmath} = \bm{K}/K_0$. 
To express the HBT radius parameters $R_{i}^2(\bm{K})$ 
in terms of space-time variances 
$\langle \tilde{x}_\mu\tilde{x}_\nu \rangle({K})$, we insert the
on-shell constraint 
$q^0 = \mbox{\boldmath${\beta}$\unboldmath} \cdot \bm{q}$ into the correlator 
(\ref{eq11}), and compare the result with the Gaussian parametrisation 
(\ref{eq5}) of the correlator. One then finds that the HBT radius
parameters measure different combinations of the spatial and temporal 
extent of the collision system\cite{Chapman:yv,Herrmann:1994rr}:
  \begin{eqnarray}   
          R_s^2({K}) &=& \langle \tilde{y}^2 \rangle({K}) \, ,
        \label{eq14}\\
          R_o^2({K}) &=& 
        \langle (\tilde{x} - \beta_\perp \tilde t)^2 \rangle({K}) \, ,
        \label{eq15}\\
          R_l^2({K}) &=& 
        \langle (\tilde{z} - \beta_l \tilde t)^2 \rangle({K}) \, ,
        \label{eq16}\\
          R_{ol}^2({K}) &=& 
        \langle (\tilde{x} - \beta_\perp \tilde t)
           (\tilde{z} - \beta_l \tilde t) \rangle({K}) \, ,
        \label{eq17} \\
          R_{os}^2({K}) &=& 0 \, ,
        \label{eq18} \\
          R_{sl}^2({K}) &=& 0 \, .
        \label{eq19}
  \end{eqnarray}
The $y \to -y$ reflection symmetry of the emission function implies 
that the three space-time variances
$\langle \tilde{x}_\mu\tilde{x}_\nu \rangle({K})$ 
linear in $\tilde{y}$ vanish. For the azimuthally
symmetric case, there are only seven non-vanishing 
space-time variances which combine to four non-vanishing 
HBT-radius parameters $R_{ij}^2({K})$. This information is
tabulated in the first column of Table~\ref{table1}.

In general, symmetries of the emission function translate
into constraints for the space-time variances and thus 
lead to simpler expressions for the HBT radius parameters. 
Of particular relevance is the limit $K_\perp \to 0$. In
this limit, no transverse vector allows to distinguish
between the {\it out}- and {\it side}-components, and
space-time variances thus become invariant under the
exchange $\tilde{x} \to \tilde{y}$. This implies that
$\langle \tilde{x}^2 \rangle (K_\perp = 0) =
\langle \tilde{y}^2 \rangle (K_\perp = 0)$. Also, the
off-diagonal terms $\langle \tilde{z}\tilde{x} \rangle (K_\perp = 0)$
and $\langle \tilde{t}\tilde{x} \rangle (K_\perp = 0)$ vanish.
As a consequence, the HBT radius parameters show the following
limiting behaviour
  \begin{eqnarray}
    \lim_{K_\perp \to 0} R_o^2({K}) &=& 
    \lim_{K_\perp \to 0} R_s^2({K})\, ,
    \label{eq20} \\
    \lim_{K_\perp \to 0} R_{ol}^2({K}) &=& 0\, .
    \label{eq21}
  \end{eqnarray}
Further symmetries are sometimes helpful to gain physical intuition.
For example, longitudinal boost-invariance implies a $\tilde{z} \to 
-\tilde{z}$ reflection symmetry of the emission function
which is approximately satisfied near mid-rapidity. This implies
that the space-time variances linear in $\tilde{z}$ vanish.
In the longitudinally comoving system (LCMS)
\begin{equation}
\label{LCMS}
\mbox{LCMS:}\qquad \beta_l = 0\, ,
\end{equation}
this leads to the further simplifications $R_l^2({K}) = 
\langle \tilde{z}^2 \rangle({K})$ and $R_{ol}^2({K}) = 0$.
%

\subsubsection{Interpretation of space-time variances}
\label{sec2b3}
The space-time variances $\langle \tilde x_\mu \tilde x_\nu \rangle(K)$
depend on the pair momentum $K$. To understand their physical meaning,
consider an observer who views a strongly expanding collision 
region. Some parts of the collision region move towards the observer
and the particle spectrum emitted from those parts will appear
blue-shifted. Other parts move away from the observer and appear red-shifted.
Thus, if the observer looks at the collision system with a wavelength
filter of some frequency, he sees only part of the collision region.
Adopting a notion coined by Sinyukov, the observer sees a
``region of homogeneity''.
%
\begin{figure}[t]\epsfxsize=12cm 
\centerline{\epsfbox{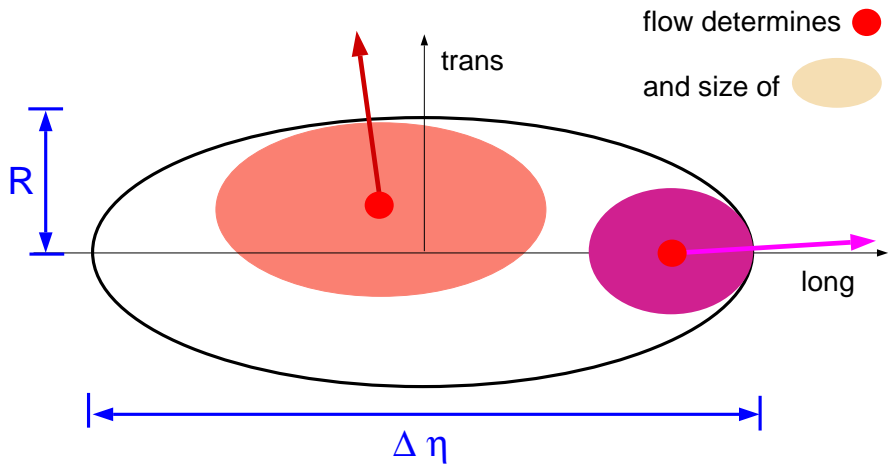}}
\caption{
Sketch of the particle emission region for a collision with 
collective dynamical expansion. The size and orientation of the average
pair momentum $\bm{K}$ serves as wavelength filter: depending on $\bm{K}$
(indicated by arrows),
HBT measurements are sensitive to different parts of the collision
region. 
}\label{fig2}
\end{figure}
%
In HBT interferometry, the role of the wavelength filter is played
by the pair momentum $\bm{K}$. The direction of $\bm{K}$ corresponds
to the direction from which the collision region is viewed.
The modulus of $\bm{K}$ characterises the central velocity 
$\bm{\beta} = \bm{K}/K_0$ of that part of the collision region which
is seen through the wavelength filter. This situation is illustrated
in Fig.~\ref{fig2}. Depending on the direction and modulus of $\bm{K}$,
different parts of the collision region are measured. The centre of a
region of homogeneity $\langle x_\mu \rangle$ depends on $\bm{K}$ and lies 
typically between the center of the collision region and the observer.
In the Gaussian approximation, the region of homogeneity is 
described by a four-dimensional space-time ellipsoid centered 
around $\langle \tilde x_\mu \rangle(\bm{K})$ and characterised
by $S(x,K)$ in eq. (\ref{eq7}). The widths of this
region of homogeneity correspond to the space-time variances 
$\langle \tilde x_\mu \tilde x_\nu \rangle(\bm{K})$, see
eq. (\ref{eq10}). Thus, HBT radius parameters give access to
the space-time variances $\langle \tilde x_\mu \tilde x_\nu \rangle(\bm{K})$
but they do not depend on the ``effective source center''
$\langle x_\mu \rangle$.

\subsubsection{The Yano-Koonin-Podgoretski\u\i\ pa\-ra\-me\-tri\-sa\-tion}
\label{sec2b4}

The Yano-Koonin-Podgoretski\u\i\ pa\-ra\-me\-tri\-sa\-tion eliminates
the mass-shell constraint $K\cdot q = 0$ in (\ref{eq11}) by
choosing the relative momentum components $q_\perp =
\sqrt{q_o^2 + q_s^2}$, $q^0$ and $q_l$ as independent.
The corresponding Gaussian ansatz 
reads\cite{Yano:gk,Pratt:su,Chapman:1995nz,Heinz:1996qu}
 \begin{eqnarray}
   C(\bm{q},\bm{K}) &=&
   1 +  \lambda(\bm{K})\,\exp\Bigl[ - R_\perp^2(\bm{K})\, q_{\perp}^2 
                            - R_\parallel^2(\bm{K}) (q_l^2 - 
                                             {(q^0)}^2)
 \nonumber \\
    &&\qquad\qquad - \left(R_0^2(\bm{K}) + R_\parallel^2(\bm{K})\right)
                         \left(q{\cdot}U(\bm{K})\right)^2
      \Bigr]  \, .
   \label{3.35}
 \end{eqnarray}
Here, the 4-velocity $U(\bm{K})$ has only one $\bm{K}$-dependent  
longitudinal spatial component,
 \begin{eqnarray}
   U(\bm{K}) &=& \frac{1}{\sqrt{1 - v^2(\bm{K})}}
   \left(1, 0, 0, v(\bm{K}) \right)\, .
   \label{3.37}
 \end{eqnarray}
The combinations of relative momenta $(q_l^2 - {(q^0)}^2)$, 
$(q{\cdot}U(\bm{K}))^2$ and $q_\perp^2$ appearing in (\ref{3.35})
are scalars under longitudinal boosts. Therefore, the three
YKP fit parameters $R_\perp^2(\bm{K})$, $R_0^2(\bm{K})$, and 
$R_\parallel^2(\bm{K})$ are longitudinally boost-invariant
and do not depend
on the longitudinal velocity of the measurement frame. 
The fourth YKP parameter is the Yano-Koonin (YK)
velocity $v(\bm{K})$. The corresponding rapidity
  \begin{equation}
    Y_{_{\rm YK}}({\bm{K}}) = \frac{1}{2} 
    \ln \left(\frac{1+v(\bm{K})}{1-v(\bm{K})} \right)
    \label{3.38}
  \end{equation}
transforms additively under longitudinal boosts.

Identifying the combinations of relative momenta $(q_l^2 - {(q^0)}^2)$, 
$(q{\cdot}U(\bm{K}))^2$ and $q_\perp^2$ in the Gaussian 
correlator (\ref{eq11}), one can relate the YKP parameters
to space-time variances in complete analogy to the Cartesian
case. Explicit expressions can be found in the original 
papers\cite{Heinz:1996qu,Wu:1996wk}.
Since the ansatz (\ref{3.35}) uses four Gaussian parameters, it 
provides a complete parametrisation for azimuthally symmetric 
collisions which is equivalent to the Cartesian Bertsch-Pratt
parametrisation. The Gaussian HBT radius parameters of the 
Cartesian parametrisation can be related to the YKP fit
parameters via\cite{Heinz:1996qu,Wu:1996wk} 
 \begin{eqnarray}
 \label{3.46}
    R_s^2 &=& R_\perp^2 \, , \\
 \label{3.47}
   R_{\rm diff}^2 &=& R_o^2 - R_s^2 = 
   \frac{\beta_\perp^2}{1-v^2(\bm{K})} 
             \left( R_0^2 + v^2 R_\parallel^2 \right) \, ,
 \\
 \label{3.48}
   R_l^2 &=& \left( 1 - \beta_l^2 \right) R_\parallel^2 
             + \frac{\left(\beta_l-v \right)^2}{1-v^2(\bm{K})} 
             \left( R_0^2 + R_\parallel^2 \right)\, ,
 \\
 \label{3.49}
   R_{ol}^2 &=& \beta_\perp \left( -\beta_l R_\parallel^2 
             + \frac{\left( \beta_l-v \right)}{1-v^2(\bm{K})} 
             \left( R_0^2 + R_\parallel^2 \right) \right)\, .
 \end{eqnarray}
The space-time interpretation of YKP parameters is
studied in detail\cite{Heinz:1996qu,Wu:1996wk,Tomasik:1999kz}.
The consistency check \eqref{3.46}-\eqref{3.49} of  
measured YKP and Cartesian parameters is mandatory for a
space-time interpretation of YKP parameters since known
technical problems with the YKP parametrization can affect
fit parameters significantly\cite{Tomasik:1999kz}.

\subsection{Collisions at finite impact parameter}
\label{sec2c}

\subsubsection{Choice of coordinate system}
\label{sec2c1}

For central collisions, azimuthal symmetry of the collision region
allows to use the {\it osl}-coordinate system which is oriented 
differently for each particle pair, such that its {\it out}-axis is
parallel to the  average pair momentum. 

%
\begin{figure}[t]\epsfxsize=9cm 
\centerline{\epsfbox{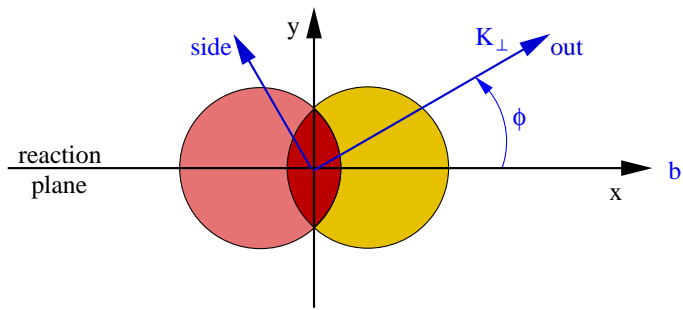}}
\caption{The impact parameter fixed coordinate system is obtained
by rotating the out-side-longitudinal coordinate system by the 
azimuthal angle $\Phi = \angle({K}_\perp,{b})$.  
}\label{impactsystem}
\end{figure}

At finite impact parameter, HBT radius parameters $R_{ij}^2(\bm{K})$
depend not only on the modulus of the transverse pair momentum $|\bm{K}|$,
but also on its orientation in the transverse plane which is 
defined with respect to some pair-independent direction in the 
laboratory system\cite{Voloshin:1995mc,Voloshin:1996ch,Wiedemann:1997cr}. 
In practice, one introduces the {\it impact
parameter fixed} coordinate system, for which the transverse 
coordinate $x$ lies within the reaction plane; the transverse 
coordinate $y$ lies orthogonal to it, see Fig.~\ref{impactsystem}.
The out-side-longitudinal and the impact parameter fixed system 
are related by a rotation with respect to the angle 
$\Phi = \angle({K}_\perp,{b})$
 \begin{equation}
        {\cal D}_\Phi = 
        \left ( \begin{array}{ccc}
        \cos\Phi  & \sin\Phi & 0 \\
        -\sin\Phi & \cos\Phi & 0 \\
        0         &     0    & 1 
        \end{array} \right )\, ,
 \end{equation} 
such that after rotating to the {\em osl}-system we have
 \begin{eqnarray}
          &&{\cal D}_\Phi \bm{\beta} =  \left( \begin{array}{c}
                 \beta_\perp \\ 0\\ \beta_l \end{array} \right)\, ,
             \qquad
           {\cal D}_\Phi \tilde{\bm{x}} =  \left( \begin{array}{c}
                 \tilde{x} \cos\Phi + \tilde{y} \sin\Phi  \\ 
                 -\tilde{x} \sin\Phi + \tilde{y} \cos\Phi \\
                 \tilde{z} \end{array} \right) \, .
   \label{eq23} 
 \end{eqnarray}
For non-central collisions, the emission function is written in
the impact parameter fixed coordinate system. 

\subsubsection{HBT radius parameters from space-time variances}
\label{sec2c2}
\begin{figure}[t]\epsfxsize=10.7cm 
\centerline{\epsfbox{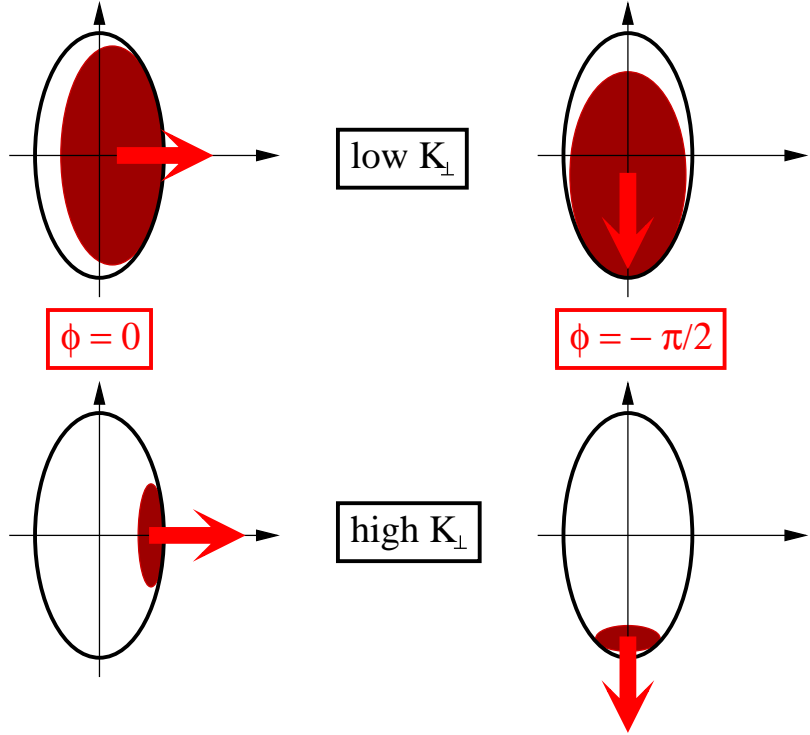}}
\caption{Schematic picture of the role of explicit and implicit
$\phi$-dependence in the transverse plane. At small transverse
pair momentum, the orientation of the region of homogeneity follows
the global geometry, and the explicit $\Phi$-dependence is
expected to dominate. At large transverse pair momentum, the
orientation and shape of the region of homogeneity can deviate
significantly from the global geometry. In this case, the
implicit $\Phi$-dependence of HBT radius parameters becomes
significant, see text. 
}\label{inexazi}
\end{figure}

In eq. (\ref{eq2}), the emission function entering 
the two-particle correlator is written in the
{\it osl}-system. For an emission function in the impact
parameter fixed system, the correlator reads:
  \begin{eqnarray}
     C(\bm{K},\bm{q})
      &=& 1 + \vert\langle e^{i\bm{q}{\cdot}({\cal D}_\Phi\bm{x}
                                         -{\cal D}_\Phi\bm{\beta}t)}
            \rangle\vert\, .
            \label{eq25}
  \end{eqnarray}
From this, the HBT radius parameters can be calculated\cite{Wiedemann:1997cr}:
 \begin{eqnarray}
   R_{ij}^2(\bm{K}) &=& - \left. \frac{\partial^2 C(\bm{q},\bm{K})} 
                           {\partial q_i\, \partial q_j}
                           \right\vert_{ \bm{q} = 0}\, \nonumber \\
   &=& \langle  \lbrack ({\cal D}_\Phi \tilde{\bm{x}})_i 
                        - ({\cal D}_\Phi \bm{\beta})_i\tilde{t}\rbrack 
               \lbrack ({\cal D}_\Phi \tilde{\bm{x}})_j 
                        - ({\cal D}_\Phi \bm{\beta})_j\tilde{t}\rbrack 
                        \rangle\, .
   \label{eq26}
 \end{eqnarray}
As for the azimuthally symmetric case, the
HBT radius parameters show implicit and explicit 
${K}$-dependences\cite{Wiedemann:1997cr}:
  \begin{eqnarray}
    && R_s^2(K_\perp,\Phi,Y) = \langle \tilde{x}^2\rangle \sin^2\Phi
                  + \langle \tilde{y}^2\rangle \cos^2\Phi
                  - \langle \tilde{x}\tilde{y}\rangle 
                       \sin 2\Phi \, ,
                      \label{eq27} \\
    && R_o^2(K_\perp,\Phi,Y) = \langle \tilde{x}^2\rangle \cos^2\Phi
                  + \langle \tilde{y}^2\rangle \sin^2\Phi
                  + \beta_\perp^2 \langle \tilde{t}^2\rangle
                      \nonumber \\
    && \qquad \qquad - 2\beta_\perp 
                       \langle \tilde{t}\tilde{x} \rangle \cos\Phi 
                     - 2\beta_\perp 
                       \langle \tilde{t} \tilde{y} \rangle \sin\Phi 
                     + \langle \tilde{x}\tilde{y}\rangle 
                       \sin 2\Phi \, ,
                      \label{eq28} \\
    && R_{os}^2(K_\perp,\Phi,Y) = 
                  \langle \tilde{x}\tilde{y}\rangle \cos 2\Phi 
                  + \textstyle{\frac{1}{2}} \sin 2\Phi 
                  (\langle \tilde{y}^2\rangle - \langle \tilde{x}^2\rangle)
                  \nonumber \\
    && \qquad \qquad + \beta_\perp \langle \tilde{t}
                       \tilde{x}\rangle \sin\Phi
                     - \beta_\perp \langle \tilde{t}
                       \tilde{y}\rangle \cos\Phi \, ,
                      \label{eq29} \\
    && R_{l}^2(K_\perp,\Phi,Y) = 
                  \langle (\tilde{z} -\beta_l\tilde{t})^2 \rangle \, ,
                      \label{eq30} \\
    && R_{ol}^2(K_\perp,\Phi,Y) = 
                  \langle (\tilde{z} -\beta_l\tilde{t})
                     (\tilde{x}\cos\Phi + \tilde{y}\sin\Phi 
                     - \beta_\perp\tilde{t}) \rangle\, ,
                      \label{eq31} \\
    && R_{sl}^2(K_\perp,\Phi,Y) = 
                  \langle (\tilde{z} -\beta_l\tilde{t})
                     (\tilde{y}\cos\Phi - \tilde{x}\sin\Phi) 
                     \rangle\, .
    \label{eq32}
  \end{eqnarray}
On the r.h.s.\ of these equations we only displayed 
the {\it explicit} $\Phi$-dependences which are a geometrical
consequence of rotating the {\it out}-axis from the direction of 
the impact parameter $\bm{b}$ to the direction of the average
pair momentum $\bm{K}_\perp$. The {\it implicit} 
$\Phi$-dependence of the space-time variances,
\begin{eqnarray}
 \langle \tilde{x}_{\mu}\tilde{x}_{\nu}\rangle
 = \langle \tilde{x}_{\mu}\tilde{x}_{\nu}\rangle(K_\perp,\Phi,Y)\, , 
 \label{eq33}
\end{eqnarray}
characterises
the dynamical correlations between the size of the effective emission 
region and the azimuthal direction in which particles are emitted. 
The role of explicit and implicit contributions is illustrated in
Fig.~\ref{inexazi}.
%

\subsubsection{Symmetries satisfied by space-time variances}
\label{sec2c3}
The implicit $\Phi$-dependence of space-time variances can be
characterised in terms of the Fourier coefficients 
$\langle \dots \rangle^c_n$ and $\langle \dots \rangle^s_n$ 
\begin{eqnarray}
 \langle \tilde{x}_{\mu}\tilde{x}_{\nu}\rangle (\Phi)
 &=& \sum_{n=0}^{\infty} 
 \langle \tilde{x}_{\mu}\tilde{x}_{\nu}\rangle^c_n \cos(n\Phi) +
 \langle \tilde{x}_{\mu}\tilde{x}_{\nu}\rangle^s_n \sin(n\Phi)\, . 
 \label{eq34}
\end{eqnarray}
Several symmetries allow to constrain this most general implicit 
$\Phi$-dependence\cite{Wiedemann:1997cr,Lisa:2000ip,Heinz:2002au}:
\begin{enumerate}
\item
\underline{Mirror symmetry with respect to the reaction plane:}\\
This symmetry of the emission function
holds at both mid and forward rapidity. It implies
that space-time variances linear in $\tilde y$ change sign under
$\Phi \to -\Phi$, while space-time variances which are not linear
in $\tilde y$ are invariant under the mirror symmetry,
\begin{eqnarray}
  \langle \tilde x_\mu \tilde x_\nu \rangle (\Phi)
  &=& - \langle \tilde x_\mu \tilde x_\nu \rangle (-\Phi)\, ,
  \quad \hbox{for either}\,\, \tilde x_\mu = \tilde y\,\,  \hbox{or}
  \,\,  \tilde x_\nu = \tilde y\, ,
   \label{eq35} \\
  \langle \tilde x_\mu \tilde x_\nu \rangle (\Phi)
  &=& \langle \tilde x_\mu \tilde x_\nu \rangle (-\Phi)\, ,
  \qquad \hbox{for}\quad \tilde x_\mu\, , \tilde x_\nu  \not= \tilde y\, ,  
   \label{eq36} \\
  \langle {\tilde y}^2 \rangle (\Phi)
  &=& \langle  {\tilde y}^2  \rangle (-\Phi)\, .
  \label{eq37}
\end{eqnarray}
\item
\underline{Point symmetry at mid rapidity:}\\
For collisions between equal mass nuclei, the total emission region 
is point symmetric under 
$(\tilde x,\tilde y,\tilde z) \to (-\tilde x,-\tilde y,-\tilde z)$
at mid-rapidity.
This is a symmetry under $\Phi \to \Phi + \pi$. At mid-rapidity,
we thus have the additional requirement
\begin{eqnarray}
  \langle \tilde x_\mu \tilde x_\nu \rangle (\Phi)
  &=& -\langle \tilde x_\mu \tilde x_\nu \rangle (\Phi+\pi)\, ,
  \quad \hbox{for either}\,\, \tilde x_\mu = \tilde t\,\,  \hbox{or}
  \,\, \tilde x_\nu = \tilde t\, ,
   \label{eq38} \\
  \langle \tilde x_\mu \tilde x_\nu \rangle (\Phi)
  &=& \langle \tilde x_\mu \tilde x_\nu \rangle (\Phi+\pi)\, ,
  \qquad \hbox{for}\quad \tilde x_\mu\, , \tilde x_\nu  \not= \tilde t\, ,
   \label{eq39} \\
  \langle {\tilde t}^2 \rangle (\Phi)
  &=& \langle  {\tilde t}^2  \rangle (\Phi+\pi)\, .  
  \label{eq40}
\end{eqnarray}
At forward rapidity, this point symmetry does not constrain
the azimuthal dependence since the same symmetry maps
$K_l \to -K_l$ on which the space-time variances depend
implicitly.
\end{enumerate}
Both symmetries imply that many of the Fourier coefficients
vanish in the harmonic expansion (\ref{eq34}) of the space-time
variances. A complete overview is given in Table~\ref{table1}.
We discuss now the implications of these
symmetries for a harmonic analysis of HBT radius parameters
whose $\Phi$-dependences are characterised in terms of
harmonic coefficients
  \begin{eqnarray}
    {R_{ij,m}^c}^2 &=& \frac{1}{2\pi} 
           \int_{-\pi}^\pi R_{ij}^2\, \cos(m\Phi)\, d\Phi\, ,
    \label{eq41} \\
    {R_{ij,m}^s}^2 &=& \frac{1}{2\pi} 
           \int_{-\pi}^\pi R_{ij}^2\, \sin(m\Phi)\, d\Phi\, .
    \label{eq42}
  \end{eqnarray}
%
\begin{table}
\tbl{Space-time variances for central and non-central
collisions \label{table1}}
{\begin{tabular}{@{}|c|cccc||ccc|c|@{}}
\hline
 & & & & & & & & \\[-2ex]
$\langle x_\mu x_\nu \rangle$ & $\bm{b}=0$ & $\bm{b}=0$ & $\bm{b}\not=0$ &
$\bm{b}\not=0$ & $\bm{b}\not=0$ & $\bm{b}\not=0$ & $\bm{b}\not=0$
& {}\\
{} & $K_\perp \not= 0$ & $K_\perp =0$ & $Y=0$ & $Y\not= 0$ & $Y=0$ &
$Y=0$ & $Y\not= 0$  & {}\\
{} &{} &{} &$n=0$ & n = 0 & $n$ even & $n$ odd  & $n\ge 1$ & {} \\
\hline
\hline
 & & & & & & & & \\[-2ex]
$\langle \tilde{x}\, \tilde{x}\rangle$ & $\times$ & 
$\times$ & $\times$ & $\times$ & $\times$ & 0 & $\times$ & $\cos$ \\
{} & {} & {}
& {} & {} & 0 & 0 & 0 & $\sin$\\
\hline 
$\langle \tilde{x}\, \tilde{y}\rangle$ & 0 & 0 & 0 & 0 & 0
& 0 & 0 & $\cos$ \\ 
{} & {} & {}
& {} & {} & $\times$ & 0  & $\times$ & $\sin$\\ \hline
$\langle \tilde{y}\, \tilde{y}\rangle$ & $\times$ & 
$\times$  & $\times$ & $\times$ & $\times$
& 0 & $\times$ & $\cos$\\ 
{} & {} & {}
& {} & {} & 0 & 0 & 0 & $\sin$\\
\hline
$\langle \tilde{x}\, \tilde{z}\rangle$ & $\times$ & 0 & $\times$ 
& $\times$ & $\times$
& 0 & $\times$ & $\cos$ \\ 
{} & {} & {} & {} & {}
& 0 & 0 & 0 & $\sin$\\ \hline
$\langle \tilde{y}\, \tilde{z}\rangle$ & 0 & 0 & 0 & 0 & 0
& 0 & 0 & $\cos$\\ 
{} & {} & {} & {} & {}
& $\times$ & 0 & $\times$ & $\sin$\\
\hline
$\langle \tilde{z}\, \tilde{z}\rangle$ & $\times$ & $\times$ 
& $\times$ & $\times$ & $\times$ 
& 0 & $\times$ & $\cos$ \\ 
{} & {} & {} & {} & {}
& 0 & 0 & 0 & $\sin$\\ \hline
$\langle \tilde{x}\, \tilde{t}\rangle$ & $\times$ & 0 & 0
& $\times$
& 0 & $\times$ & $\times$ & $\cos$\\ 
{} & {} & {} & {} & {}
& 0 & 0 & 0 & $\sin$\\
\hline
$\langle \tilde{y}\, \tilde{t}\rangle$ & 0 & 0 & 0 & 0 & 0
& 0 & 0 & $\cos$\\ 
{} & {} & {} & {} & {}
& 0 & $\times$ & $\times$ & $\sin$\\
\hline
$\langle \tilde{z}\, \tilde{t}\rangle$ & $\times$ & $\times$ & 0 & $\times$ 
& 0 & $\times$ & $\times$ & $\cos$\\ 
{} & {} & {} & {} & {}
& 0 & 0 & 0 & $\sin$ \\
\hline
$\langle \tilde{t}\, \tilde{t}\rangle$ & $\times$ & $\times$ & $\times$ 
& $\times$ & $\times$
& 0 & $\times$ & $\cos$ \\
{} & {} & {} & {} & {}
& 0 & 0 & 0 & $\sin$ \\ 
\hline \hline
\end{tabular}}
\begin{tabnote}
The non-vanishing ($\times$) and vanishing ($0$) space-time variances
for central ($\bm{b} = 0$) and non-central ($\bm{b} \not= 0$)
collision systems at mid-rapidity ($Y=0$) and away from mid-rapidity.
For non-central collisions, the harmonic coefficients $\langle\dots
\rangle^c_n$ and $\langle\dots \rangle^s_n$ are listed in the three
rightmost columns. Away from mid-rapidity, no symmetry ensures a 
different behaviour of odd and even harmonics.
\end{tabnote}
\end{table}
%
\subsubsection{Explicit azimuthal dependences at mid-rapidity}
\begin{figure}[t]\epsfxsize=11cm 
\centerline{\epsfbox{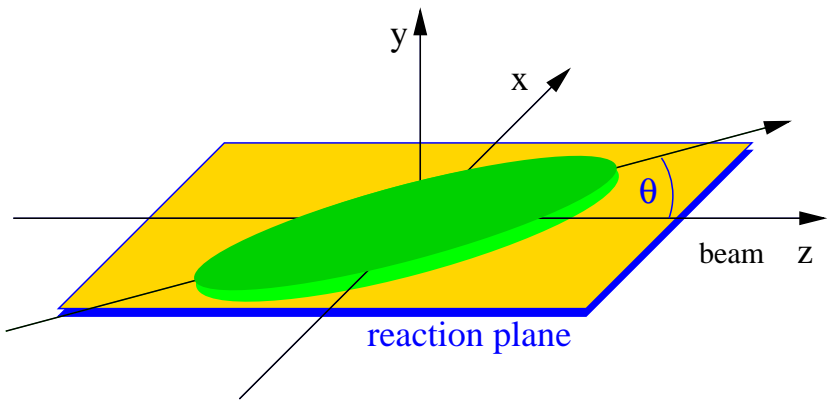}}
\caption{In general, the main axis of the emission ellipsoid 
is tilted for finite impact parameter collisions by an angle
$\Theta$ with respect to the beam axis. This tilt can be
measured via the first harmonics of the off-diagonal
HBT radius parameters $R_{ol}^2$ and $R_{sl}^2$ at
mid-rapidity. 
}\label{tilttilt}
\end{figure}

As long as $\Phi$-dependent position-momentum correlations 
in the source are weak compared to the global essentially geometrical 
asymmetry of the finite impact parameter collision, the implicit
$\Phi$-dependence can be neglected 
compared to the explicit one. In this case, all space-time
variances (\ref{eq34}) are characterised by their zeroth Fourier 
components. Inserting the zeroth components listed in Table~\ref{table1}
into the expressions for the HBT radius parameters 
(\ref{eq27})--(\ref{eq32}), one finds\cite{Wiedemann:1997cr,Lisa:2000ip}:
  \begin{eqnarray}
    \hspace{-0.2cm}
    && R_s^2(K_\perp,\Phi,Y) = {\textstyle\frac{1}{2}}
                  \left( \langle\tilde{x}^2\rangle 
                         + \langle\tilde{y}^2\rangle\right)
                  + {\textstyle\frac{1}{2}} 
                  \left( \langle\tilde{y}^2\rangle 
                          - \langle \tilde{x}^2\rangle\right)
                       \cos 2\Phi \, ,
                      \label{eq43} \\
    \hspace{-0.2cm}
    && R_o^2(K_\perp,\Phi,Y) = {\textstyle\frac{1}{2}}
                  \left( \langle\tilde{x}^2\rangle 
                         + \langle\tilde{y}^2\rangle\right)
                  - {\textstyle\frac{1}{2}} 
                  \left( \langle\tilde{y}^2\rangle 
                          - \langle \tilde{x}^2\rangle\right)
                  \cos 2\Phi
                  + \beta_\perp^2 \langle \tilde{t}^2\rangle\, ,
                      \label{eq44} \\
    \hspace{-0.2cm}
    && R_{os}^2(K_\perp,\Phi,Y) = 
                  {\textstyle\frac{1}{2} }
                  (\langle \tilde{y}^2\rangle - \langle \tilde{x}^2\rangle)
                  \sin 2\Phi\, , 
                  \label{eq45} \\
    \hspace{-0.2cm}
    && R_{l}^2(K_\perp,\Phi,Y) = 
                  \langle \tilde{z}^2\rangle + 
                  \beta_l^2 \langle \tilde{t}^2 \rangle \, ,
                      \label{eq46} \\
    \hspace{-0.2cm}
    && R_{ol}^2(K_\perp,\Phi,Y) = 
                  \langle \tilde{x} \tilde{z}\rangle\, \cos \Phi\, ,
                      \label{eq47} \\
    \hspace{-0.2cm}
    && R_{sl}^2(K_\perp,\Phi,Y) = 
                 - \langle \tilde{x} \tilde{z}
                     \rangle\, \sin\Phi \, .
    \label{eq48}
  \end{eqnarray}
Two $\Phi$-dependent geometrical informations are contained in these
equations:
\newpage
\begin{enumerate}
\item \underline{Eccentricity of the emission 
ellipsoid in the transverse plane}\cite{Wiedemann:1997cr}\\
The HBT radius parameters $R_o^2$, $R_s^2$ and $R_{os}^2$ are 
sensitive to the extension of the emission region in the transverse
plane. They show second harmonic oscillations of the same
strength $\frac{1}{2} \left( \langle\tilde{x}^2\rangle 
- \langle \tilde{y}^2\rangle\right)$ and thus measure the 
eccentricity of the emission ellipsoid in the transverse plane.
The second harmonic coefficients of HBT radius parameters thus 
satisfy the rules
  \begin{gather}
    {R_{o,2}^s}^2 = 0\, ,\quad  
    {R_{s,2}^s}^2 = 0\, ,\quad   
    {R_{os,2}^c}^2 = 0\, , 
    \label{eq50}\\
    {R_{o,2}^c}^2 = - {R_{s,2}^c}^2 = -{R_{os,2}^s}^2\, . 
    \label{eq49}
  \end{gather}
While \eqref{eq50} must be always satisfied at mid-rapidity 
because of the symmetry requirements
(see also next section),
deviations from \eqref{eq49} are an unambiguous sign for implicit
$\Phi$-dependences of space-time variances, i.e. for 
azimuthally dependent position-momentum correlations in the 
source.
\item \underline{Tilt of the emission ellipsoid in 
the reaction plane}\cite{Lisa:2000ip}\\
The HBT radius parameters $R_{ol}^2$ and $R_{sl}^2$ are sensitive to the 
off-diagonal component  $\langle \tilde{x} \tilde{z}\rangle$. 
They measure the angle $\Theta$ by which the
major longitudinal axis of the emission ellipsoid is tilted away
from the beam direction, see Fig.~\ref{tilttilt}, 
\begin{equation}
  \Theta = \frac{1}{2} \tan^{-1} \left(
    \frac{2 \langle \tilde{x} \tilde{z}\rangle}{
      \langle \tilde{z} \tilde{z}\rangle - \langle \tilde{x} \tilde{x}\rangle}
    \right)\, .
    \label{eq51}
\end{equation}
The tilt $\Theta$ 
 is measured from the first harmonic oscillations at
mid-rapidity. Any deviation from the rule
\begin{equation}
  {R_{ol,1}^c}^2 =  - {R_{sl,1}^s}^2\, .
  \label{eq52} 
\end{equation}
is an unambiguous sign for strong azimuthally dependent 
position-momentum correlations in the source. Again, symmetry 
constraints generally require 
\begin{eqnarray}
  {R_{ol,1}^s}^2 &=& 0\, ,
  \label{eq52b}\\
  {R_{sl,1}^c}^2 &=& 0\, .
  \label{eq53} 
\end{eqnarray}

\end{enumerate}

%

\subsubsection{Implicit azimuthal dependence at mid-rapidity}
To access the effect of an implicit $\Phi$-dependence of
space-time variances, we expand the radius parameters 
(\ref{eq27})--(\ref{eq32}) in power of the harmonic 
coefficients\cite{Wiedemann:1997cr,Heinz:2002au}. The
following discussion includes all harmonic contributions
allowed by the symmetries. It may simplify 
further if the emission function is sufficiently smoothly
shaped for higher harmonic coefficients to be negligible,
\begin{equation}
  \langle \tilde x_\mu \tilde x_\nu \rangle_n^{s/c} \gg
  \langle \tilde x_\mu \tilde x_\nu \rangle_{2+n}^{s/c}\, .
  \label{eq54}
\end{equation}
Using the Fourier expansion (\ref{eq34}) of space-time variances
at mid-rapidity, one finds:\\
1. \underline{The zeroth moments of the HBT radius parameters}:
\begin{eqnarray}
  R_{s,0}^2 &=& \frac{1}{2} \left( \langle {\tilde x}^2\rangle_0
                                   + \langle {\tilde y}^2\rangle_0\right)
            + \frac{1}{4} \langle {\tilde y}^2\rangle_2^c
              - \frac{1}{2} \langle {\tilde x} \tilde y \rangle_2^s
              - \frac{1}{4} \langle {\tilde x}^2\rangle_2^c\, ,
              \label{eq55} \\
  R_{o,0}^2 &=& \frac{1}{2} \left( \langle {\tilde x}^2\rangle_0
                                   + \langle {\tilde y}^2\rangle_0 \right)
              + \beta_\perp^2 \langle {\tilde t}^2\rangle_0
             + \frac{1}{4} \langle {\tilde x}^2\rangle_2^c
              + \frac{1}{2} \langle {\tilde x} \tilde y \rangle_2^s
              - \frac{1}{4} \langle {\tilde y}^2\rangle_2^c
              \nonumber \\
            && - \beta_\perp \left( \langle {\tilde x} \tilde t\rangle_1^c
              + \langle {\tilde y} \tilde t\rangle_1^s \right)\, ,
              \label{eq56} \\
  R_{l,0}^2 &=& \langle {\tilde z}^2\rangle_0\, .
               \label{eq57}
\end{eqnarray}
These expressions show that contributions to the absolute size of 
azimuthally averaged HBT radius parameters stem not only from 
the azimuthal averages $\langle \tilde x_\mu \tilde x_\nu \rangle_0$,
but also from the harmonic oscillations of these space-time variances.
Nevertheless, only those Fourier components of the HBT radius parameters
are non-zero, which were found to be non-zero in the case without
implicit $\Phi$-dependence. This continues to be the case for

\noindent
\underline{2. The first moments of the HBT radius parameters}:\\
The only non-vanishing first moments are
\begin{eqnarray}
  {R_{ol,1}^c}^2 &=& \langle \tilde x \tilde z\rangle_0
  + \frac{1}{2} \left(\langle \tilde x \tilde z\rangle_2^c
                     + \langle \tilde y \tilde z\rangle_2^s\right)
     - \beta_\perp\, \langle \tilde z \tilde t\rangle_1^c\, ,
     \label{eq58} \\
  {R_{sl,1}^c}^2 &=& - \langle \tilde x \tilde z\rangle_0
  + \frac{1}{2} \left(\langle \tilde x \tilde z\rangle_2^c
                     + \langle \tilde y \tilde z\rangle_2^s\right)\, .
   \label{eq59}
\end{eqnarray}
For the case of a static source, these quantities
determine the tilt angle (\ref{eq51}) and satisfy
${R_{ol,1}^c}^2 =  - {R_{sl,1}^s}^2$. According to eqs.
(\ref{eq58}) and (\ref{eq59}), a non-zero value of
\begin{equation}
  {R_{ol,1}^c}^2 + {R_{sl,1}^c}^2 = \langle \tilde x \tilde z\rangle_2^c
                     + \langle \tilde y \tilde z\rangle_2^s
                      - \beta_\perp\, \langle \tilde z \tilde t\rangle_1^c
                      \label{eq60}
\end{equation}
is an unambiguous sign of dynamically generated $\Phi$-dependent
correlations in the source.

\noindent
\underline{3. The second moments of the HBT radius parameters}:\\
The only non-vanishing second moments are 
\begin{eqnarray}
  {R_{s,2}^c}^2 &=& \frac{1}{2} \left( \langle {\tilde y}^2 \rangle_0
                           - \langle {\tilde x}^2 \rangle_0\right)
                  + \frac{1}{2} \left( \langle {\tilde y}^2 \rangle_2^c
                           + \langle {\tilde x}^2 \rangle_2^c\right)
                \nonumber \\ &&
                {} - \frac{1}{4} \left ( \la\tx^2\ra_4^c - \la\ty^2\ra_4^c\right )
                   - \frac{1}{2} \la \tx\ty\ra_4^s \, ,
                   \label{eq61} \\
  {R_{o,2}^c}^2 &=& -\frac{1}{2} \left( \langle {\tilde y}^2 \rangle_0
                           - \langle {\tilde x}^2 \rangle_0\right)
                  + \frac{1}{2} \left( \langle {\tilde y}^2 \rangle_2^c
                           + \langle {\tilde x}^2 \rangle_2^c\right)
                         \nonumber \\
             && {} - \beta_\perp \left( \langle \tilde x \tilde t\rangle_1^c
                             -  \langle \tilde y \tilde t\rangle_1^s \right)
               + \beta_\perp^2  \langle {\tilde t}^2 \rangle_2^c
                        \nonumber \\
             && {} - \beta_\perp\left ( \la \tx\ttt\ra_3^c + \la\ty\ttt\ra_3^s \right )
                 + \frac{1}{4} \left(\la \tx^2\ra_4^c - \la\ty^2\ra_4^c\right )
                + \frac{1}{2} \la \tx\ty\ra_4^s \, ,
               \label{eq62} \\
  {R_{os,2}^s}^2 &=& \frac{1}{2} \left( \langle {\tilde y}^2 \rangle_0
                           - \langle {\tilde x}^2 \rangle_0\right)
               + \frac{\beta_\perp}{2} 
                 \left( \langle \tilde x \tilde t\rangle_1^c
                         -  \langle \tilde y \tilde t\rangle_1^s \right)
                \nonumber \\ 
                && {} - \frac{\beta_\perp}{2} 
                        \left ( \la \tx\ttt\ra_3^c  + \la\ty\ttt\ra_3^s \right)
                - \frac{1}{4} \left(\la\tx^2\ra_4^c - \la\ty^2\ra_4^c\right)
                + \frac{1}{2} \la \tx\ty \ra_4^s\, ,
                         \label{eq63} \\
  {R_{l,2}^c}^2 &=& \langle {\tilde z}^2 \rangle_2^c\, .
  \label{eq64}
\end{eqnarray}
The equations (\ref{eq50}) remain true in the presence of $\Phi$-dependent
position-momentum gradients in the source, while deviations from the
${R_{o,2}^c}^2 = - {R_{s,2}^c}^2 = -{R_{os,2}^s}^2$ rule (\ref{eq49}) 
are an unambiguous sign for an implicit $\Phi$-dependence, namely:
\begin{eqnarray}
  {R_{o,2}^c}^2 + {R_{s,2}^c}^2 &=&
  \langle {\tilde x}^2 \rangle_2^c + \langle {\tilde y}^2 \rangle_2^c
  + \beta_\perp^2 \langle {\tilde t}^2 \rangle_2^c
  \nonumber \\ && 
  - \beta_\perp \left( \langle \tilde x \tilde t\rangle_1^c
                      - \langle \tilde y \tilde t\rangle_1^s 
                      + \la \tx\ttt \ra_3^c + \la \ty\ttt\ra_3^s\right)\, ,
                    \label{eq65} \\
      {R_{o,2}^c}^2 + {R_{os,2}^s}^2 &=& \frac{1}{2} \left(
  \langle {\tilde x}^2 \rangle_2^c + \langle {\tilde y}^2 \rangle_2^c \right)
  + \beta_\perp^2 \langle {\tilde t}^2 \rangle_2^c
  - \frac{\beta_\perp}{2}  \left( \langle \tilde x \tilde t\rangle_1^c
                      - \langle \tilde y \tilde t\rangle_1^s \right)
  + \la \tx\ty \ra_4^s
  \nonumber \\ &&
  {} - \frac{3}{2} \beta_\perp\left(\la\tx\ttt\ra_3^c+\la\ty\ttt\ra_3^s\right)
  + \frac{1}{2}\left( \la\tx^2\ra_4^c - \la\ty^2\ra_4^c\right )
   \, ,
                    \label{eq66} \\
      {R_{s,2}^c}^2 - {R_{os,2}^s}^2 &=& \frac{1}{2} \left(
  \langle {\tilde x}^2 \rangle_2^c + \langle {\tilde y}^2 \rangle_2^c \right)
  - \frac{1}{2} \beta_\perp \left( \langle \tilde x \tilde t\rangle_1^c
                      - \langle \tilde y \tilde t\rangle_1^s 
                      - \la \tx\ttt\ra_3^c -\la \ty\ttt \ra_3^s\right)
   \nonumber \\ &&
   {} - \frac{1}{2} \left(\la \tx^2\ra_4^c - \la\ty^2\ra_4^c\right)
   - \la \tx\ty \ra_4^s \, .
                    \label{eq67}
\end{eqnarray}
If the fourth order moments $\la \tx^2\ra_4^c - \la\ty^2\ra_4^c$
and $\la \tx\ty \ra_4^s$ are negligible and if the transverse pair 
momentum $K_\perp$ is sufficiently small, then 
\begin{equation}
  {R_{s,2}^c}^2 - {R_{os,2}^s}^2 \approx
  {R_{o,2}^c}^2 + {R_{os,2}^s}^2 
  \, .
  \label{eq67b}
\end{equation}

\noindent
\underline{4. The $n$-th moments of the HBT radius parameters ($n>2$)}:\\
In general, the properties of the $\sin$ and $\cos$ functions
imply that the $n$-th order harmonics ${R_{ij,n}^s}^2$,
${R_{ij,n}^c}^2$ in eqs. \eqref{eq41}, \eqref{eq42} are built
up of $m$-th order harmonics $\la \tilde{x}_\mu \tilde{y}_\nu\ra_m^s$,
$\la \tilde{x}_\mu \tilde{y}_\nu\ra_m^c$ with $n-2 \leq m \leq n+2$.
This limits the number of terms appearing in $n$-th order expressions.
Since it is an open question whether the approximation (\ref{eq54})
applies for realistic sources\footnote{We thank Mike Lisa for drawing 
our attention to this issue.}, we give these expressions here for
completeness.

For the {\it side-long} and {\it out-long} HBT radius parameters, 
only odd harmonic terms appear. For $n = 3,\, 5,\, 7,\, \dots$, 
we have
\begin{eqnarray}
{R^c_{ol,n}}^2 & = & \frac{1}{2} \left ( \la\tilde x\tilde z\ra_{n-1}^c 
  - \la \tilde y \tilde z \ra_{n-1}^s \right )
  - \beta_\perp \la \tilde z \tilde t \ra_n^c  
\nonumber \\ &&
  + \frac{1}{2} \left ( \la\tilde x\tilde z\ra_{n+1}^c 
  + \la \tilde y \tilde z \ra_{n+1}^s \right ) \, ,
\label{Rol-n}\\
{R^s_{sl,n}}^2 & = & \frac{1}{2} \left ( \la \tilde y \tilde z\ra_{n-1}^s -
\la \tilde x \tilde z \ra_{n-1}^c \right ) 
+ \frac{1}{2} \left ( \la \ty \tz \ra_{n+1}^s + \la \tx\tz\ra_{n+1}^c \right )
\, .
\label{Rsl-n} 
\end{eqnarray}
The Fourier decomposition of the other HBT radius parameters contains
only even harmonic terms. For $n=4,\, 6,\, 8,\, \dots$, we have 
\begin{eqnarray}
{R^c_{s,n}}^2 & = & 
-\frac{1}{4} \left( \la \tx^2 \ra_{n-2}^c - \la \ty^2\ra_{n-2}^c \right )
+ \frac{1}{2}\la \tx\ty \ra_{n-2}^s 
\nonumber \\ &&
+ \frac{1}{2} \left ( \la \tx^2\ra_n^c + \la \ty^2\ra_n^c \right ) 
\nonumber \\ &&
{} - \frac{1}{4} \left( \la \tx^2 \ra_{n+2}^c - \la \ty^2 \ra_{n+2}^c \right )
- \frac{1}{2} \la \tx\ty \ra_{n+2}^s \, , 
\label{Rs-n} \\ 
{R^c_{o,n}}^2 & = & 
\frac{1}{4} \left ( \la \tx^2 \ra_{n-2}^c - \la \ty^2 \ra_{n-2}^c \right )
- \frac{1}{2} \la \tx\ty \ra_{n-2}^s
\nonumber \\ &&
{}-\beta_\perp \left(\la \tx\ttt \ra_{n-1}^c - \la \ty\ttt\ra_{n-1}^s \right)
\nonumber \\ &&
{} + \frac{1}{2} \left ( \la \tx^2 \ra_{n}^c + \la \ty^2 \ra_{n}^c \right )
+ \beta_\perp^2 \la \ttt^2 \ra_n^c
\nonumber \\ &&
{}-\beta_\perp \left(\la\tx\ttt\ra_{n+1}^c + \la \ty\ttt\ra_{n+1}^s \right)
\nonumber \\ &&
{} + \frac{1}{4} \left(\la \tx^2 \ra_{n+2}^c - \la \ty^2 \ra_{n+2}^c \right)
+ \frac{1}{2}  \la \tx\ty \ra_{n+2}^s\, ,
\\
\label{Ro-n}
{R^s_{os,n}}^2 & = & 
-\frac{1}{4} \left (\la\tx^2\ra_{n-2}^c - \la\ty^2\ra_{n-2}^c\right ) 
+ \frac{1}{2} \la \tx \ty \ra_{n-2}^s
\nonumber \\ &&
{} + \frac{1}{2} \beta_\perp \left(\la\tx\ttt \ra_{n-1}^c -\la\ty\ttt\ra_{n-1}^s\right)
\nonumber \\ &&
{} - \frac{1}{2} \beta_\perp \left(\la\tx\ttt \ra_{n+1}^c +\la\ty\ttt\ra_{n+1}^s\right)
\nonumber \\ &&
{} - \frac{1}{4} \left (\la\tx^2\ra_{n+2}^c - \la\ty^2\ra_{n+2}^c\right )
+ \frac{1}{2} \la\tx\ty\ra_{n+2}^s \, , 
\label{Ros-n} \\ 
{R^c_{l,n}}^2 & = & \la \tz^2 \ra_n^c \, .
\label{Rl-n}
\end{eqnarray}
Whether these higher harmonic coefficients are
numerically important remains to be established experimentally.

\subsubsection{Implicit azimuthal dependence at forward rapidity}

At non-central rapidity $Y\not= 0$, additional Fourier components of 
the space-time variances can contribute to HBT radius parameters, as 
is seen from Table~\ref{table1}. The reason is that at $Y\not=0$, the
symmetries (\ref{eq38})--(\ref{eq40}) do not hold. It is remarkable 
that the zeroth and second moments of the out, side and out-side
radius parameters, given in eqs. (\ref{eq55}), (\ref{eq56}),
(\ref{eq61}), (\ref{eq62}) and (\ref{eq63}), do not receive extra
contributions at forward rapidity. In particular, this implies that
deviations from the ${R_{o,2}^c}^2 = - {R_{s,2}^c}^2 = {R_{os,2}^s}^2$
rule of (\ref{eq49}) remains an unambiguous test for the presence of 
angular dependent position-momentum correlations at all rapidities.
In addition, the out, side and out-side radius parameters
acquire first harmonic oscillations away from mid-rapidity, 
\begin{eqnarray}
  {R_{s,1}^c}^2 &=& \frac{1}{4} \left(
    \langle {\tilde x}^2 \rangle_1^c - 2 \langle \tilde x
    \tilde y \rangle_1^s + 3 \langle {\tilde y}^2\rangle_1^c \right)
        \nonumber \\ &&
    - \frac{1}{4} \left(
     \la \tx^2\ra_3^c + 2 \la \tx\ty \ra_3^s - \la \ty^2\ra_3^c\right )\, ,
  \label{eq68} \\
  {R_{o,1}^c}^2 &=& \frac{1}{4} \left(
    3\langle {\tilde x}^2 \rangle_1^c + 2 \langle \tilde x
    \tilde y \rangle_1^s + \langle {\tilde y}^2\rangle_1^c \right)
  \nonumber \\
   && {}- \beta_\perp \left( 2\langle {\tilde x} \tilde t \rangle_0 
    + \langle \tilde x \tilde t \rangle_2^c 
    + \langle \tilde y \tilde t \rangle_2^c  \right)
    + \beta_\perp^2 \langle {\tilde t}^2\rangle_1^c
        \nonumber \\ &&
    {} + \frac{1}{4} \left ( 
    \la \tx^2\ra_3^c + 2 \la \tx\ty\ra_3^s - \la \ty^2 \ra_3^c \right)\, ,
    \label{eq69} \\
  {R_{os,1}^s}^2 &=& \frac{1}{4} \left(
    - \langle {\tilde x}^2 \rangle_1^c - 2 \langle \tilde x
    \tilde y \rangle_1^s + \langle {\tilde y}^2\rangle_1^c \right)
  \nonumber \\
   && + \beta_\perp \langle {\tilde x} \tilde t \rangle_0 
    - \frac{1}{2} \beta_\perp \left(  \langle \tilde x \tilde t \rangle_2^c 
    + \langle \tilde y \tilde t \rangle_2^c  \right)
        \nonumber \\ &&
    {} + \frac{1}{4} \left (
    \la \tx^2\ra_3^c + 2 \la \tx\ty\ra_3^s - \la \ty^2\ra_3^c \right )\, .
    \label{eq70} 
\end{eqnarray}
In the so-called blast wave model\cite{star02-HBT}),
there is no correlation between the transverse 
position and the time at which particles are emitted. Hence, the
space-time variances linear in $\tilde t$ vanish. Also,
the emission duration $\langle {\tilde t}^2\rangle$ does not depend
on the azimuthal angle. In addition, if the source shows a
sufficiently smooth azimuthal dependence for the third order terms 
$\la \tx^2\ra_3^c$, $\la \tx\ty\ra_3^s$,  $\la \ty^2\ra_3^c$ to be
negligible, then
\begin{eqnarray}
  {R_{s,1}^c}^2 - {R_{o,1}^c}^2 \approx 2 {R_{os,1}^s}^2\, .
  \label{eq71}
\end{eqnarray}

Moreover, it was observed in a class of model 
studies\cite{Wiedemann:1997cr}
that $\langle {\tilde x}^2 \rangle_1^c$ should be much larger than
$\langle {\tilde x} \tilde y \rangle_1^s$ and 
$\langle {\tilde y}^2 \rangle_1^c$, since asymmetries with respect to 
the beam axis will occur predominantly in the direction of the 
impact parameter. This translates into the rule
\begin{eqnarray}
  {R_{o,1}^c}^2 \, :\, {R_{s,1}^c}^2 \, :\, {R_{os,1}^s}^2
  = 3\, :\, 1\, :\, -1\, .
  \label{eq72}
\end{eqnarray}
A test of (\ref{eq71}), (\ref{eq72}) allows to establish whether
these additional model-dependent assumptions are satisfied.

The three HBT radius parameters $R_l^2$, $R_{ol}^2$ and $R_{sl}^2$
involve longitudinal information and depend on the longitudinal 
velocity $\beta_l$. This leads to additional contributions
away from mid-rapidity. For completeness, we
list here the non-vanishing first and second moments:
\begin{eqnarray}
  {R_{l,1}^c}^2 &=& \langle {\tilde z}^2 \rangle_1^c
                     - 2\beta_l \langle {\tilde z}\tilde t \rangle_1^c
                     + \beta_l^2 \langle {\tilde t}^2 \rangle_1^c\, ,
                     \label{eq73} \\
  {R_{l,2}^c}^2 &=& \langle {\tilde z}^2 \rangle_2^c
                     - 2\beta_l \langle {\tilde z}\tilde t \rangle_2^c
                     + \beta_l^2 \langle {\tilde t}^2 \rangle_2^c\, ,
                     \label{eq74} \\
  {R_{ol,1}^c}^2 &=& \langle \tilde x {\tilde z} \rangle_0
       + \frac{1}{2}\left( \langle \tilde x {\tilde z} \rangle_2^c
                          + \langle \tilde y {\tilde z} \rangle_2^s\right)
       - \beta_\perp \langle \tilde z {\tilde t} \rangle_1^c
       \nonumber \\
       && + \frac{1}{2} \beta_l \left( 2\beta_\perp 
         \langle {\tilde t}^2 \rangle_1^c 
         - 2 \langle \tilde x {\tilde t} \rangle_0
         - \langle \tilde x {\tilde t} \rangle_2^c
         - \langle \tilde y {\tilde t} \rangle_2^s \right)\, ,
       \label{eq75} \\
  {R_{ol,2}^c}^2 &=& \frac{1}{2} \left(
                     \langle \tilde x {\tilde z} \rangle_1^c
                    - \langle \tilde y {\tilde z} \rangle_1^s\right)
                  - \beta_\perp \langle \tilde z {\tilde t} \rangle_2^c
       \nonumber \\
       && + \frac{1}{2} \beta_l \left( 2\beta_\perp 
         \langle {\tilde t}^2 \rangle_2^c 
         - \langle \tilde x {\tilde t} \rangle_1^c
         + \langle \tilde y {\tilde t} \rangle_1^s \right)
        \nonumber \\ &&
        {} + \frac{1}{2} \left (
        \la \tx\tz\ra_3^c + \la\ty\tz\ra_3^s \right) - \frac{\beta_l}{3}
        \left ( \la \tx\ttt\ra_3^c + \la \ty\ttt\ra_3^s \right)\, ,
       \label{eq76} \\
  {R_{sl,1}^s}^2 &=& - \langle \tilde x {\tilde z} \rangle_0
       + \frac{1}{2}\left( \langle \tilde x {\tilde z} \rangle_2^c
                          + \langle \tilde y {\tilde z} \rangle_2^s\right)
       \nonumber \\ 
       && {} + \frac{1}{2} \beta_l \left( 
          2 \langle \tilde x {\tilde t} \rangle_0
         - \langle \tilde x {\tilde t} \rangle_2^c
         - \langle \tilde y {\tilde t} \rangle_2^s \right)\, ,
       \label{eq77} \\  
  {R_{sl,2}^s}^2 &=& 
       \frac{1}{2}\left( \langle \tilde y {\tilde z} \rangle_1^s
                          - \langle \tilde x {\tilde z} \rangle_1^c\right)
       + \frac{1}{2} \beta_l \left( 
         \langle \tilde x {\tilde t} \rangle_1^c
         - \langle \tilde y {\tilde t} \rangle_1^s \right)
        \nonumber \\ && {}
        + \frac{1}{2} \left ( \la \tx\tz\ra_3^c + \la \ty\tz\ra_3^s \right)
        - \frac{\beta_l}{2} \left (
        \la \tx\ttt\ra_3^c + \la \ty\ttt\ra_3^s \right )\, .
       \label{eq78}
\end{eqnarray}
%
\subsubsection{Reconstruction of the reaction plane}
The above analysis of HBT radius parameters for non-central collisions
requires the measurement of the angle $\Phi$ and thus assumes knowledge 
about the event-wise orientation of the reaction plane. This orientation
$\Psi_R$ is usually measured from the azimuthal dependence of 
single-particle transverse momentum 
spectra\cite{Ollitrault:1997vz,Poskanzer:1998yz}
  \begin{eqnarray}
    E\frac{dN}{d^3p} &=&
    \frac{d^3N}{p_t\, dp_t\, dy\, d\phi} = \int d^4x S(x,p)
    \nonumber \\
    &=& {\frac{1}{2\pi}} \frac{d^2N}{p_t\, dp_t\, dy}
        \lbrack 1 + 2\sum_{n=1}^\infty v_n\cos n(\phi -\psi_R) 
                \rbrack\, .
    \label{eq79}
  \end{eqnarray}      
However, the orientation of the true reaction plane can only 
be measured with limited accuracy. 
Since fluctuations in a finite multiplicity environment result
in azimuthal anisotropies without geometrical origin, this
limited accuracy arises largely as a consequence of the basic
statistical properties of a mesoscopic system, and cannot be
reduced by larger event samples or refined measurements. 
The resulting uncertainty has to be corrected for if one
aims for a geometrical interpretation of the $\Phi$-dependence
of HBT radius parameters. Such corrections are discussed in
literature\cite{Wiedemann:1997cr,Heinz:2002au}.

\section{Two-particle correlations from model calculations}
\label{sec-models}
The emission function $S(x,K)$ is not determined uniquely by
the correlator $C(\bm{q},\bm{K})$. This is a consequence of
the on-shell constraint (\ref{eq6}) which implies that only
a specific time-average over the emission function, the
so-called relative distance distribution $S_{\bm{K}}(\bm{x})$,
is uniquely measurable\cite{Wiedemann:1999qn}
\begin{eqnarray}
   S_{\bm{K}}(\bm{x}) &=& \int dt\, d(\bm{x}+{\beta}t, t; K)\, ,
   \label{reldist}\\
    d(x,K) &=& \int d^4X\, 
   \frac{S(X+{\textstyle\frac{x}{2}},K)}{\int d^4y\, S(y,K)}\, 
   \frac{S(X-{\textstyle\frac{x}{2}},K)}{\int d^4y\, S(y,K)}\, .
    \label{4.01} 
\end{eqnarray}
The direct reconstruction of $S_{\bm{K}}(\vert \bm{x}\vert)$ 
from experimental data has been pursued 
successfully\cite{Brown:1997sn,Brown:2000aj,Panitkin:2001qb}. However, 
due to statistical uncertainties the numerical inversion of 
$C(\bm{q},\bm{K}) - 1 = \int d^3\bm{x}\, \cos(\bm{q}\cdot \bm{x})\,
S_{\bm{K}}(\bm{x})$ is complicated. In practice, it requires additional
model assumptions to achieve convergence. Thus most data analyses
proceed via model studies. Either they start from a model 
parametrisation of the emission function $S(x,K)$, or they start 
from a dynamical calculation
of $S(x,K)$ based on a hydrodynamic or particle-cascade
based simulation. In this section, we review the main features
of these approaches.
%
\subsection{Model parametrisations of the emission function}
\label{secmodel}
The main features of the collision region at freeze-out can be 
characterised by its width in the different spatial and temporal
extensions, its collective dynamical gradients (usually ascribed to
a collective flow field $u_\mu(x)$ which determines the position-momentum
correlations in the source) and its random dynamical component
(usually ascribed to a local temperature $T$). Model parametrisations 
of $S(x,K)$ implement these main features in a (minimal)
analytical ansatz for $S(x,K)$. The model parameters 
are then extracted from a fit to one- and two-particle spectra.

\noindent
\underline{Example of a model emission function:}
For illustration, consider a source in local thermal equilibrium
at temperature $T$ whose extension is given by the
transverse width $R$, space-time rapidity width $\Delta\eta$ and
proper longitudinal emission time $\tau_0$ smeared with a width
$\Delta\tau$.
The transverse and longitudinal expansion of the collision
system results in a longitudinally boost-invariant flow profile
at freeze-out with a transverse component $\eta_t(r) = \eta_f \frac{r}{R}$
characterised by the transverse gradient $\eta_f$,
 \begin{eqnarray}
   u_\mu(x) &=& \left( \cosh\eta\, \cosh\eta_t, \textstyle\frac{x}{r}
     \sinh\eta_t, {\textstyle\frac{y}{r}} \sinh\eta_t, 
     \sinh\eta\, \cosh\eta_t\right)\, .
     \label{5.6}
 \end{eqnarray}
In longitudinal proper time $\tau = \sqrt{t^2 - z^2}$ and rapidity
$\eta = \textstyle\frac{1}{2} \ln{\lbrack{ (t+z)/(t-z) }\rbrack}$,
this source can be written 
as\cite{Akkelin:sg,Chapman:1994ax,Csorgo:1995bi,Wiedemann:1996ig}: 
 \begin{eqnarray}
   S_r(x,p) &=& \frac{2J_r + 1}{(2\pi)^3 \sqrt{2\pi}\, \Delta\tau}\,
   m_\perp \cosh({\rm y}-\eta) 
   \exp\left[- \frac{p \cdot u(x) - \mu_r}{T} \right]\,  
   \nonumber \\
   && \times \exp\left[ - \frac{r^2}{2 R^2} 
                     - \frac{\eta^2}{2 (\Delta\eta)^2}
                     - \frac{(\tau-\tau_0)^2}{2 (\Delta\tau)^2}
                 \right] \, . 
 \label{5.1}
 \end{eqnarray}
Here, $r$ labels the particle species which are produced in
thermal abundances with chemical potentials $\mu_r$.
The model emission function (\ref{5.1}) is completely
specified by the model parameters $T\, ,\eta_f\, ,R\, ,\Delta\eta\, 
,\Delta\tau\, ,\tau_0\, ,\mu_r$. This basic model allows for a 
satisfactory fit to experimental data from 
the CERN SPS\cite{Tomasik:1999cq}.

\noindent
\underline{Overview of models and model extensions:}
In what follows, we review the physics arguments which motivated
the study of modifications and extensions of the parameterisation 
(\ref{5.1}):
\begin{enumerate}
  \item Varying transverse density and flow profiles:\\
    The functional shape of (\ref{5.1}) was varied by
    replacing the Gaussian transverse density distribution with 
    a box profile\cite{Tomasik:1999cq} or varying the functional 
    dependence of the transverse flow profile\cite{Wiedemann:1995au}. 
    This gives further support to the general statement
    that HBT radius parameters are mainly sensitive to the
    average r.m.s. width of $S(x,K)$. However, details in
    the functional shape of $S(x,K)$ can leave observable
    traces in the $K_\perp$ dependence of the HBT radii: 
    in particular, experimental data from the SPS
    favour a transverse box profile over a Gaussian one.
  \item Surface dominated versus bulk dominated emission:\\
    Model (\ref{5.1}) implements bulk emission, i.e., 
    particles decouple at the same average proper freeze-out 
    time $\tau_0$ from all spatial positions in the source
    with a probability proportional to the source density.
    However, if reabsorption of particles by the surrounding
    matter is significant, hadronic freeze-out may proceed
    via surface evaporation. In analytical parametrisations of
    $S(x,K)$, such ``opaque sources'' have been modelled via
    addition of absorption factors\cite{Heiselberg:1997vh}. 
    The main outcome of these
    studies is that surface-dominated emission can imply 
    $\langle {\tilde x}^2 \rangle \ll \langle {\tilde y}^2 \rangle$
    and this makes it possible 
    that $R_o^2 < R_s^2$ at sufficiently
    large $K_\perp$. 
  \item Temperature gradients:\\
    Models which include a spatially varying local temperature
    attribute position-momentum correlations in the source to 
    a combination of two different effects: $T(x)$ and $u_\mu(x)$.
    This typically introduces two additional fit 
    parameters\cite{Csorgo:1995bi,Tomasik:1997eq} which
    characterise the longitudinal and transverse dependence of $T(x)$.
    It removes to some extent the constraint between temperature
    and transverse flow which can be exploited when fitting (\ref{5.1})
    to a combination of one- and two-particle spectra. Such a model
    was fitted successfully to first data from RHIC\cite{Csorgo:2002ry}.
  \item Model emission functions for finite impact parameter collisions:\\
    The model (\ref{5.1}) has been extended to
    azimuthally asymmetric transverse
    flow and asymmetric transverse Gaussian\cite{Wiedemann:1997cr}  
    or box geometry\cite{Retiere:2001ed}.
    For the first and second harmonics of HBT radius parameters
    calculated in these models,
    explicit $\Phi$-dependence dominates over
    the implicit one and deviations from the purely geometrical 
    identities (\ref{eq49}) and (\ref{eq72}) satisfy 
    the dynamical identities (\ref{eq67b}) and (\ref{eq71}),
    respectively. The latter point, however, is mainly a consequence of 
    studying a class of models for which $t$-dependent space-time
    variances are $\Phi$-independent and azimuthal deformations
    are essentially elliptic. 
\end{enumerate}

\noindent
\underline{Generic properties:}

The analytical parametrisations reviewed above were instrumental in
establishing how the main properties of the two-particle
correlator translate into specific geometric or dynamical 
features of the collision region.
The following generic properties emerge:
\begin{enumerate}
  \item Size and transverse momentum slopes of HBT radii:\\
  Qualitatively, the main information contained in the absolute
  size and $M_\perp$-dependence of HBT radius parameters can be
  illustrated in terms of pocket formulas 
  derived in a saddle point approximation of (\ref{5.1}).
  For the  side radius parameter one
  finds\cite{Chapman:1994ax} for $\Delta \eta = \infty$ and $\Delta \tau = 0$
 \begin{equation}  
    R_s^2(K_\perp) \approx \frac{R^2}{1 + \frac{M_\perp}{T} \eta_f^2} \, ,
    \label{eq106} 
 \end{equation}
The size of this radius is proportional to the source size, but
it is also sensitive to the transverse flow strength $\eta_f$ of the 
source. This illustrates that HBT radii characterise only that part
of a dynamically expanding source which can be viewed through a filter 
of wavelength $K$. This shrinking effect increases for increasing 
$M_\perp$ proportional to the ratio $\eta_f^2/T$. The
Makhlin-Sinyukov formula\cite{Makhlin:1987gm} for the longitudinal radius,
 \begin{eqnarray}  
    R_l^2 &\approx& {\tau}_0^2 \frac{T}{M_{\perp}}\, ,
    \label{eq107}
 \end{eqnarray}
shows compared to (\ref{eq106}) a stronger 
$M_\perp$-dependence consistent with the stronger longitudinal 
expansion implemented in (\ref{5.1}). Its dependence on
$\tau_0$ is a direct consequence of the assumed longitudinal
boost-invariance and receives corrections for sources of 
finite longitudinal extension. 

While quantitative corrections to these analytical expressions
can be significant\cite{Wiedemann:1995au}, these pocket formulas illustrate
qualitatively the interplay of geometry and dynamics in 
determining HBT radius parameters. This picture is supported
by numerous numerical studies. 
\item The difference $R_o^2 - R_s^2$:\\
The main interest in this observable\cite{Pratt:zq} lies in its 
sensitivity to the emission duration [see also discussion of eq.
(\ref{eq22})]
\begin{equation}
 R_o^2 - R_s^2 \approx \beta_\perp^2\, \la \ttt^2 \ra \, . 
 \label{eqt}
\end{equation}
Numerical calculations with a Gaussian density profile typically 
result in a small but positive signal for 
$R_o^2 - R_s^2$. For steeper transverse 
profiles and particle emission at sufficiently large $K_\perp$,
or for opaque source\cite{Heiselberg:1997vh,McLerran:2002dt} 
models with surface dominated emission, 
also negative values can be found for $R_o^2 - R_s^2$. 
Equation~(\ref{eqt}) ignores the contribution from the
$\langle \tilde x \tilde t \rangle$ correlation term which vanishes in the 
model~(\ref{5.1}) but is present in hydrodynamic models and
Monte Carlo event generators, see below. 
\item Influence of resonance decay contributions:\\
Pions from resonance decays have a tendency to be emitted at 
later times and larger 
distances\cite{Grassberger:1976au,Csorgo:1994in,Wiedemann:1996ej}. 
For models showing bulk
emission, their effect on the size of HBT radius parameters
is however small\cite{Wiedemann:1996ej}. This is due to a combination
of three effects: i) in models of the type (\ref{5.1}), the
emission region of the heavier resonances is smaller than
that of direct pions, ii) the large decay widths of the most 
abundant resonances like $\rho$'s and $\Delta$'s and their
non-relativistic velocities imply that these decays occur
within the emission region of the direct pions, iii) resonances
with large lifetime ($\eta$, $\eta'$) decay so far outside
that their decay pions interfere with the directly produced 
ones on a very small relative momentum scale 
($\vert {\bf q}\vert < 1$ MeV) only. This produces a peak of the 
correlation function at $\vert {\bf q}\vert < 1$ MeV which is narrower 
than the experimental resolution 
and thus leads to an apparently reduced intercept $\lambda$ 
of the correlator $C(\bm{q},\bm{K})$ without affecting its shape.
Only pions from $\omega$ decays stem from a resonance
which is neither sufficiently short-lived nor sufficiently long-lived
and thus can affect the shape of the correlator. This spoils a
naive core-halo interpretation and contributes to non-Gaussian deviations
of the two-particle correlator\cite{Wiedemann:1996ig}.
\end{enumerate}

\subsection{Hydrodynamic models}
Hydrodynamic behaviour is an idealised but well-defined 
limiting case of the realistic dynamical evolution of the
collision region in heavy ion collisions. It emerges as the 
zero mean free path limit of a particle cascade. In this limit,
matter in the collision region is treated as an ideal, locally 
thermalised fluid whose dynamics is governed by the relativistic 
hydrodynamic equations. 

\noindent
\underline{Input for simulations:} 
A hydrodynamic model is fully specified by the equation 
of state and the initial conditions. Typically, the 
parametrisation of the latter models the outcome
of an initial pre-equilibrium stage with initial energy density
estimated from the Glauber approach to entropy and energy production
in nucleon-nucleon collisions\cite{Kolb:2000sd}.

\noindent
\underline{Freeze-out criterion:} 
The freeze-out criterion, according to which the hydrodynamic 
simulation is terminated, is another important input in hydrodynamic
model studies. Usually, the freeze-out criterion is set by a
critical energy density or temperature. If the criterion is
satisfied in a fluid cell, the cell is immediately assumed to 
freeze-out. Local properties of this cell
are converted into a thermal ideal gas distribution of hadronic 
resonances with temperature and chemical potential set by the
local energy and baryon density of the simulation. This leads
to a sharp freeze-out along a three-dimensional hypersurface and 
specifies the emission function $S(x,K)$ entering the calculation
of two-particle correlation functions. 
There are ``hybrid models'' in which the earlier hot stage 
of the collision is treated hydrodynamically but the hadronic phase
is modelled with a Monte Carlo event generator code\cite{Soff:2000eh}.
An event generator naturally leads to an emission function in 
a finite four-volume, see next subsection.

\noindent
\underline{Successes and problems at RHIC:} 
At RHIC, hydrodynamic simulations compare in general well 
with the hadronic one-particle transverse momentum spectra
up to $\approx 2$ GeV. The major success of this approach is 
the prediction of the size of the measured elliptic flow $v_2$, 
as well as the correct description of its $p_\perp$-dependence
for identified pion and proton spectra. This indicates that
the main contribution to elliptic flow originates in the early 
stages of the collision where the system is very dense and the
mean free path is close to the hydrodynamic limit zero.
However, two-particle correlations 
are determined at freeze-out, where the mean free path (or rather the mean
scattering time) is grown 
and a hydrodynamic picture becomes questionable. This may
be one of the reasons why so far hydrodynamic simulations 
have significant problems in calculating two-particle correlators
which are at least in qualitative agreement with experimental
data\cite{Heinz:2002un}, see the following discussion.

\begin{table}[t]
\tbl{The main hydrodynamic model calculations with published
results on HBT correlation functions. Codes follow either the full
three-dimensional expansion or the two-dimensional expansion in the
transverse plane (with assumed 
boost-invariance in the remaining longitudinal direction).
\label{T:hydro}}
{\begin{tabular}{|clcc|}
\hline\hline  && &  \\[-1.5ex]
Authors${}^{\mbox{\small ref.}}$ & energies & HBT data & Dim. \\[1ex]
   & studied & compared to &  \\[1ex]
\hline & & &  \\[-1.5ex]
HYLANDER\cite{Schlei:1992jj,Bolz:1992hc} & SPS & NA44\cite{Schlei:1998zy} & (3+1)-dim  \\[2ex]
Rischke, Gyulassy\cite{Rischke:1996em} & SPS, RHIC & RHIC prediction & (2+1)-dim \\[2ex]
Zschiesche {\it et al.}\cite{Zschiesche:2001dx} & SPS, RHIC & NA49, STAR & (2+1)-dim \\[2ex]
Kolb, Heinz\cite{Heinz:2002un} & RHIC & STAR, PHENIX\cite{Heinz:2002un} 
& (2+1)-dim \\[2ex]
Hirano, Morita {\it et al.} & SPS, RHIC & NA49\cite{Morita:1999vj,Morita:2002av}, 
STAR\cite{Hirano:2001yi,Morita:2002av} & (3+1)-dim\\[2ex]
\hline\hline
\end{tabular}}
\end{table}

\noindent
\underline{Generic properties of hydrodynamic simulations for HBT:}
An overview of hydrodynamic model calculations is given
in Table \ref{T:hydro}. This list is limited to studies 
which include beyond the calculation of one-particle spectra
also two-particle correlation functions. 
\begin{enumerate}
  \item Freeze-out hyper-surface shows strong outward-temporal 
        correlations:\\
        The freeze-out criterion implemented in hydrodynamic 
        simulations amounts to a sudden switch from a zero mean
        free path to an infinite mean free path approximation.
        This tends to favour sharp geometrical correlations along
        the freeze-out hyper-surface. In comparison
        to model sources with emission from a finite four-volume, 
        the $\langle \tilde x\tilde t \rangle$ variance is
        significantly stronger\cite{Rischke:1995cm}. Thus,
        the difference
        $R_o^2 - R_s^2$ does not measure a lifetime effect only.
        This consequence of a sharply localised freeze-out
        hyper-surface may be tamed in hybrid models in which
        hydrodynamic evolution is followed by a hadronic
        rescattering phase\cite{Soff:2000eh}.
  \item Large ``lifetime'' effect and $R_o^2 \gg R_s^2$:\\
        Hydrodynamic simulations lead to sources with
        very large emission durations. The size of
        this lifetime signal depends on the equation of state (EOS):
        a softer EOS results in a more delayed pressure
        build-up and a longer lifetime\cite{Rischke:1995cm,Rischke:1996em}.
        Irrespective of model details,
        the resulting values for $R_o/R_s$ are
        generically much larger than the measured result\cite{Heinz:2002un}.
        This is the main problem of hydrodynamic simulations.
  \item Resonance decays contribute significantly to HBT radii:\\ 
        In contrast to model sources of the type (\ref{5.1}), resonance
        decay contributions added to the freeze-out of hydrodynamic
        simulations were reported to increase the size of HBT radii
        significantly\cite{Bolz:1992hc}. This may be attributed to the 
        different shapes of the freeze-out hyper-surfaces. The homogeneity
        regions for direct pion and resonance emission are the same in
        hydrodynamic simulations whereas the latter are smaller in the
        model (\ref{5.1}). Thus even short-lived resonance
        decay contributions tend to increase the hydrodynamic pion source.
\end{enumerate}

\begin{table}[t]
\tbl{Event generator model calculations with published
results on HBT correlation functions. \label{T:evgen}}
{\begin{tabular}{|clc|}
\hline\hline  & &  \\[-1.5ex]
Code${}^{\mbox{\small ref.}}$ & energies & HBT data  \\[1ex]
   & studied & compared to  \\[1ex]
\hline & &  \\[-1.5ex]
RQMD\cite{Sorge:vt,Sorge:dy}${}^{,a}$ & SPS & NA35\cite{Sullivan:wb}  \\[2ex]
hydro\cite{Rischke:1996em}+URQMD\cite{Bass:1998ca}${}^{,b}$ & 
SPS, RHIC\cite{Soff:2000eh,Soff:2001hc} & STAR\cite{Soff:2002qw}   \\[2ex]
Humanic\cite{Humanic:ya}${}^{,c}$ & AGS, SPS, RHIC & E859/866, 
NA44\cite{Humanic:vd} 
STAR\cite{Humanic:2002iw}  \\[2ex]
AMTP\cite{Lin:2002gc}${}^{,d}$ & RHIC & STAR\cite{Lin:2002gc}  \\[2ex]
MPC\cite{Molnar:2002bz} & RHIC & STAR, PHENIX\cite{Molnar:2002bz}  \\[2ex]
\hline\hline
\end{tabular}}
\begin{tabnote}
${}^a$ Code not maintained any more, no more recent studies available. \\
${}^b$ Hadronic rescattering phase dominates HBT 
radii\cite{Soff:2002qw}.\\
${}^c$ This code models final state rescattering only. \\
${}^d$ This is a multi-phase transport model which includes initial 
partonic and final state hadronic interactions.
\end{tabnote}
\end{table}

\subsection{Monte Carlo event generators}
Event generators are widely used to simulate particle production in 
ultra-relativistic heavy ion collisions. In particular, they allow 
to study how global collective dynamical properties emerge 
in a mesoscopic system from microscopic (2-to-2 or 2-to-3 body) interactions.
In principle, each event generator output defines an emission function
from which two-particle correlations can be calculated. 
However, since the event generator output is not 
a wave-function with proper quantum-mechanical symmetrisation,
an additional prescription is needed of how to relate it to the
emission function. There is an extensive literature on the
conceptual problem\cite{Wiedemann:1999qn}. In practice, the
afterburner program of Scott Pratt\cite{crab} is most frequently used.

So far, there are only very few calculations of HBT
correlation functions from event generators, see Table ~\ref{T:evgen}. 
The main conclusion
from these calculations is that the late hadronic rescattering 
phase largely determines the size of HBT radius parameters. 
Also, in contrast to hydrodynamic models, the generation of
models with relatively small lifetime, satisfying
$R_o/R_s \sim 1$, does not appear to be a fundamental problem. 
While some simulations find a ratio $R_o/R_s$ which for large $K_\perp$
lies between 1.4 and 2.0, inconsistent with experimental 
data\cite{Soff:2002qw}, 
other simulations\cite{Humanic:2002iw,Lin:2002gc} are consistent
with $R_o/R_s \sim 1$.

\section{HBT measurements}
\label{Sec3}

Two-particle correlations have been measured at all energies from
AGS to RHIC. Here we give an overview of the experimental situation.

\subsection{Coulomb final state corrections}
\label{sec3-4}

Data on two-particle momentum correlations between
identical charged pions are usually corrected by the experimentalists
for the pairwise Coulomb repulsion. These corrections are
difficult since Coulomb interaction and Bose-Einstein interference
effects are of similar size and affect the two-particle correlator
on similar relative momentum scales. Moreover, the used correction
methods differ between experiments and sometimes even
between different publications in one experiment. Differences
between the used correction techniques can change the
resulting size of the HBT radius parameter by more than
1 fm and they may affect the $K_\perp$-slope of HBT
radii. Thus Coulomb final state corrections are a major
source of systematic uncertainty in the space-time analysis
of correlation measurements.

The final state Coulomb interaction between two charged
particles is described by the relative Coulomb wave-function
of the particle pair, written in terms of
the confluent hypergeometric function $F$ \cite{Lednicky:1981su,Pratt:ev},
  \begin{eqnarray}
    \Phi_{{\bm{q}}/2}^{\rm coul}({\bm{r}}) &=& \Gamma(1+i\eta)\,
    e^{-\frac{1}{2}\pi\eta}\,
    e^{\frac{i}{2}{\bm{q}}\cdot {\bm{r}}}\,
    F \left(-i\eta; 1; z_-\right)\, ,
    \label{c.1}\\
    z_{\pm} &=& {\textstyle\frac{1}{2}}(q r \pm {\bm{q}}\cdot {\bm{r}})
    = {\textstyle\frac{1}{2}}q r (1 \pm \cos\theta)\, .
    \label{c.2}
  \end{eqnarray}
Here, $r = |\bm{r}|$, $q = |\bm{q}|$, and $\theta$ denotes the
angle between these vectors. The Sommerfeld parameter
$\eta = \alpha/ (v_{\rm rel}/c) $ depends on the
particle mass $m$ and the electro-magnetic coupling strength $e$.
We write
  \begin{equation}
    \eta_\pm =
    \pm \frac{e^2}{4\pi} \frac{\mu}{q/2}
           = \pm \frac{m\, e^2}{4\pi q}\, ,
    \label{c.3}
  \end{equation}
where $\mu$ is the reduced mass and  
the plus (minus) sign is for pairs of unlike-sign (like-sign)
particles. If particle pairs are emitted from a {\it static} source
at initial relative distance $\bm{r}$ with a probability
$S_{\rm stat}(\bm{r};\bm{K})$, then the corresponding correlation
is given by an average over the squared wave-function (\ref{c.1}),
  \begin{equation}
    C(\bm{q},\bm{K}) = \int d^3r\, S_{{\rm stat}}(\bm{r};\bm{K})\,
                 \vert \Phi_{\bm{q}/2}^{\rm coul}(\bm{r})\vert^2\, .
    \label{c.4}
  \end{equation}
In the case of identical particles, the two-particle symmetrised
version of (\ref{c.1}) should enter equation (\ref{c.4}). 
The $\bm{K}$-dependence of the pair emission probability is
often neglected when calculating Coulomb corrections.

The following Coulomb correction methods are based on this
starting point:
\begin{enumerate}
  \item \underline{Point-like Gamow Correction}\\
   For a point-like source $S_{\rm stat}(\bm{r}) =
   \delta^{(3)}(\bm{r})$, the correlator (\ref{c.4}) is given by the
   Gamow factor $G(\eta)$
  \begin{equation}
    G(\eta) = \Bigg\vert \Phi_{\bm{q}/2}^{\rm coul}(0)\Bigg\vert^2
    = \frac{2\pi\eta}{e^{2\pi\eta} - 1}\, .
    \label{c.5}
  \end{equation}
  Early studies constructed the corrected like-sign two-particle
  correlation by
  dividing the measured correlator by this Gamow factor
  \begin{equation}
    C_{\rm corr}^{(--)}(\bm{q},\bm{K}) =
    C_{\rm meas}^{(--)}(\bm{q},\bm{K})/G(\eta_-) \, .
    \label{c.6}
  \end{equation}
  \item \underline{Static Finite Size Correction}\\
    The point-like Gamow correction (\ref{c.5}) largely overestimates
    the real effect of Coulomb corrections since particles are emitted
    in reality with finite separation $\bm{r}$ which leads to a
    weaker Coulomb interaction. An improvement advocated 
    repeatedly\cite{Pratt:ev,Bowler:vx,Baym:1996wk} is to calculate
    the correction factor for a finite size static source. Typically,
    a Gaussian ansatz $S_{{\rm stat}}(\bm{r}) \propto \exp\left[
      -\bm{r}^2/ 4R^2\right]$ is chosen in (\ref{c.4}),
  \begin{eqnarray}
    F_{\rm corr}^{\rm stat}(\bm{q}) &=& \int d^3r\, S_{{\rm stat}}({\bf r})\,
                 \vert \Phi_{\bm{q}/2}^{\rm coul}(\bm{r})\vert^2\, ,
    \label{c.7}\\
    C_{\rm corr}^{(--)}(\bm{q},\bm{K}) &=&
    C_{\rm meas}^{(--)}(\bm{q},\bm{K})/ F_{\rm corr}^{\rm stat}(\bm{q})\, .
    \label{c.8}
  \end{eqnarray}
  Instead of the analytical emission function in eq.~\eqref{c.7},
  the particle-emitting source can be characterized in terms of a
  discrete set of phase-space points obtained e.g. from a Monte Carlo
  simulation. In this case, the correlation due to particle symmetrization
  and final state interactions is usually calculated with a 
  so-called afterburner routine. The most widely used
  afterburner is Scott Pratt's CRAB\cite{crab} (CoRrelation After-Burner).

  In practice, the value for the source width $R$ in eq.~\eqref{c.7}
  is determined iteratively
  from the extracted HBT radius parameter in the fitting procedure.
  A finite purity of the sample due to misidentified particle leads to
  an overall correlation strength $\Lambda < 1$. This can be taken into
  account by generalising\cite{Sinyukov:1998fc} eq.~(\ref{c.8}) to
  \begin{eqnarray}
    C_{\rm meas}^{(--)}(\bm{q},\bm{K}) &=& \left( 1 - \Lambda\right)
    + \Lambda\, C_{\rm corr}^{(--)}(\bm{q},\bm{K})\,
    F_{\rm corr}^{\rm stat}(\bm{q}) \, .
    \label{c.9}
  \end{eqnarray}
  \item \underline{Correction of like-sign by unlike-sign correlations}\\
    Rather than to calculate Coulomb corrections for finite size sources,
    one can make use of the fact that unlike-sign correlations receive
    no contribution from Bose-Einstein symmetrisation effects but depend
    on Coulomb correlations of the same magnitude (but opposite sign),
  \begin{equation}
    C_{\rm corr}^{(--)}({\bm{q}},{\bm{K}}) =
    C_{\rm meas}^{(+-)}({\bm{q}},{\bm{K}})\,
    C_{\rm meas}^{(--)}({\bm{q}},{\bm{K}})\, .
    \label{c.10}
  \end{equation}
    Theoretical support for this procedure comes from the fact, that
    like-sign and unlike-sign Coulomb correlations calculated from
    (\ref{c.4}) compensate largely. For point-like sources, e.g., the
    product of the Gamow factors deviates from unity by less than
    five percent for relative momenta $q > 8 \textstyle\frac{m}{137}$
  \begin{equation}
    G(\eta_+)\, G(\eta_-) =
    {1\over {1 + (\pi^2/ 3)\eta^2 + O(\eta^4)}}\, .
    \label{c.11}
  \end{equation}
A further improvement over (\ref{c.10}) is to take this
deviation into account\cite{Sinyukov:1998fc},
  \begin{equation}
    C_{\rm corr, improved}^{(--)}({\bm{q}},{\bm{K}}) =
    {C_{\rm meas}^{(+-)}({\bm{q}},{\bm{K}}) \,
      C_{\rm meas}^{(--)}({\bm{q}},{\bm{K}})
    \over G(\eta_+)\, G(\eta_-)} \, .
    \label{c.12}
  \end{equation}
This was shown to work with excellent accuracy for a wide range
of source parameters\cite{Sinyukov:1998fc}.

The effects of finite momentum resolution reduce both 
$C_{\rm meas}^{(+-)}({\bm{q}},{\bm{K}})$
and $C_{\rm meas}^{(--)}({\bm{q}},{\bm{K}})$. As an unwanted
consequence, these effects are amplified in the product
defining the corrected like-sign correlation functions
\eqref{c.10} and \eqref{c.12}.
An empirical parametrization which takes into account finite 
momentum resolution is discussed below; see eq.~\eqref{c.14}.
  \item \underline{Experimental parametrisations of Coulomb corrections}\\
       Unlike-sign correlations $C_{\rm meas}^{(+-)}(\bm{q},\bm{K})$
       were parametrised by the function\cite{na35-S}
       \begin{equation}
         \label{c.13}
         F(q_{\rm inv}) = 1 + \bigl(G(\eta_+) -1\bigr)\,
         e^{-q_{\rm inv}/Q_0}\, ,
       \end{equation}
       which depends on 
       \begin{equation}
         \label{qinv}
         q_{\rm inv} = \sqrt{\bm{q}^2 - (q^0)^2}\, .
       \end{equation}
The parameter $Q_0$ is extracted from the fit. It quantifies a 
phenomenological finite-size correction for large relative momentum. 
The function $F(q_{\rm inv})$ approaches the Gamow factor (\ref{c.5}) 
for a point-like source, $Q_0 \to \infty$.

In order to take into account the imperfect purity 
of the sample and the finite experimental momentum resolution, 
CERES\cite{Adamova:2002wi}  parametrised the Coulomb 
correction by
  \begin{equation}
    C_{\rm meas}^{(--)}(\bm{q},\bm{K}) = 
      \left( 1-\lambda\right)
      + \lambda\, C_{\rm corr}^{(--)}
      \left[ w_{K_\perp}( F_{\rm coul}(q_{\rm inv}) - 1) + 1\right]
      \, .
    \label{c.14}
  \end{equation}
In a Monte Carlo simulation of the final momentum resolution, the
Coulomb correction function $F_{\rm coul}(q_{\rm inv})$ was obtained
by evaluating eq.~\eqref{c.7} and reducing 
$F_{\rm corr}^{\rm stat}(\bm{q})$ accordingly. The same Monte
Carlo simulation determines $w_{K_\perp}$
which accounts for the depletion of the parameter $\lambda$ due to 
finite momentum resolution effects. This is chosen such  that 
$\lambda w_{K_\perp}$ gives the ``true'' corrected 
intercept parameter, which then multiplies the correction factor 
$( F_{\rm coul}(q_{\rm inv}) - 1)$. For perfect momentum resolution, 
$w_{K_\perp} \to 1$ and 
$F_{\rm coul}(q_{\rm inv}) \to F_{\rm corr}^{\rm stat}(q_{\rm inv})$,
and the prescription \eqref{c.14} agrees with \eqref{c.9}.

\end{enumerate}

The above discussion is mainly for static sources and
involves $q_{\rm inv}$-dependent correction factors only. In principle,
Coulomb correction effects are different for the different relative
momentum components, as seen from eq.~\eqref{c.7}. 
For dynamically expanding sources, a formalism
for the calculation of Coulomb corrections exists\cite{Anchishkin:1997tb},
but it has not been used so far in comparison to data. Only
one of the correction methods listed above, eq. (\ref{c.10}), 
contains some information about expansion effects since it uses the 
measured unlike-sign correlation as correction factor. 
%

\subsection{Experiments at the Alternating Gradient Synchrotron (AGS)}
\label{sec3-1}

\begin{sidewaystable}
\tbl{BE correlation data available from AGS experiments. Data are
taken at mid-rapidity, are finite size Coulomb corrected and are in 
Bertsch-Pratt parametrisation, unless noted otherwise. The fourth 
column indicates the analysis frame and the fifth column denotes 
whether the cross-term was included in the fits (yes), not included 
(no), checked to be 0 and then not included (0) or was not applicable
since a simpler parametrisation was used (N/A).
\label{T:AGS}}
{\begin{tabular}{|cclccl|}
\hline\hline &&&&& \\[-1.5ex]
\multicolumn{6}{|l|}{\bf Au+Au collisions} \\[1ex]
\hline
impact energy & collab.${}^{\mbox{\small ref.}}$ &  $K_\perp$ and binning &
frame & cross-term & note \\
\hline
\parbox{2.4cm}{2 $A$GeV/\\
$\sqrt{s} = 2.4\,A\mbox{GeV}$} &
E895\cite{e895-edep} & 0.1-0.3 GeV/$c$, 3 bins & CMS & 0 &  
at midrapidity CMS=LCMS\\  \hline

\parbox{2.4cm}{4 $A$GeV\\
$\sqrt{s} = 3.1\,A\mbox{GeV}$} &
E895\cite{e895-edep} & 0.1-0.3 GeV/$c$, 3 bins & CMS & 0 &
at midrapidity CMS=LCMS\\ \hline

 &
E895\cite{e895-edep} & 0.1-0.3 GeV/$c$, 3 bins & CMS & 0 & 
at midrapidity CMS=LCMS\\
\rb{\parbox{2.4cm}{6 $A$GeV\\
$\sqrt{s} = 3.6\,A\mbox{GeV}$}} & E917\cite{e917-atlanta} &   & -- & N/A &
preliminary data of $R_{\rm inv}$ \\ \hline

 &
E895\cite{e895-edep} & 0.1-0.3 GeV/$c$, 3 bins & CMS & 0 & 
at midrapidity CMS=LCMS\\
\rb{\parbox{2.4cm}{8 $A$GeV\\
$\sqrt{s} = 4.1\,A\mbox{GeV}$}} & E917\cite{e917-atlanta} &   & -- & N/A &
preliminary data of $R_{\rm inv}$ \\ \hline

\parbox{2.4cm}{10.6 $A$GeV\\
$\sqrt{s} = 4.7\,A\mbox{GeV}$} &
E917\cite{e917-atlanta} &   & -- & N/A &
preliminary data of $R_{\rm inv}$ \\ \hline

 &
E877\cite{Barrette:1997fj} & \parbox{2.5cm}{\flushleft 0-0.5 GeV/$c$, 1 bin 
$\langle p_\perp\rangle = 0.1\, \mbox{GeV/}c $} & 
\parbox{2cm}{beam rapidity} & yes &
data at beam rapidity \\
\raisebox{2.4ex}[2.4ex]{\parbox{2.4cm}{10.8 $A$GeV\\
$\sqrt{s} = 4.7\,A\mbox{GeV}$}}
 & E877\cite{e877-qm96hbt} & 1 bin & -- & N/A &
prelim. $R_{\rm inv}$ from $K^+K^+$ \\ \hline

\parbox{2.4cm}{11.6 $A$GeV\\
$\sqrt{s} = 4.9\,A\mbox{GeV}$} &
E802\cite{e802-arxiv} & \parbox{3cm}{\flushleft 3 bins, 
$\langle m_\perp \rangle =$\\ 0.29, 0.4, 0.54 {$A$GeV}/$c$} & YK frame
& yes &  \\ \hline\hline &&&&& \\[-1.5ex]

\multicolumn{6}{|l|}{\bf Si+X collisions at 14.6 $A$GeV} \\[1ex] \hline

target & collab.${}^{\mbox{\small ref.}}$ &  $K_\perp$ and binning & 
frame & cross-term & note \\ \hline

Pb & E814\cite{e814-plb1994} & 1 bin & -- & N/A &
$R_{\rm inv}$, beam rap., Gamow \\ \hline

Al & E802\cite{e802-arxiv}&  1 bin & YK frame & yes & \\ \hline

 & & & & & \\

 & E802\cite{e802-arxiv}&  \raisebox{1.3ex}[1.3ex]{
\parbox{2.4cm}{\flushleft 3 for $\pi^-$:\\
2 for $\pi^+$}} & YK frame  & yes & \\ 
\raisebox{1.9ex}[1.9ex]{Au}& E802/E859\cite{e802-qm95} & 1 bin & CMS & no & 
prelim. $K^+K^+$, Gamow corr.  \\

\hline\hline
\end{tabular}}
\end{sidewaystable}

At the AGS of the Brookhaven National Laboratory (BNL) three series of 
collaborations have measured and published results on Bose-Einstein 
correlations in fixed target experiments with beam energies 
varying between 2 $A$GeV and 11.6 $A$GeV. Due to lack of statistics,
measured correlation functions were parametrised often by a 
1-dimensional parametrisation
\begin{equation}
  C(q_{\rm inv}) = 1 + \lambda\, e^{-q_{\rm inv}^2 R^2_{\rm inv}}\, ,
  \label{qinv-par}
\end{equation}
where the invariant momentum difference $q_{\rm inv}$ is defined 
in eq.~(\ref{qinv}).

\subparagraph{E802/\-E859/\-E866/\-E917}
These experiments use a rotating spectrometer (the ``Henry
Higgins'' Spectrometer) which in the E866 upgrade  
was supplemented by
a Forward Spectrometer. The acceptance is at or close to mid-rapidity. 
The E802/E859/E866 Collaboration has 
published a systematic study of the dependence of HBT radius parameters
on transverse mass $M_\perp = \sqrt{m^2 + \bm{K}_\perp^2}$, system size 
and centrality dependences\cite{e802-arxiv}. The last of the
series, the E917 experiment, collected HBT data on beam energy dependence
in Au+Au collisions from 6 to 10.6 $A$GeV. At the time of this writing, 
these data are still preliminary\cite{e917-atlanta}.

\subparagraph{E895}
This experiment uses the EOS time projection chamber inherited from 
Bevalac\cite{EOS-TPC}. BE correlations were measured in 2, 4, 6, and 
8 $A$GeV Au+Au collisions. Published data exist for the transverse mass 
dependences of correlation radii\cite{e895-edep}, and the azimuthal 
dependences in non-central collisions for the three lower 
energies\cite{Lisa:2000xj}. Results on 
average phase-space density as a function of $K_\perp$ and beam energy were
shown at the Quark Matter 2001 conference\cite{e895-qm01}. All published
data are at mid-rapidity.

\subparagraph{E814/E877}
Particle correlations at projectile rapidity were measured with the
Forward Magnetic Spectrometer of the E814/E877 Collaboration. The E814 
is a Si-beam 
experiment\cite{e814-plb1994} while E877 measured  Au+Au at 
10.8 $A$GeV\cite{Barrette:1997fj}. E877 determined the pion 
phase-space density at freeze-out 
for the latter system\cite{Barrette:1997fj}. At the Quark Matter 97 conference,
they showed a direct fit of a Gaussian core-halo source function to 
$K_\perp$ and $y$-binned correlation function and extracted source radii 
for pions and for kaons\cite{e877-qm97}.
\vspace{0.5cm}

A summary of AGS experiments is given in Table~\ref{T:AGS}.

\subsection{Experiments at the CERN Super Proton Synchrotron (SPS)}
\label{sec3-2}

The CERN SPS was used first to accelerate ${}^{16}$O 
and $^{32}$S nuclei to 200 $A$GeV. Then it was upgraded to accelerate
a 158 $A$GeV $^{208}$Pb beam. Recently, the CERN SPS delivered Pb-beams
at lower energies:  40 and 80 $A$GeV. The step-by-step improvement
of analysis tools during the CERN SPS heavy ion program is clearly
seen in the available data. The correlation measurements for
oxygen and sulphur beams were parametrised first in terms of 
$q_{\rm inv}$ only. The later three-dimensional fits do not include 
the cross-term (\ref{eq17}). Also,
the Coulomb repulsion was corrected initially by multiplying
with a Gamow factor which overestimates the repulsion significantly. 
Improved Coulomb corrections, based on averaging the squared
Coulomb wave function over a finite source size, were only introduced
approximately with the arrival of the Pb beam. 

\begin{table}[t]
\tbl{BE correlations data for central oxygen-induced reactions
at 200~$A$GeV/$c$. All correlation functions are constructed from
hadron pairs and are corrected for final state interactions 
by the Gamow factor. The parametrisations used by both experiments
are listed in the last column. They include an additional 
factor 1/2 in the exponent of the correlation function.
\label{T:O}}
{\begin{tabular}{|cccccl|}
\hline\hline &&&&& \\[-1.5ex]
target & collab.${}^{\mbox{\small ref.}}$ & rapidity & frame & $K_\perp$ &
 param. \\[1ex]
\hline 
C & & &  & & \\
Cu & &  & & & $q_{\rm inv}$ and \\
Ag &  \rb{WA80\cite{wa80-1995}} & \rb{$-1<y_{\rm lab}<1$} & \rb{lab} & 
\rb{40 -- 200 MeV/$c$} & 2d: $q_\perp, q_l$ \\
Au & & & & & \\[2ex]
Au & NA35\cite{na35-S} & 
\parbox{2cm}{\flushleft $0.5\! <\! y_{\rm lab}\! <\! 3.5$\\ 3 bins} & CMS & 
50 -- 600 MeV/$c$ & \parbox{1.8cm}{\flushleft BP\\ no cross-term} \\[4ex]
\hline\hline
\end{tabular}}
\end{table}

\begin{sidewaystable}
\tbl{BE correlation data for central sulphur-induced reactions
at 200~$A$GeV. In all cases, final state interactions are
corrected for by the Gamow factor. None of the parametrisations
listed in the sixth column includes the cross term in the
BP parametrisation.
\label{T:S}}
{\begin{tabular}{|ccccccl|}
\hline\hline &&&&& & \\[-1.5ex]
target & collab.${}^{\mbox{\small ref.}}$ & rapidity & frame & $K_\perp$ &
 param.& note \\[1ex]
\hline 
&&&&& & \\[-1.5ex]
C & NA35\cite{na35-S} & $2.5<y_{\rm lab}<4.5$, 2 bins & 
CMS & 50 -- 600 MeV/$c$ & BP & factor 1/2 in the exponent \\[1.5ex]

 & NA35\cite{na35-S} & $0.5<y_{\rm lab}<3.5$, 3 bins & CMS & 
50 -- 600 MeV/$c$ ${}^a$ 
& BP & factor 1/2 in the exponent \\
\rb{S} & NA44\cite{na44-mult} & $y_{\rm lab} \approx 3$ & LCMS & 
$K_\perp \le 400\,\mbox{MeV}/c$ & BP & \\[1.5ex]

Al & WA80\cite{wa80-1995} & $-1<y_{\rm lab}<1$ & lab & 40 -- 200 MeV/$c$ & 
2~dim: $q_\perp, q_l$ & factor 1/2 in the exponent\\[1.5ex]

Cu & NA35\cite{na35-S} & $2.5<y_{\rm lab}<4.5$, 2 bins &
CMS & 50 -- 600 MeV/$c$ ${}^{a,b}$ & BP & factor 1/2 in the exponent \\[1.5ex]

 & NA35\cite{na35-S} & $0.5<y_{\rm lab}<4.5$, 4 bins & CMS & 
50 -- 600 MeV/$c$ ${}^c$  
& BP & factor 1/2 in the exponent \\
\rb{Ag} & NA44\cite{na44-mult} & $y_{\rm lab} \approx 3$ & LCMS &
$K_\perp \le 400\,\mbox{MeV}/c$ & 2~dim: $q_\perp, q_l$ & \\[1.5ex]

 & NA44\cite{NA44-mperp} & $y_{\rm lab} \approx 3$ & LCMS &
2 bins${}^d$ &  BP & \\
\rb{Pb} & NA44\cite{na44-kaon94} & $y_{\rm lab} \approx 3$ & LCMS &
2 bins${}^e$ & BP${}^f$ & kaon interferometry \\[1.5ex]

 & NA35\cite{na35-S} & $0.5<y_{\rm lab}<4.5$, 4 bins & CMS & 
50 -- 600 MeV/$c$ ${}^{a,b}$ & BP & factor 1/2 in the exponent \\
\rb{Au} & WA80\cite{wa80-1995} & $-1<y_{\rm lab}<1$ & lab & 
40 -- 200 MeV/$c$ & 2~dim: $q_\perp, q_l$ & factor 1/2 in the exponent
\\[0.5ex]

\hline
\end{tabular}}
\begin{tabnote}
${}^a$ The $K_\perp$ dependence of correlation radii was measured for 
$3.5<y_{\rm lab}<4.5$.\\
${}^b$ The $K_\perp$ dependence of correlation radii for $2.5<y_{\rm lab}<3.5$
was also published\cite{na35-kt}.\\
${}^c$ The $K_\perp$ dependences of correlation radii were measured in all
rapidity bins. \\
${}^d$ The two NA44 $K_\perp$ bins correspond to
$\langle p_\perp \rangle \approx 150$ and $450\, \mbox{MeV}/c$
respectively.\\
${}^e$ For the BP parametrisation, the two bins correspond to 
$\langle p_\perp \rangle \approx 163$ and 
$246\, \mbox{MeV}/c$ respectively.\\
${}^f$ Results from other parametrisations are also 
available\cite{na44-kaon94}.
\end{tabnote}
\end{sidewaystable}
\begin{sidewaystable}
\tbl{BE correlation data for Pb+Pb collisions at the SPS. Correlation radii 
were measured as functions
of $K_\perp$ and the cross-term was included in the parametrisations. 
The fifth column (FSI) specifies how final state interactions are 
corrected for by referring to the corresponding relation in 
Section~\ref{sec3-4}. The column on particle identification (PID)
specifies whether measurements are for
identified pions (YES) or charged hadrons (NO).
\label{T:Pb-list}}
{\begin{tabular}{|ccccccl|}
\hline\hline  & & & & & &  \\[-1.5ex]
collab.${}^{\mbox{\small ref.}}$ & frame & rapidity & param. & FSI & 
PID & note \\[1ex]
\hline & & & & & &  \\[-1.5ex]
NA44\cite{na44-98,na44-01} & LCMS & central${}^{a}$ & BP & 
(\ref{c.8}) & yes & also kaon interferometry\cite{na44-01} \\[3ex]

NA45\cite{Adamova:2002wi} & LCMS & central & BP & 
(\ref{c.14}) & no &
\parbox{3.cm}{\flushleft measured for Pb+Au\\ 
\symbol{64} 40, 80, 158 $A$GeV/$c$} \\[6ex]

NA49\cite{NA49-app,NA49-ganz,Seyboth:2002wu} & FLCMS & 
\parbox{2.3cm}{$0.1\!<\! y_{\rm CMS}\! <\! 2.1$ 4 bins} &
YKP, BP & \parbox{2.4cm}{\center (\ref{c.13})\cite{NA49-app,NA49-ganz}, 
(\ref{c.10})\cite{NA49-ganz}, (\ref{c.14})\cite{Seyboth:2002wu}} & no & 
\parbox{3.43cm}{\flushleft  
40, 80, 158 $A$GeV/$c$\cite{Seyboth:2002wu}\\
also kaon interferometry\cite{Afanasiev:2002fv}} 
\\[6ex]

WA97\cite{Antinori:2001yi} & LCMS & 
\parbox{2.3cm}{$-0.3\!\!<\!\! y_{\rm CMS}\!\! <\!\! 0.9$ 4 bins} &
YKP, BP & iterative (\ref{c.8})${}^{b}$ &
no & \\[5ex]

WA98\cite{Aggarwal:2000zs,wa98inprep} & LCMS & $-0.8\!<\! y_{\rm CMS}\! <\! 0.2$${}^{c}$ &
BP, YKP & iterative (\ref{c.8}) & yes & \\[1.5ex]

\hline\hline
\end{tabular}}
\begin{tabnote}
${}^a$ NA44 has a ``banana-shaped'' acceptance around midrapidity.
Small-$p_\perp$ data are at slightly forward, high-$p_\perp$ data 
at slightly backward rapidity.\\
${}^b$ In determining the Coulomb correction $F_{\rm corr}(\bm{q})$
from eq.~\eqref{c.7} WA97 iteratively used the emission function 
\eqref{5.1} instead of a static Gaussian distribution.\\
${}^{c}$ WA98 has a ``banana-shaped'' acceptance at slightly backward 
rapidity. High-$p_\perp$ data are further away from midrapidity.
\end{tabnote}
\end{sidewaystable}

\subparagraph{NA35}
Originally, this experiment
used a large streamer chamber in a magnetic field to measure tracks of 
charged particles and their momenta in O+Au collisions. 
The large volume of the detector allowed for the study of
three rapidity windows between --2.4 and 1.6 in the CMS of the nucleon-nucleon 
collision\cite{na35-O}. 
Due to small statistics, the correlation function
was parametrised in $q_l=q_{\rm long}$ and $q_\perp = \sqrt{q_o^2+q_s^2}$,
\begin{equation}
  \label{2dparam} 
  C(q)=1+\lambda
  \exp\left[-\frac{1}{2}q_l^2R_l^2 - \frac{1}{2}q_\perp^2 R_\perp^2\right]\, .
\end{equation}
In contrast to other experiments, this parametrisation used by NA35 
has an additional factor 1/2 in the exponent. 
The NA35 detector does not identify pions. Thus
NA35 always studied hadron-hadron correlations. With the O-induced collisions 
a measurement of ``single-event interferometry''\cite{na35-hum} was
attempted. For measurements with the sulphur beam, the detector
was upgraded with a time projection chamber (TPC) which was crucial
in gaining good statistics for correlation analysis.
In a comprehensive
study\cite{na35-S} of S-induced reactions with C, S, Cu, Ag, and Au targets,
NA35 measured the rapidity, $K_\perp$, and multiplicity dependence 
of Bertsch-Pratt correlation radii. O+Au results were reanalysed 
in this work with better statistics. The $K_\perp$ dependence of 
correlation radii was also reported in a letter\cite{na35-kt}.
Although the existence of a sizeable cross-term was first confirmed by
NA35\cite{na35-qm95}, no results with the cross-term were published.

\begin{table}[t]
\tbl{Data on the centrality dependence of correlation 
radii. 
\label{T:central-list}}
{\begin{tabular}{|clc|}
\hline\hline  &&  \\[-1.5ex]
collab.${}^{\mbox{\small ref.}}$ & system and energy & rapidity \\[1ex]
\hline & &  \\[-2.5ex]
E802\cite{e802-arxiv} & 
\parbox{3.5cm}{\flushleft
Si+Al \symbol{64}~14.6~$A$GeV
Si+Au \symbol{64}~14.6~$A$GeV
Au+Au \symbol{64}~11.6~$A$GeV} & mid-rapidity \\[6ex]

NA35\cite{na35-S} & S+Ag \symbol{64}~200~$A$GeV & 
$-2.5<y_{\rm CMS}<1.5$, 4 bins \\[1ex]

NA44\cite{na44-mult} & \parbox{2.5cm}{\flushleft 
S+S, S+Ag, S+Pb \symbol{64}~200~$A$GeV} &
mid-rapidity \\[4ex]

NA45\cite{Adamova:2002wi} & Pb+Au \symbol{64}~40, 80, 158~$A$GeV &
mid-rapidity \\[2ex]

NA49\cite{na49-qm99,na49QM2002} & Pb+Pb \symbol{64}~40, 80, 158~$A$GeV & 
$y_{\rm CMS} = 0$\cite{na49QM2002}, 1.3\cite{na49-qm99}  \\[2ex]

WA97\cite{Antinori:2001yi} & Pb+Pb \symbol{64}~158~$A$GeV &
$-0.3<y_{\rm CMS}<0.9$ \\[3ex]

STAR\cite{Adler:2001zd,star02-HBT} & Au+Au \symbol{64} 
\parbox{3cm}{\flushleft $\sqrt{s}=130\,A\mbox{GeV}$\cite{Adler:2001zd}\\
$\sqrt{s}=200\,A\mbox{GeV}$\cite{star02-HBT}}
& mid-rapidity \\[4ex]

PHENIX\cite{phenixQM2002} 
& Au+Au \symbol{64} $\sqrt{s}=200\,A\mbox{GeV}$ &
mid-rapidity \\[2ex]

\hline\hline
\end{tabular}}
\begin{tabnote}
\end{tabnote}
\end{table}

\subparagraph{NA49}
For the lead beam, this collaboration equipped
the NA35 detector  with four large TPCs
which allow for precise tracking of the secondaries in the rapidity region 
$2<y<5.5$ (values given in the laboratory system with 
$y_{\rm CMS} = 2.9$ for the 158~$A$GeV Pb beam). The NA49 detector is 
able to identify  particles by a combination of energy loss and
time-of-flight measurements. So far, however, only unidentified 
hadron-hadron correlations are available. For Pb+Pb collisions at
158~$A$GeV, the rapidity and $K_\perp$ dependence of YKP parameters was 
published\cite{NA49-app}. Also, a rather comprehensive compilation of 
preliminary data at all rapidities and in both YKP and Bertsch-Pratt 
is available\cite{NA49-ganz}. Preliminary 
results\cite{Seyboth:2002wu,na49QM2002} exist for $h^-h^-$ correlations
in 40, 80, and 158 $A$GeV Pb+Pb collisions with different 
centralities. Results for kaon-kaon correlations in central 158 $A$GeV 
Pb+Pb collisions were published very recently\cite{Afanasiev:2002fv}.
These latter studies are the only ones based on ``global tracking''
for which tracks from all TPCs are 
matched before constructing the correlation function.

\begin{table}[t]
\tbl{Data on three-pion interferometry.
\label{T:3pi}}
{\begin{tabular}{|clc|}
\hline\hline  &&  \\[-1.5ex]
collab.${}^{\mbox{\small ref.}}$ & system and energy & rapidity \\[1ex]
\hline & &  \\[-1.5ex]
NA44\cite{na44-3part} & S+Pb \symbol{64}~200~$A$GeV & mid-rapidity \\[2ex]

WA98\cite{Aggarwal:2000ex,wa98inprep} & Pb+Pb \symbol{64}~158~$A$GeV &
\parbox{3cm}{\flushleft $-0.8 < y_{\rm CMS} <0.2$\\ 
$\langle y_{\rm CMS}\rangle = -0.2 $} \\[4ex]

STAR\cite{Willson-3pi} & Au+Au \symbol{64} $\sqrt{s}=130 A\mbox{GeV}$ &
mid-rapidity\\[2ex]

\hline\hline
\end{tabular}}
\end{table}

\subparagraph{NA44}
This experiment is based on a focusing spectrometer. Its narrow acceptance 
is around mid-rapidity and depends in detail on the detector setting which 
may be varied. The performance is optimised for
particles with small momentum difference and good particle identification 
is achieved. For the S-beam, one-dimensional correlation functions for S+Pb 
collisions\cite{na44-93}, and a kaon interference study\cite{na44-kaon94}
were published first. Bertsch-Pratt
radii without the cross-term were published in 1995\cite{na44-3d} and
a dedicated paper was written on their $M_\perp$ dependence\cite{NA44-mperp}.
NA44 studied collisions with other targets\cite{na44-mult}: 
S+S, S+Ag, S+Pb. More recently an investigation of three-particle correlations
in S+Pb systems was published\cite{na44-3part}.

There are two papers with data from the 158 $A$GeV Pb+Pb collisions, 
one for pion\cite{na44-98}, and one for kaon\cite{na44-01} interferometry.
These study the $M_\perp$ dependence of the BP correlation radii within
the limited acceptance of the detector.

\begin{table}[t]
\tbl{Data on the azimuthally dependence of HBT radius parameters. All 
data are taken at central rapidity.
\label{T:phi}}
{\begin{tabular}{|ccl|}
\hline\hline  & &  \\[-1.5ex]
collab.${}^{\mbox{\small ref.}}$ & $\phi$-dependent radii
& system and energy \\[1ex]
\hline & &  \\[-1.5ex]

E895\cite{Lisa:2000xj} & all six radii $R_{ij}^2(\phi)$
&Au+Au \symbol{64} 2, 4, 6 $A$GeV \\[2ex]

STAR\cite{star02-HBT} & 
$R_o^2$, $R_s^2$, $R_{os}^2$, $R_l^2$ &
Au+Au \symbol{64} $\sqrt{s} = 130,\, 200\, A\mbox{GeV}$ \\[2ex]

\hline\hline
\end{tabular}}
\end{table}

\subparagraph{NA45-CERES}
In 1998, the CERES collaboration upgraded their detector with a 
time projection chamber with radial 
drift field. This allows for interferometric
studies. CERES measured correlations of non-identified hadrons
in Pb+Au collisions
at 40, 80 and 158~$A$GeV with various centralities\cite{Adamova:2002wi}.

\subparagraph{WA80}
The correlation analysis of WA80 is based on the so-called plastic ball 
detector which has coverage in the target rapidity region and at low 
$p_\perp<220\, \mbox{MeV}/c$.  
For the oxygen beam they used C, Cu, Ag, and Au targets. The sulphur beam was
collided with Al and Au targets,
the proton beam at 450 $A$GeV/$c$ was collided with 
C and Au targets. An earlier analysis of target dependence of the observed 
HBT radii for O-induced reactions\cite{wa80-1992} was superseded by a study 
including also S and p as projectiles in which the detector performance 
correction was better understood\cite{wa80-1995}. The latter paper 
also shows a simple model fit to the correlation function with Coulomb 
correction (\ref{c.8}) instead of Gamow factor multiplication (\ref{c.6}). 
They used an additional
factor of 1/2 in the correlation function like NA35 did. Another
work\cite{wa80-inter} analyses these data in the context of
intermittency.

\begin{figure}[t]\epsfxsize=11cm 
\centerline{\epsfbox{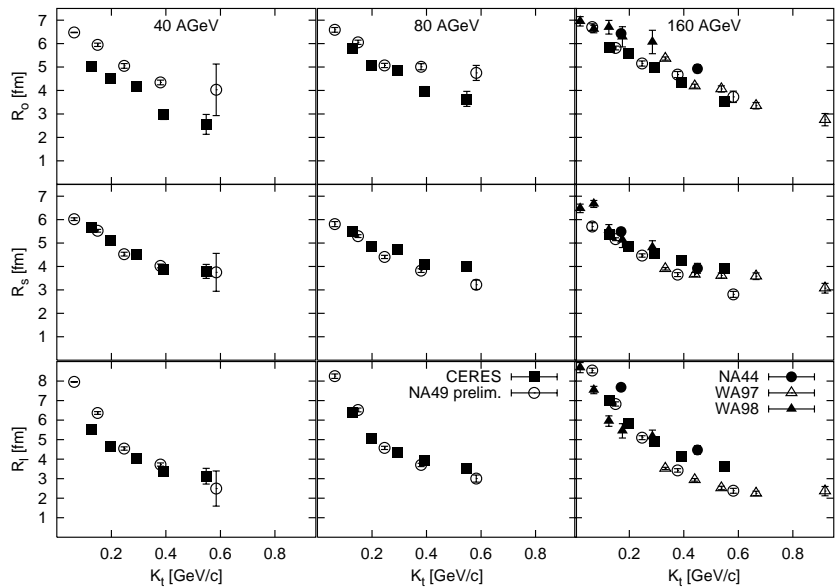}}
\caption{Summary of the $K_\perp$-dependence of correlation radii 
measured at the CERN SPS in Pb+Pb collisions at midrapidity.
Data are taken from NA44\protect\cite{na44-98}, 
CERES\protect\cite{Adamova:2002wi}, 
preliminary NA49\protect\cite{Seyboth:2002wu}, 
WA97\protect\cite{Antinori:2001yi} and WA98\protect\cite{wa98inprep}.
}\label{f:sps-hbt}
\end{figure}

\subparagraph{WA98}
For the lead beam runs, this collaboration made use of the plastic ball 
calorimeter. To measure charged particles the detector includes a two
arm tracking spectrometer with ``banana-shaped'' $\pi^-$ acceptance 
around mid-rapidity. WA98 published a study of the $K_\perp$ dependence of 
correlation radii for identified $\pi^-$ in Bertsch-Pratt and YKP 
parametrisations\cite{Aggarwal:2000zs,wa98inprep}.
Three pion interferometry\cite{Aggarwal:2000ex,wa98inprep} and pion 
phase-space density\cite{wa98inprep} were also studied.

\subparagraph{WA97}
To measure BE correlations, WA97 used a silicon telescope which
provides precise tracking of produced particles within the magnetic 
field. 
They do not identify particle species but sample the correlation 
function with $h^-h^-$ pairs. The data are mainly presented in the
YKP parametrisation but consistency checks with the Bertsch-Pratt
form were performed\cite{Antinori:2001yi}. Transverse momentum and rapidity 
dependences of the correlation radii were investigated in the 
acceptance window $-0.3<y<0.9$ (in CMS) and 
$0.2\,\mbox{GeV}/c < p_\perp < 1.3\, \mbox{GeV}/c$.

\subsection{Experiments at the Relativistic Heavy Ion Collider (RHIC)}
\label{sec3-3}

Three of the four collaborations at RHIC published results 
on BE interferometry: STAR, PHENIX, and PHOBOS. Data were taken
from Au+Au collisions at CMS energies of 130 and 200 $A$GeV.

\begin{figure}[t]\epsfxsize=11cm 
\centerline{\epsfbox{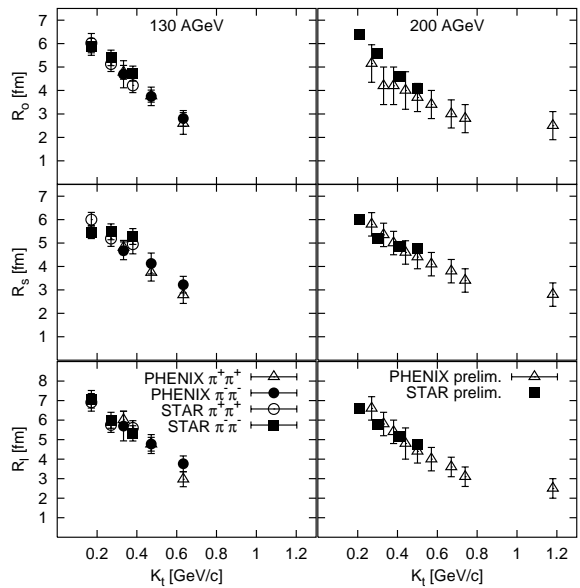}}
\caption{Summary of the $K_\perp$-dependence of correlation radii at
mid-rapidity measured at RHIC in Au+Au collisions.
Data are taken from STAR\protect\cite{Adler:2001zd,star02-HBT}
and PHENIX\protect\cite{Adcox:2002uc,phenixQM2002}. Data for 
$\sqrt{s}=200 A$GeV are taken from transparencies shown at the 
Quark Matter 2002 conference.
}\label{f:rhic-hbt}
\end{figure}

%
\subparagraph{STAR}
uses a time projection chamber in a solenoidal magnetic field.
The analysis for 
$\sqrt{s}=130\, A\mbox{GeV}$\cite{Adler:2001zd} covers in three bins
the $K_\perp$-dependence from
0.125~GeV/$c$ to 0.45~GeV/$c$ within a rapidity region 
$|y_{\rm CMS}|<0.5$. Preliminary data at $\sqrt{s}=200\, A\mbox{GeV}$
range in four bins up to 
$K_\perp \approx 0.52\, \mbox{GeV}/c$\cite{star02-HBT}. 
The centrality dependence of the HBT radii at both 
energies was studied\cite{Adler:2001zd,star02-HBT}. 
The azimuthal dependence of the $K_\perp$-integrated 
HBT radii $R_o^2$, $R_s^2$ and $R_{os}^2$ was also analysed 
at both energies in minimum bias events\cite{star02-HBT}.
So far, STAR determines the orientation of the reaction plane
but it does not determine the direction of the impact parameter.
Thus, the azimuthal dependence of $R_{sl}^2$ and
$R_{ol}^2$ cannot be measured. Finally, a first study of three pion 
correlations indicates that the source is fully chaotic\cite{Willson-3pi}.

\subparagraph{PHENIX}
extends the results of STAR in a pseudo-rapidity window $|\eta|<0.35$
to higher $K_\perp$: 
$0.2<K_\perp<1.2~\mbox{GeV}/c$ at the lower and 
$0.2<K_\perp<2~\mbox{GeV}/c$ at the higher CMS energy.
The particle momentum is measured by a drift chamber and a pad chamber. 
At $\sqrt{s}=130\, A\mbox{GeV}$, correlation radii
for both positive and negative identified pions are determined in three
$K_\perp$ bins\cite{Adcox:2002uc}. For collisions at 
$\sqrt{s}=200\, A\mbox{GeV}$, a much better statistics allowed to
split the pairs into nine $K_\perp$ bins\cite{phenixQM2002}. 
The centrality dependence of the correlation radii was also 
studied.    
Correlation radii from kaon-kaon correlations did not 
show\cite{phenixQM2002}
a simple $M_\perp$ scaling with the $\pi\pi$ radii in contrast 
to expectations from certain hydrodynamically motivated parametrisations
of the freeze-out state of the fireball\cite{Csorgo:1995bi}.

\subparagraph{PHOBOS}
presented so far two sets of 
BP correlation radii for the 15\% most central
$\sqrt{s}=200\, A\mbox{GeV}$ Au+Au collisions. Data are for one $K_\perp$ bin 
from 0.15 to 0.35 GeV/$c$ and for $0.2<y<1.5$. One set was measured with 
$\pi^+$ pairs, the other one with $\pi^-$ pairs\cite{phobosHBT}.
 
\begin{figure}[t]\epsfxsize=11.7cm 
\centerline{{\hspace*{0.7cm}\epsfbox{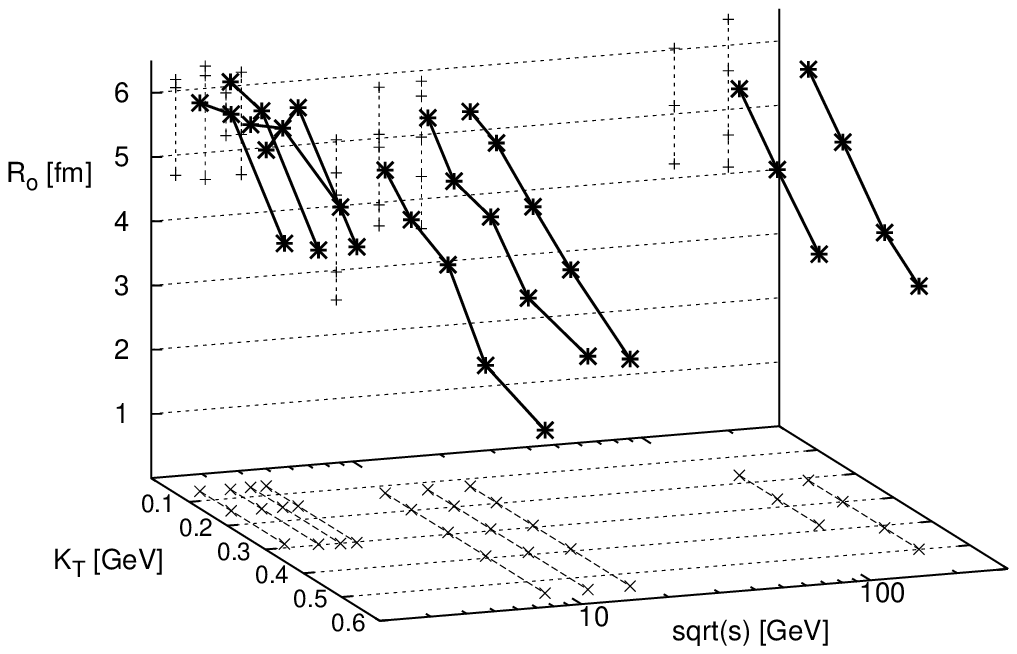}}}
\vspace{-2cm}\epsfxsize=11.7cm
\centerline{{\hspace*{0.7cm}\epsfbox{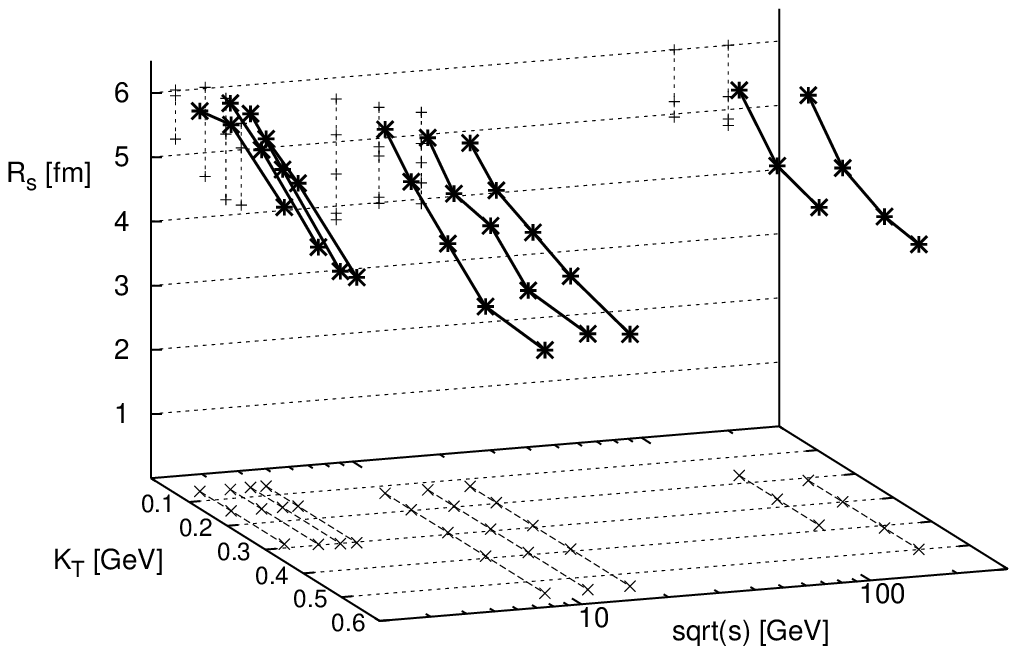}}}
\caption{The $\sqrt{s}$ and $K_\perp$ dependence of $R_o$ and $R_s$.
Data without error bars are summarised from E895\protect\cite{e895-edep} 
(Au+Au), NA45-CERES\protect\cite{Adamova:2002wi} (Pb+Au), and STAR 
$\pi^+\pi^+$\protect\cite{Adler:2001zd,star02-HBT} (Au+Au). 
STAR results for $\sqrt{s} = 200 A$GeV are taken from transparencies 
shown at the Quark Matter 2002 conference.}
\label{fig-sdep}
\end{figure}

\begin{figure}[t]\epsfxsize=12cm 
\centerline{{\hspace*{0.7cm}\epsfbox{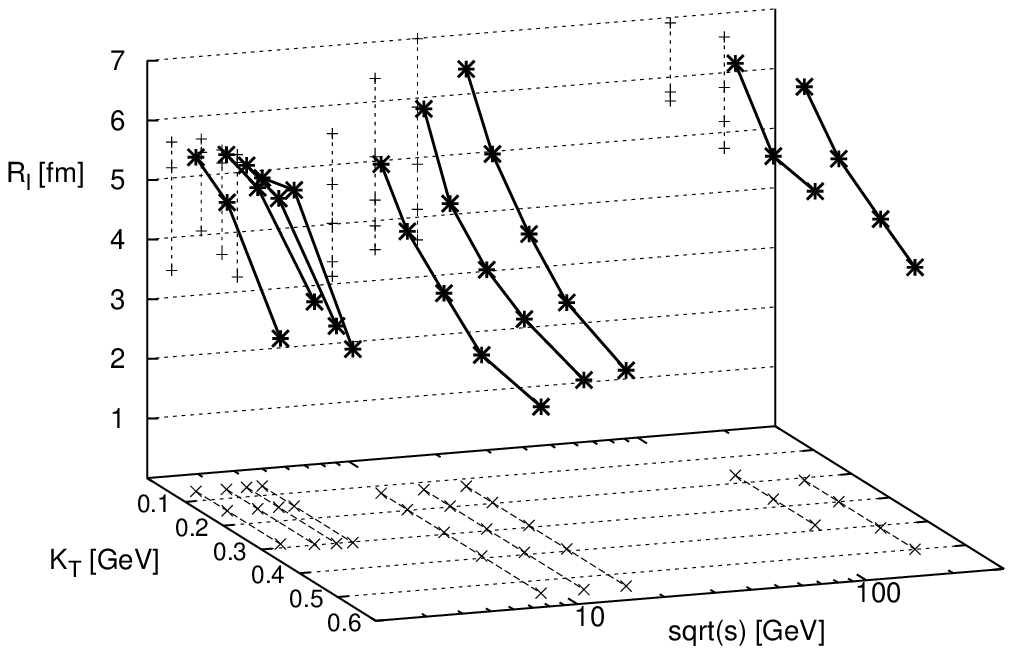}}}
\caption{The $\sqrt{s}$ and $K_\perp$ dependence of $R_l$.
Data without error bars are summarised from E895\protect\cite{e895-edep}, 
NA45-CERES\protect\cite{Adamova:2002wi}, 
and STAR $\pi^+\pi^+$\protect\cite{Adler:2001zd,star02-HBT}. 
STAR results for $\sqrt{s} = 200 A$GeV are taken from transparencies 
shown at the Quark Matter 2002 conference.}
\label{fig-sdep-Rl}
\end{figure}

\subsection{Discussion of the data}

\subsubsection{Size and transverse momentum dependence of HBT radii}
The out-, side-, and longitudinal HBT radius parameters vary
typically around $5-6$ fm at small transverse pair momentum
$K_\perp$ and decrease with increasing $K_\perp$.
Their absolute size shows no significant dependence on beam energy
(see Figs.~\ref{fig-sdep} and  ~\ref{fig-sdep-Rl}). 
For data from the CERN SPS, the $K_\perp$-slope of the 
longitudinal radius parameter is steeper (see Fig.~\ref{f:sps-hbt})
while all radius parameters measured at RHIC show approximately
the same slope (see Fig.~\ref{f:rhic-hbt}). 

As explained in section~\ref{sec2b3}, the HBT radius parameters 
of an expanding source correspond to the width of the $K_\perp$-dependent 
region of homogeneity. This is smaller than the width of the entire 
collision region. The $K_\perp$-slope of the HBT radii is
a measure of the collective dynamical expansion. This
picture can be illustrated by the pocket formulas 
(\ref{eq106}), (\ref{eq107}) and is supported by many model
comparisons. If SPS data are fitted by a model with Gaussian 
transverse density distribution\cite{Ster:1999ib,Tomasik:1999cq},
this leads to a radius of the entire collision region 
$R \approx 7$ fm. To put this number into perspective,
we relate the two-dimensional rms width of the collision region,
$r_{\rm rms}^{\rm source} = \sqrt{ \langle \tilde{x}^2 + \tilde{y}^2\rangle }
= \sqrt{2}\, R \approx 10$ fm, to the 
two-dimensional rms widths of a cold lead nucleus. 
The hard sphere radius $R_{\rm hs}^{\rm A} = 1.2\, A^{1/3}$ fm is 
for lead $R_{\rm hs}^{\rm Pb} = 7.1$ fm, and the 
corresponding two-dimensional transverse rms width
is $r_{\rm rms}^{\rm cold\, Pb} = \sqrt{ \langle \tilde{x}^2 + 
\tilde{y}^2\rangle_{\rm Pb} } =   \sqrt{3/5}\, \, R_{\rm hs}^{\rm Pb} 
\approx 4.4$ fm. SPS data favour a model with a transverse box density 
profile\cite{Tomasik:1999cq} over a Gaussian density profile. 
The box radius is $10-12$ fm. Irrespective of the transverse profile, 
one concludes that 
during the collision the system has expanded by a factor $\approx 2$ 
from the transverse size of the 
overlapping cold lead nuclei to the transverse extension at freeze-out.

All available data are subject to significant systematic uncertainties.
Additional uncertainties arise when comparing data from fixed target 
and collider experiments. In view of the rather mild changes of HBT radii 
between SPS and RHIC, this makes it difficult to assess to what extent the 
dynamical interpretation given above changes from SPS to RHIC.
A first analysis of RHIC data\cite{Csorgo:2002ry} argues in favour of
a more extended source with larger transverse
flow, thus supporting the picture of a more vigorous transverse 
expansion at higher centre of mass energies. 
 
\subsubsection{$R_o/R_s$}\label{RooRs}
The main interest in the quotient or difference of the two
transverse HBT radius parameters (\ref{eq14}) and (\ref{eq15})
lies in a model-dependent argument, that the emission 
duration $\langle \tilde{t}^2 \rangle$ can be extracted from
\cite{Bertsch:vn,Pratt:zq}
  \begin{equation}
    R_o^2({K}) - R_s^2({K}) \approx \beta_\perp^2
     \langle \tilde{t}^2 \rangle\, .
     \label{eq22}
  \end{equation}
%
\begin{figure}[t]\epsfxsize=12cm 
\centerline{{\hspace*{0.7cm}\epsfbox{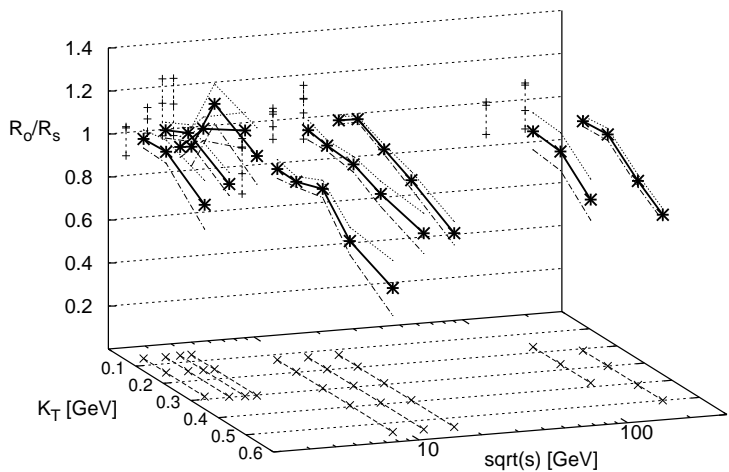}}}
\caption{The $\sqrt{s}$ and $K_\perp$ dependence of $R_o/R_s$.
Data are summarised from E895\protect\cite{e895-edep}, 
NA45-CERES\protect\cite{Adamova:2002wi}, 
and STAR $\pi^+\pi^+$\protect\cite{Adler:2001zd,star02-HBT}.
STAR results from $\sqrt{s} = 200 A$GeV are taken from transparencies 
shown at the Quark Matter 2002 conference; their error bars may be
underestimated.}
\label{fig-sdep-RoRs}
\end{figure}
%
This statement is based on two model-dependent assumptions. 
First, the term 
$-2\beta_\perp \langle \tilde{x} \tilde t \rangle$ should be
negligible compared to (\ref{eq22}). This assumption, however,
can be violated in models with strong  expansion.
Second, the difference
$\langle \tilde{x}^2 \rangle - \langle \tilde{y}^2 \rangle$ 
should be negligible compared to (\ref{eq22}). This latter
assumption can be violated at sufficiently high
$K_\perp$, in particular in models for which particle emission peaks
close to the surface due to dynamical or opacity
effects. 

Many model calculations predict $R_o/R_s \gg 1$. In particular,
Rischke and Gyulassy\cite{Rischke:1996em} emphasised that this 
would be an unambiguous signal of an equation of state which is
sufficiently soft in the phase transition region to result in
a significantly delayed build-up of transverse expansion. This would result
in a large lifetime effect, $R_o/R_s \sim 1.5$. In contrast,
data indicate values $R_o/R_s \lesssim 1.1$ even at RHIC, 
see Fig.~\ref{fig-sdep-RoRs}. As mentioned in section~\ref{secmodel},
one can think of physics effects which result in $R_o/R_s \lesssim 1$.
This issue is presently under study\cite{Lin:2002gc,McLerran:2002dt}.

\begin{figure}[t]\epsfxsize=7.6cm 
\vskip -0.5cm
\centerline{
{\epsfbox{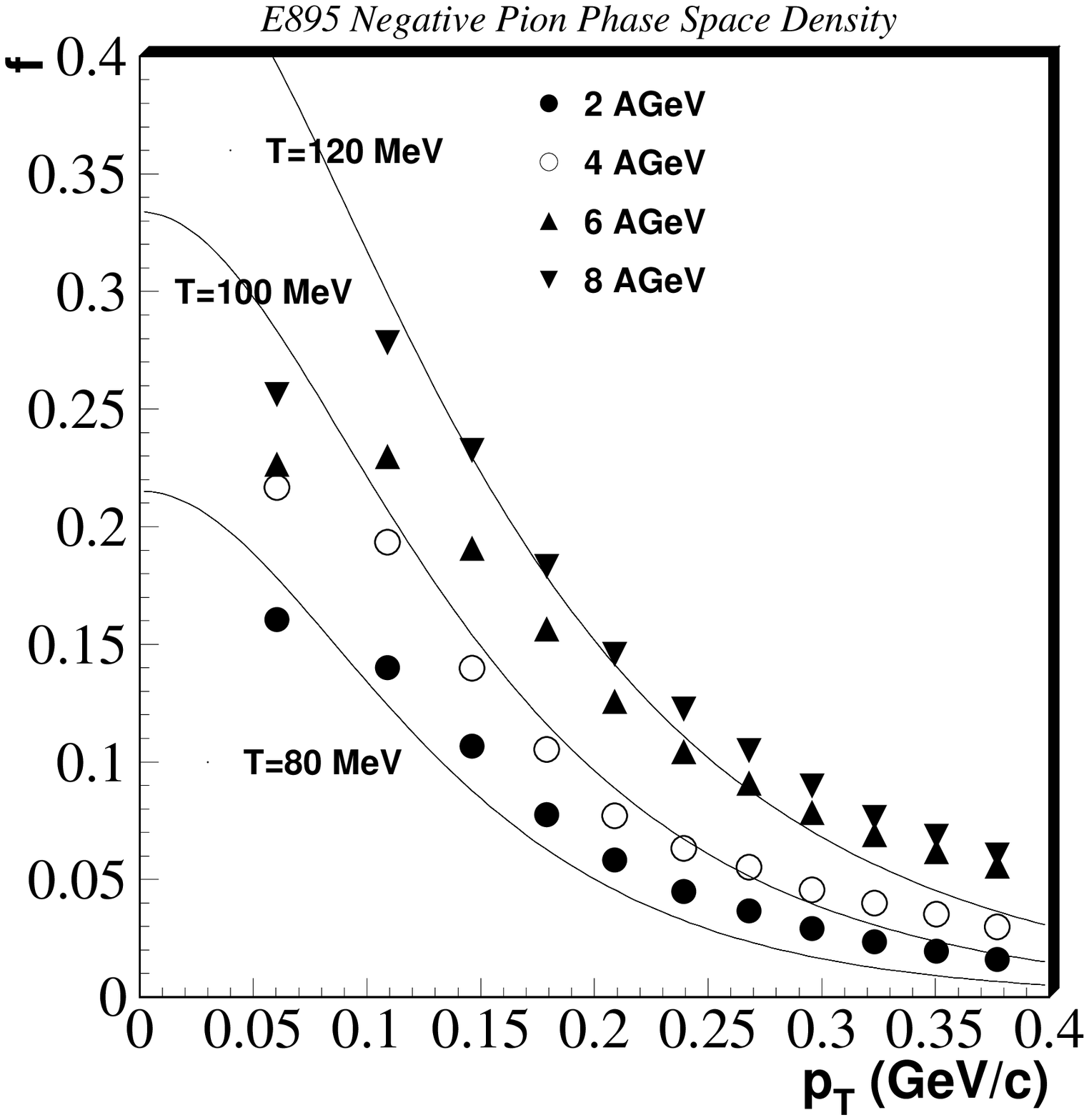}}
\begin{minipage}[b]{3.5cm}
\caption{Preliminary
data on the average phase-space density as 
measured by the E895 Collaboration for different AGS projectile 
energies as a function of $p_\perp$\protect\cite{e895-qm01}. 
The lines show Bose-Einstein distributions of given temperature.
This does not account for the effects of expansion. 
}\label{f:e895-psd}
\end{minipage}}
\end{figure}
\subsubsection{Average phase-space density}
\label{ss:psd}

\noindent
\underline{Idea and formalism:}
The spatial average of the phase-space density of pions at times
later than freeze-out $t_f$

  \begin{equation}
    \langle f\rangle(\bm{p}) = 
    {{\int d^3x\, f^2(\bm{x},\bm{p},t>t_f)}\over
    {\int d^3x\, f(\bm{x},\bm{p},t>t_f)}}\, ,
     \label{3.59b}
  \end{equation}
can be expressed in terms of the observable one- and two-particle 
spectra\cite{Bertsch:qc,e877-qm96hbt,Ferenc:1999ku,Tomasik:1999cq} 
\begin{eqnarray}
    \langle f\rangle(K_\perp,Y) & = & \frac{1}{\sqrt{\lambda}}\, 
       \frac{1}{E}\, 
       {dN\over dY\, M_\perp dM_\perp\, d\Phi}\, 
        {1\over V_{\rm eff}(K_\perp,Y)}\, ,
    \label{3.61} \\
    V_{\rm eff}(K_\perp,Y) &=&  
      \left [ \int d^3\bm{q}\, (C(\bm{q},\bm{K})-1) \right ]^{-1} \,. 
    \label{veff}
\end{eqnarray}
The phase-space density is determined by the average number of pions
with given momentum (the {\em non}-invariant spectrum) 
divided by the volume in which they are contained.
The factor $\lambda^{-1/2}$ corrects for the ``purity'' of the sample:
it ensures that only directly produced pions and not those coming
from resonance decays are taken into 
account (see discussion at the end of section~\ref{sec2aa}). 
As long as lifetime effects are small (which is consistent with all
data measured so far, see Sec.~\ref{RooRs}), the volume is given by 
(\ref{veff}). For the Cartesian BP parametrisation, it takes the form
\begin{equation}
    V_{\rm eff}(K_\perp,Y) = \frac{1}{\pi^{3/2}}
    R_s({\bm{K}})\sqrt{R_o^2({\bm{K}})\, R_l^2({\bm{K}}) - 
        (R_{ol}^2({\bm{K}}))^2}\, .
    \label{3.62}
\end{equation}
%

\begin{figure}[t]\epsfxsize=11cm 
\centerline{\epsfbox{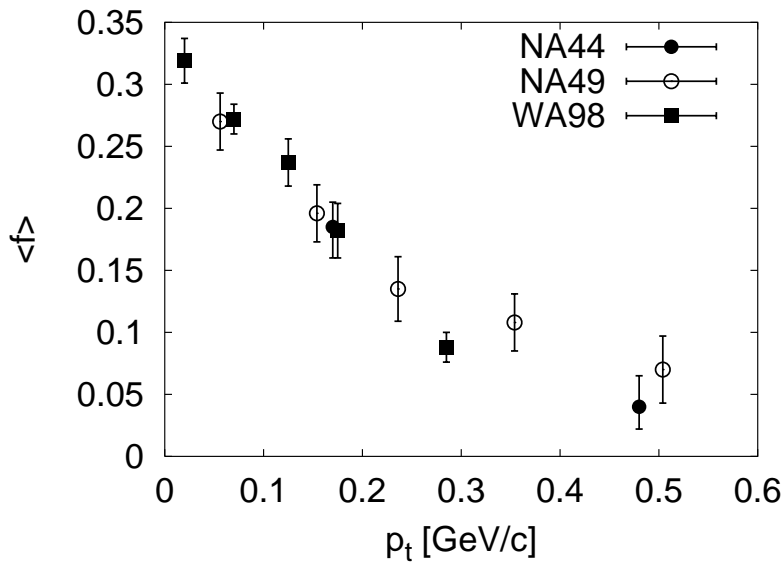}}
\caption{The average pion phase-space density measured by 
NA44\protect\cite{murray:psd}, 
NA49\protect\cite{Ferenc:1999ku}, and 
WA98\protect\cite{wa98inprep} in 158~$A$GeV Pb+Pb collisions
at mid-rapidity.
}\label{f:sps-psd}
\end{figure}
\noindent
\underline{Thermal model as a reference:}
For reference, it is customary to compare the measured $\langle f \rangle$
to that of a thermal distribution of a given temperature and chemical 
potential. Since $\langle f \rangle$ is a spatial average, it averages
for an expanding source over homogeneity regions which move relatively 
to the point of maximum emissivity. This leads to deviations from 
a naive Bose-Einstein phase-space density which for the case of a 
boost-invariant longitudinally expanding pion source with
transverse expansion can be written as\cite{Tomasik:2001uz}
 \begin{eqnarray}
    \langle f \rangle (p_\perp) &=& 
    \left( \sum_{n=1}^{\infty} A_n(p_\perp) \right)
    \Bigg / \left(\sum_{n=2}^{\infty} (n{-}1) A_n(p_\perp) \right)\, ,
 \label{E:res-main} \\
    A_n(p_\perp) &=& 4 \pi m_\perp \tau_{\rm f} \int_0^\infty r\,dr\, 
       e^{n[\mu(r)/T]}
 \nonumber\\
    &&\times {\rm I}_{0}\bigl(n\,p_\perp\sinh\zeta(r)/T\bigr)\,
       {\rm K}_{1}\bigl(n\,m_\perp\cosh\zeta(r)/T\bigr)\,.
 \label{E:res-A3}
\end{eqnarray}
%
\begin{figure}[t]\epsfxsize=11.3cm 
\centerline{\epsfbox{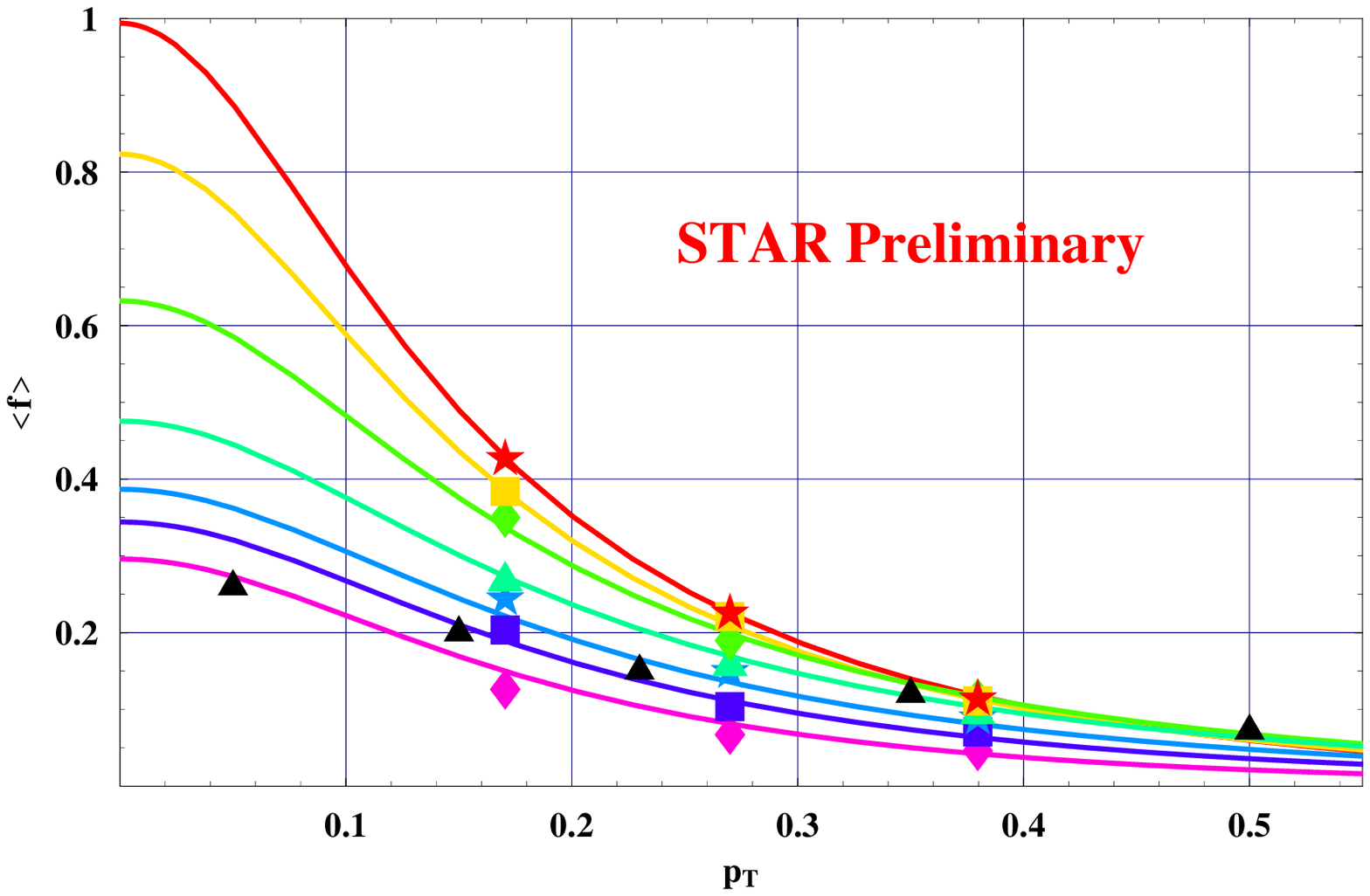}}
\caption{The average pion phase-space density as measured by STAR
Collaboration\protect\cite{STARQM02} for Au+Au collisions
at $\sqrt{s} = 130\, A\mbox{GeV}/c$. 
Different symbols correspond to centralities (from top to bottom)
0-5\%, 5-10\%, 10-20\%, 20-30\%, 30-40\%, 40-50\%, and 50-80\%
of the total cross section. Curves are fits to eq.~(\ref{E:res-main}). 
The same thermal model also reproduces the single-particle spectrum.
Triangles represent the NA49 data\protect\cite{Ferenc:1999ku} 
for 158~$A$GeV Pb+Pb collisions at mid-rapidity.
}\label{f:starpsd}
\end{figure}
%
Here, the transverse rapidity profile is characterised by $\zeta(r)$ and
the transverse density profile by $\mu(r)/T$. For illustration, 
the transverse geometry can be modelled e.g. by
a box profile
\begin{equation}
\label{E:box}
\mbox{[box]} \qquad \mu(r) = 
\left \{ 
\begin{array}{ll}
  \mu_B \quad \mbox{for} \quad & r\le R_{\rm box} \\
  - \infty \quad & \mbox{otherwise}
\end{array} , \right .
\end{equation}
or a Gaussian profile 
\begin{equation}
\label{E:Gauss}
\mbox{[Gauss]} \qquad \mu (r) = \mu_G - T \frac{r^2}{2R^2_{\rm Gauss}}\, .
\end{equation}
The chemical potential $\mu(r)$ introduced here are 
position-dependent and result in a non-uniform density profile;
$\mu_B$ or $\mu_G$ are the values in the centre of the fireball.
A certain spatial average of this chemical potential determines
the particle multiplicities\cite{Tomasik:2001uz}. For comparison
of the present formalism to AGS data, the assumption of 
boost-invariance entering (\ref{E:res-main}) has to be modified 
by a rapidity cut-off.

\noindent
\underline{Data:}
The average phase-space density was studied first by the
E877 experiment for Au+Au collisions at the 
AGS\cite{e877-qm96hbt,Barrette:1997fj}
in the projectile fragmentation region where it was found to
decrease with increasing rapidity. The E895 collaboration 
measured the average phase-space density at mid-rapidity at 
2,~4,~6, and 8~$A$GeV and observed its increase with 
the collision energy\cite{e895-qm01} (Fig.~\ref{f:e895-psd}).

At SPS energies there is an extensive compilation of phase-space
densities for various collision systems based on data 
of the NA35/NA49 collaboration\cite{Ferenc:1999ku}, the 
NA44 measurements of S+S, S+Pb, and Pb+Pb 
collisions\cite{murray:psd}, and the Pb+Pb data measured 
by WA98. For Pb+Pb collisions at 158~$A$GeV, 
the results of these experiments are mutually consistent, 
see Fig.~\ref{f:sps-psd}.

At RHIC energy, preliminary data indicate that the average pion 
phase-space density is significantly higher than at SPS. Moreover, the 
STAR collaboration reported a strong dependence of the phase-space
density on centrality\cite{STARQM02} (Fig.~\ref{f:starpsd}).
The origin of this increase, as well as the apparently different
$p_\perp$-dependence of $\langle f\rangle$ at RHIC and SPS, is 
currently under study. 
%

\begin{figure}[t]\epsfxsize=11cm 
\begin{center}
\psfig{bbllx= 110pt, bblly=265pt, bburx=400pt, bbury=515pt,
file=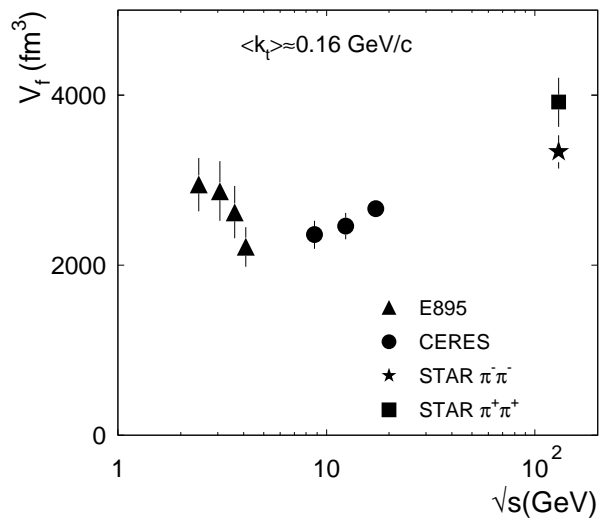, height= 7cm , clip=}
\end{center}
\caption{Dependence
of the ``freeze-out volume'' $V_f = (2\pi)^{3/2}R_s^2R_l$
on $\sqrt{s}$. 
Data are measured by E895\protect\cite{e895-edep},
CERES\protect\cite{Adamova:2002wi} and STAR\protect\cite{Adler:2001zd}
for low-$p_\perp$ pions at mid-rapidity.
}\label{f:volpart}
\end{figure}
\subsubsection{Energy and multiplicity dependence}
Despite the weak dependence of HBT radius parameters on 
centre of mass energy, the volume $V_f$ from which particles decouple 
shows an interesting
non-monotonous behaviour, see Fig.~\ref{f:volpart}. 
Here the volume is estimated as 
$V_f\propto R_s^2 R_l$. The radius $R_o$ is not used in calculating 
$V_f$ because it contains contributions from the temporal extent
of the source, see eq.~(\ref{eq15}). The estimate $V_f\propto R_s^2 R_l$
does not take into account the effects of the expansion on
the HBT radii. Figure~\ref{f:volpart} shows that the volume decreases
gradually within the AGS energy range, reaches a minimum around
the highest AGS energies and then increases monotonously by almost
a factor 2 up to the highest RHIC energy. 

\begin{figure}[t]\epsfxsize=11cm 
\begin{center}
\psfig{bbllx= 30pt, bblly=310pt, bburx=270pt, bbury=540pt,
file=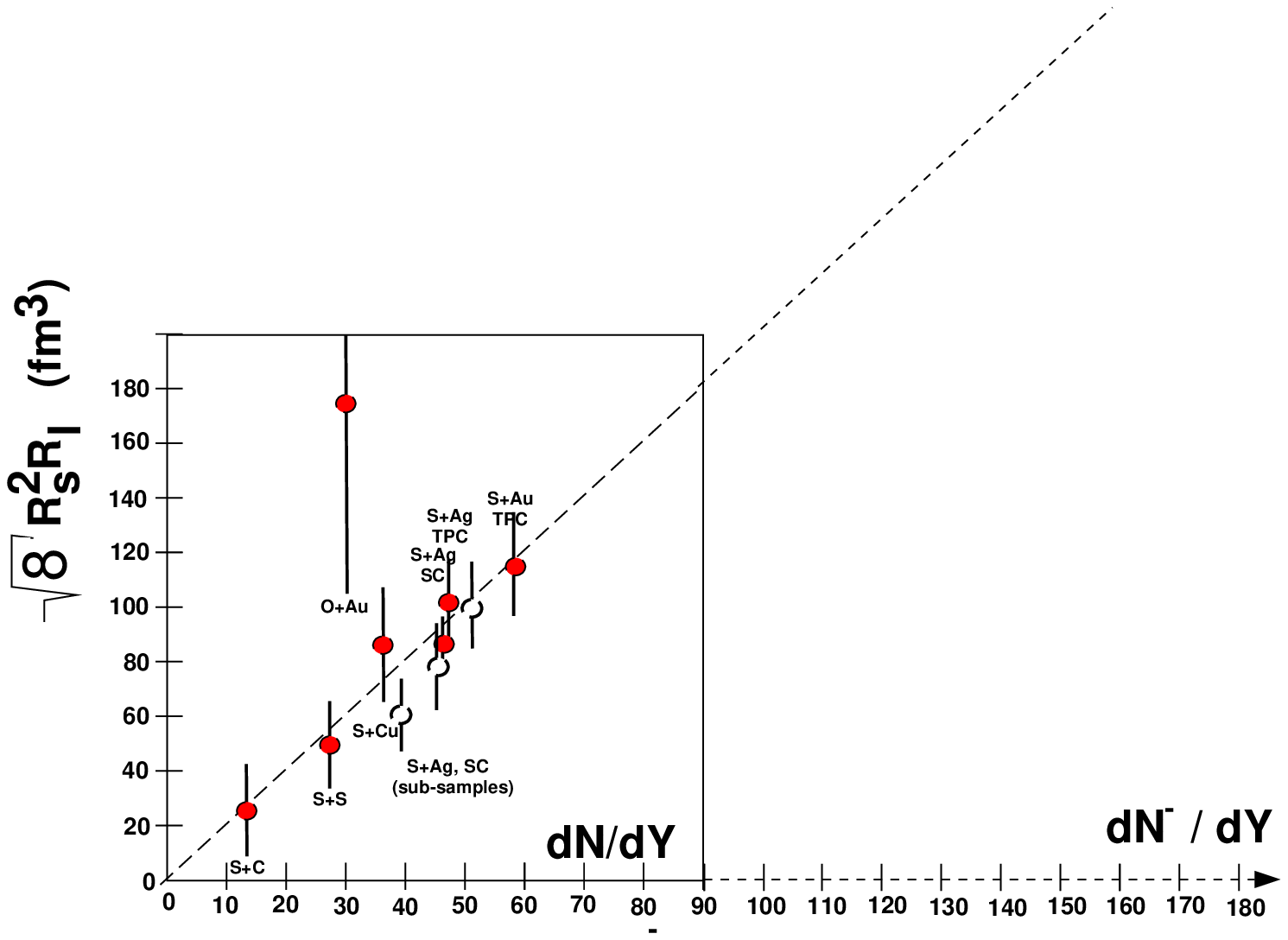, height= 7cm , clip=}
\end{center}
\caption{The ``freeze-out volume'' ($V_f \propto \sqrt{8}R_s^2R_l$)
as a function of the number of produced negative particles per unit
of rapidity. Data are by NA35\protect\cite{na35-S} 
for different S-induced reactions  at 200~$A$GeV beam energy.
}\label{na35volume}
\end{figure}

At approximately fixed
centre of mass energy at the SPS, the same freeze-out volume 
was found previously to grow linearly with the charged particle
multiplicity per unit rapidity, see Fig.~\ref{na35volume}. Consistent
with this finding is the centrality dependence of the freeze-out
volume at SPS energies which grows linearly with the number of
participants\cite{Adamova:2002wi}. These two observations support
the conjecture that freeze-out occurs at a fixed particle density.
This is, however, contradicted by the non-monotonous energy 
dependence of $V_f$. A linear relation between freeze-out volume 
and particle multiplicity does not hold. 

Particle density, and thus particle multiplicity is certainly
important in characterising the freeze-out condition, since 
it affects the hadronic escape probability from the medium.
However, chemical composition, collective expansion and the
momentum of the escaping particle are other factors which
determine this escape probability\cite{Tomasik:2002qt}.
To illustrate this, one can consider e.g. the mean free path of 
a pion at freeze-out\cite{Adamova:2002ff}
\begin{equation}
\label{mfp}
  \lambda_{\rm mfp}^{-1} = 2n_{p+\bar p} \sigma_{\pi N} +   
  3 n_{\pi^-}\sigma_{\pi\pi} = 2\frac{N_{p+\bar p}}{V_f} \sigma_{\pi N}
  +  3 \frac{N_{\pi^-}}{V_f} \sigma_{\pi\pi}\, .
\end{equation}
While the freeze-out volume $V_f$ and the numbers $N_i$ of particles of
species $i$ contained in $V_f$ both depend significantly on the centre of
mass energy, the mean free path (\ref{mfp}) is approximately
$\sqrt{s}$-independent\cite{Adamova:2002ff}.

To understand in more detail how the interplay of different 
properties of the collision region determines the freeze-out volume, 
a realistic freeze-out criterion is required. A good  
starting point is the particle escape probability from the hot
and dense but rapidly expanding collision 
region\cite{Grassi:1994ng,Sinyukov:2002if}
\begin{equation}
     {\cal P}(x,p,\tau) = 
     \exp \left ( - \int_\tau^\infty \, d\bar{\tau} \, 
         {\cal R}(x+v\bar{\tau},p) \right )\, .
     \label{freezeoutprob}
\end{equation}
Here, $v$ is the velocity of the escaping particle and ${\cal R}(x,p)$ 
denotes the scattering rate which is defined as the inverse of the 
mean time between collisions for a particle at position $x$ with 
momentum $p$. Freeze-out at different centre of mass energies is 
then assumed to occur when the probability ${\cal P}$ reaches 
a characteristic value. 

Recently, the scattering rate ${\cal R}(x,p)$ was calculated
for a full hadron resonance gas with chemical composition corresponding 
to SPS and RHIC energies\cite{Tomasik:2002qt}. The main observation
is that the scattering rate shows a significant momentum dependence
suggesting that particles of different momenta are emitted at
different times. This effect is neglected in hydrodynamic simulations
which are based on the Cooper-Frye prescription\cite{Cooper:1974mv} for 
freeze-out along a sharp three-dimensional hypersurface. Deviations
from this Cooper-Frye prescription, i.e. freeze-out along finite
four volumes may affect the transverse momentum slope of HBT radius 
parameters\cite{Grassi:2000ke}, and the 
momentum dependence of the freeze-out volume. The role of a change 
in the chemical composition from SPS to RHIC was found to be relatively
small in spite of the large increase of pion phase-space density
(Fig.~\ref{f:starpsd}). This is a consequence of the small pion
contribution to the total scattering rate, resulting from the
comparatively small pion-pion cross-section. 
In contrast, collective transverse expansion gradients affect
the freeze-out volume significantly. The reason is that an increase in the 
scattering rate at freeze-out can be compensated by stronger transverse 
flow gradients which lead to a faster density decrease in the 
collision region thus keeping the opacity integral in the
exponent of (\ref{freezeoutprob}) constant.
The possible effect of the flow gradients on the $\sqrt{s}$-dependence
of the freeze-out volume (Fig.~\ref{f:volpart}) remains to be studied.

\subsubsection{Azimuthal dependence of HBT radius parameters}
\label{ss-azdep}
%
\begin{figure}[t]\epsfxsize=11cm 
\centerline{\epsfbox{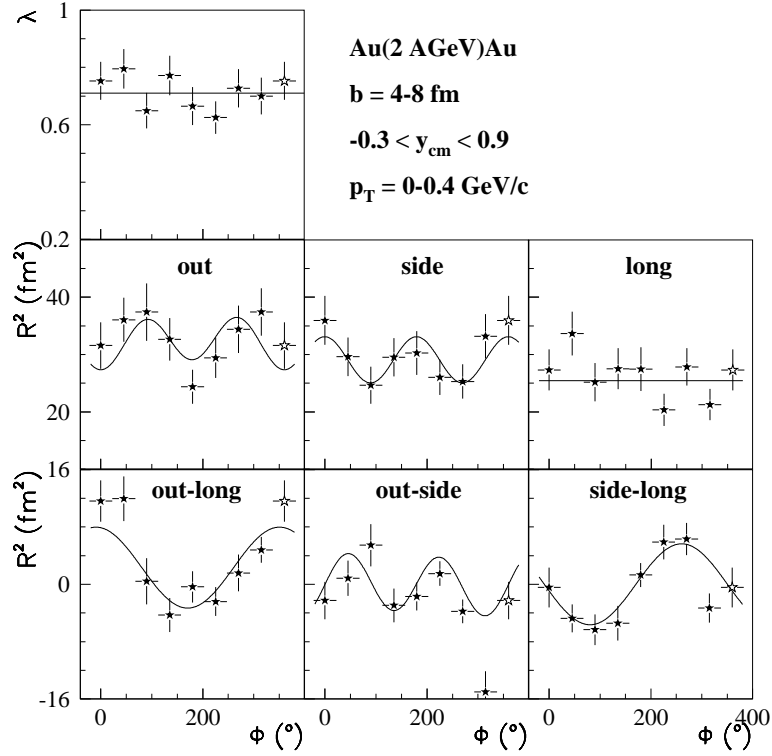}}
\caption{Azimuthal dependence of HBT radius parameters as published
by the E895 Collaboration~\protect\cite{Lisa:2000xj} for Au+Au 
collisions 
at 2 $A$GeV. Curves correspond to a static source 
(\protect$\langle {\tilde t}^2 \rangle = 0$) according to the equations
(\protect\ref{eq43})-(\protect\ref{eq48}).
}\label{e895azi}
\end{figure}
%
Two years ago, the first measurements of the $\Phi$-dependence
of HBT radius parameters were published by the E895 
Collaboration\cite{Lisa:2000xj} for beam energies of 2, 4 and 6 $A$GeV in 
semi-peripheral Au+Au collisions at the AGS. Results at all three
energies show a sizeable first order harmonics in the $\Phi$-dependence
of $R_{ol}^2$ and $R_{sl}^2$, and comparatively small second order 
harmonics in the fully transverse HBT radii $R_o^2$, $R_s^2$ and
$R_{os}^2$. The longitudinal radius parameter and the intercept
$\lambda$ are consistent with a $\Phi$-independent ansatz.
According to (\ref{eq51}), the first order harmonics allow
to reconstruct the angle $\Theta$ by which the emission ellipsoid 
is tilted out of the beam axis, see Fig.~\ref{tilttilt}. 
At lower AGS energies, this angle is with $\Theta \approx 30^\circ$
surprisingly large. This value is consistent with RQMD transport model
simulations\cite{Lisa:2000ip}. Interestingly, the spatial tilt
is found to point in the direction opposite to the directed flow in 
momentum space. This indicates that at lower AGS energies pion 
reflection from the bulk of the matter rather than pion absorption
by this matter is at the root of the observed direct flow signal.

With increasing centre of mass energy, a longitudinally approximately
boost-invariant region develops around mid-rapidity.
As a consequence, the tilt angle $\Theta$ of the emission ellipsoid
is expected to decrease with increasing $\sqrt{s}$. However, there is
so far no measurement of $R_{ol}^2$ and $R_{sl}^2$ at higher energies,
which would be required to establish this effect experimentally.

The $\Phi$-dependence of the fully transverse radius parameters 
$R_o^2$, $R_{os}^2$ and $R_s^2$ is easier to measure than that of
$R_{ol}^2$ and $R_{sl}^2$: while the former require the event-wise
reconstruction of the orientation of the reaction plane, the latter
require in addition the direction in which the impact parameter
points. Due to this complication, at RHIC first preliminary data 
are available for the fully transverse radius parameters 
only\cite{star02-HBT}. These data are expected to contain information
about whether the spatial orientation of the source at freeze-out
is in-plane or out-of-plane. However, statistical and systematic
uncertainties in these preliminary data are still too large to 
draw conclusions. A significant improvement in statistics 
is expected within the next run.

\section*{Acknowledgements}
\addcontentsline{toc}{section}{Acknowledgements}

We are indebted to many colleagues for giving helpful informations.
Thanks go in particular to Jan Pi\v s\'ut for a critical reading of 
the manuscript. We thank Giuseppe Bruno, John Cramer, Mike Lisa, and 
Piotr Skowro\'nski for their comments to the manuscript.  We
profited from discussions with Harry Appelsh\"auser, Ulrich Heinz,
and Mike Lisa.
We are also very grateful to Harry Appelsh\"auser, Giuseppe Bruno,
John Cramer, Peter Filip, Mike Lisa, Michael Murray, Laurent Rosselet, 
and Peter Seyboth who provided us with some of the data 
reviewed here.


\begin{thebibliography}{000}

\centerline{\underline{Theoretical papers:}}

\vspace{0.5cm}
\bibitem{Aichelin:1996iu}
J.~Aichelin,
Nucl.\ Phys.\ A {\bf 617} (1997) 510.
%
\bibitem{Akkelin:sg}
S.~V.~Akkelin and Y.~M.~Sinyukov,
Z.\ Phys.\ C {\bf 72} (1996) 501.
%
\bibitem{Akkelin:2001wv}
S.~V.~Akkelin, P.~Braun-Munzinger and Y.~M.~Sinyukov,
Nucl.\ Phys.\ A {\bf 710} (2002) 439.
%
\bibitem{Anchishkin:1997tb}
D.~Anchishkin, U.~W.~Heinz and P.~Renk,
Phys.\ Rev.\ C {\bf 57} (1998) 1428.
%
\bibitem{Barz:gr}
H.~W.~Barz,
Phys.\ Rev.\ C {\bf 53} (1996) 2536.
%
\bibitem{Bass:1998ca}
S.~A.~Bass {\it et al.},
Prog.\ Part.\ Nucl.\ Phys.\  {\bf 41} (1998) 225.
%
\bibitem{Baym:1996wk}
G.~Baym and P.~Braun-Munzinger,
Nucl.\ Phys.\ A {\bf 610} (1996) 286C.
%
\bibitem{Bertsch:1988db}
G.~F.~Bertsch, M.~Gong and M.~Tohyama,
Phys.\ Rev.\ C {\bf 37} (1988) 1896.
%
\bibitem{Bertsch:vn}
G.~F.~Bertsch,
Nucl.\ Phys.\ A {\bf 498} (1989) 173C.
%
\bibitem{Bertsch:1993nx}
G.~F.~Bertsch, P.~Danielewicz and M.~Herrmann,
Phys.\ Rev.\ C {\bf 49} (1994) 442.
%
\bibitem{Bertsch:qc}
G.~F.~Bertsch,
Phys.\ Rev.\ Lett.\  {\bf 72} (1994) 2349
[Erratum-ibid.\  {\bf 77} (1994) 789].
%
\bibitem{Boal:yh}
D.~H.~Boal, C.~K.~Gelbke and B.~K.~Jennings,
Rev.\ Mod.\ Phys.\  {\bf 62} (1990) 553.
%
\bibitem{Bolz:1992hc}
J.~Bolz, U.~Ornik, M.~Pl\"umer, B.~R.~Schlei and R.~M.~Weiner,
Phys.\ Rev.\ D {\bf 47} (1993) 3860.
%
\bibitem{Bowler:vx}
M.~G.~Bowler,
Phys.\ Lett.\ B {\bf 270} (1991) 69.
%
\bibitem{Brown:1997sn}
D.~A.~Brown and P.~Danielewicz,
Phys.\ Rev.\ C {\bf 57} (1998) 2474.
%
\bibitem{Brown:2000aj}
D.~A.~Brown and P.~Danielewicz,
Phys.\ Rev.\ C {\bf 64} (2001) 014902.
%
\bibitem{Chapman:xa}
S.~Chapman and U.~W.~Heinz,
Phys.\ Lett.\ B {\bf 340} (1994) 250.
%
\bibitem{Chapman:yv}
S.~Chapman, P.~Scotto and U.~W.~Heinz,
Phys.\ Rev.\ Lett.\  {\bf 74} (1995) 4400.
%
\bibitem{Chapman:1994ax}
S.~Chapman, P.~Scotto and U.~W.~Heinz,
Heavy Ion Phys.\  {\bf 1} (1995) 1.
%
\bibitem{Chapman:1995nz}
S.~Chapman, J.~R.~Nix and U.~W.~Heinz,
Phys.\ Rev.\ C {\bf 52} (1995) 2694.
%
\bibitem{Cocconi:1974pr}
G.~Cocconi,
Phys.\ Lett.\ B {\bf 49} (1974) 459.
%
\bibitem{Cooper:1974mv}
F.~Cooper and G.~Frye,
Phys.\ Rev.\ D {\bf 10} (1974) 186.
%
\bibitem{Csorgo:1994in}
T.~Cs\"org\H o, B.~L\"orstad and J.~Zim\'any,
Z.\ Phys.\ C {\bf 71} (1996) 491.
%
\bibitem{Csorgo:1995bi}
T.~Cs\"org\H o and B.~L\"orstad,
Phys.\ Rev.\ C {\bf 54} (1996) 1390.
%
\bibitem{Csorgo:1999sj}
T.~Cs\"org\H o,
arXiv:hep-ph/0001233.
%
\bibitem{Csorgo:2002ry}
T.~Cs\"org\H o and A.~Ster,
arXiv:nucl-th/0207016.
%
\bibitem{Goldhaber:sf}
G.~Goldhaber, S.~Goldhaber, W.~Y.~Lee and A.~Pais,
Phys.\ Rev.\  {\bf 120} (1960) 300.
%
\bibitem{Grassberger:1976au}
P.~Grassberger,
Nucl.\ Phys.\ B {\bf 120} (1977) 231.
%
\bibitem{Grassi:1994ng}
F.~Grassi, Y.~Hama and T.~Kodama,
Z.\ Phys.\ C {\bf 73}, 153 (1996).
%
\bibitem{Grassi:2000ke}
F.~Grassi, Y.~Hama, S.~S.~Padula and O.~J.~Socolowski,
Phys.\ Rev.\ C {\bf 62} (2000) 044904.
%
\bibitem{Gyulassy:yi}
M.~Gyulassy, S.~K.~Kauffmann and L.~W.~Wilson,
Phys.\ Rev.\ C {\bf 20} (1979) 2267.
%
\bibitem{Gyulassy:xb}
M.~Gyulassy and S.~K.~Kauffmann,
Nucl.\ Phys.\ A {\bf 362} (1981) 503.
%
\bibitem{Ferenc:1999ku}
D.~Ferenc, U.~W.~Heinz, B.~Tom\'a\v sik, U.~A.~Wiedemann and J.~G.~Cramer,
Phys.\ Lett.\ B {\bf 457} (1999) 347.
%
\bibitem{HanburyBrown:1954wr}
R.~Hanbury Brown and R.~Q.~Twiss,
Phil.\ Mag.\  {\bf 45} (1954) 663.
%
\bibitem{HanburyBrown:1956pf}
R.~Hanbury Brown and R.~Q.~Twiss,
Nature {\bf 178} (1956) 1046.
%
\bibitem{Heinz:1996qu}
U.~W.~Heinz, B.~Tom\'a\v sik, U.~A.~Wiedemann and Y.~F.~Wu,
Phys.\ Lett.\ B {\bf 382} (1996) 181.
%
\bibitem{Heinz:1999rw}
U.~W.~Heinz and B.~V.~Jacak,
Ann.\ Rev.\ Nucl.\ Part.\ Sci.\  {\bf 49} (1999) 529.
%
\bibitem{Heinz:2002un}
U.~W.~Heinz and P.~F.~Kolb,
arXiv:hep-ph/0204061.
%
\bibitem{Heinz:2002au}
U.~W.~Heinz, A.~Hummel, M.~A.~Lisa and U.~A.~Wiedemann,
Phys.\ Rev.\ C {\bf 66} (2002) 044903.
%
\bibitem{Heiselberg:1997vh}
H.~Heiselberg and A.~P.~Vischer,
Eur.\ Phys.\ J.\ C {\bf 1} (1998) 593.
%
\bibitem{Heiselberg:1998ik}
H.~Heiselberg,
Phys.\ Rev.\ Lett.\  {\bf 82} (1999) 2052.
%
\bibitem{Heiselberg:1998es}
H.~Heiselberg and A.~M.~Levy,
Phys.\ Rev.\ C {\bf 59} (1999) 2716.
%
\bibitem{Herrmann:1994rr}
M.~Herrmann and G.~F.~Bertsch,
Phys.\ Rev.\ C {\bf 51} (1995) 328.
%
\bibitem{Hirano:2001yi}
T.~Hirano, K.~Morita, S.~Muroya and C.~Nonaka,
Phys.\ Rev.\ C {\bf 65} (2002) 061902.
%
\bibitem{Humanic:ya}
T.~J.~Humanic,
Phys.\ Rev.\ C {\bf 50} (1994) 2525.
%
\bibitem{Humanic:vd}
T.~J.~Humanic,
Phys.\ Rev.\ C {\bf 53} (1996) 901.
%
\bibitem{Humanic:2002iw}
T.~J.~Humanic,
arXiv:nucl-th/0205053.
%
\bibitem{Kittel:2001zw}
W.~Kittel,
Acta Phys.\ Polon.\ B {\bf 32} (2001) 3927.
%
\bibitem{Kolb:2000sd}
P.~F.~Kolb, J.~Sollfrank and U.~W.~Heinz,
Phys.\ Rev.\ C {\bf 62} (2000) 054909.
%
\bibitem{Koonin:fh}
S.~E.~Koonin,
Phys.\ Lett.\ B {\bf 70} (1977) 43.
%
\bibitem{Kopylov:qw}
G.~I.~Kopylov and M.~I.~Podgoretsky,
Sov.\ J.\ Nucl.\ Phys.\  {\bf 15} (1972) 219
[Yad.\ Fiz.\  {\bf 15} (1972) 392].
%
\bibitem{Kopylov:qq}
G.~I.~Kopylov and M.~I.~Podgoretsky,
Sov.\ J.\ Nucl.\ Phys.\  {\bf 18} (1974) 336
[Yad.\ Fiz.\  {\bf 18} (1974) 656].
%
\bibitem{Lednicky:1981su}
R.~Lednick\'y and V.~L.~Lyuboshits,
Sov.\ J.\ Nucl.\ Phys.\  {\bf 35} (1982) 770
[Yad.\ Fiz.\  {\bf 35} (1982) 1316].
%
\bibitem{Lednicky:hp}
R.~Lednick\'y and V.~L.~Lyuboshits,
Heavy Ion Phys.\  {\bf 3} (1996) 93.
%
\bibitem{Lednicky:vk}
R.~Lednick\'y, V.~L.~Lyuboshits, B.~Erazmus and D.~Nouais,
Phys.\ Lett.\ B {\bf 373} (1996) 30.
%
\bibitem{Lin:2002gc}
Z.W.~Lin, C.M.~Ko and S.~Pal,
arXiv:nucl-th/0204054.
%
\bibitem{Lisa:2000ip}
M.~A.~Lisa, U.~W.~Heinz and U.~A.~Wiedemann,
Phys.\ Lett.\ B {\bf 489} (2000) 287.
%
\bibitem{Makhlin:1987gm}
A.~N.~Makhlin and Y.~M.~Sinyukov,
Z.\ Phys.\ C {\bf 39} (1988) 69.
%
\bibitem{McLerran:2002dt}
L.~McLerran and S.~S.~Padula,
arXiv:nucl-th/0205028.
%
\bibitem{Molnar:2002bz}
D.~Moln\'ar and M.~Gyulassy,
arXiv:nucl-th/0211017.
%
\bibitem{Morita:1999vj}
K.~Morita, S.~Muroya, H.~Nakamura and C.~Nonaka,
Phys.\ Rev.\ C {\bf 61} (2000) 034904.
%
\bibitem{Morita:2002av}
K.~Morita, S.~Muroya, C.~Nonaka and T.~Hirano,
Phys.\ Rev.\ C {\bf 66} (2002) 054904.
%
\bibitem{Ollitrault:1997vz}
J.~Y.~Ollitrault,
Nucl.\ Phys.\ A {\bf 638} (1998) 195C.
%
\bibitem{Podgoretsky:1982xu}
M.~I.~Podgoretsky,
Sov.\ J.\ Nucl.\ Phys.\  {\bf 37} (1983) 272
[Yad.\ Fiz.\  {\bf 37} (1983) 455].
%
\bibitem{Poskanzer:1998yz}
A.~M.~Poskanzer and S.~A.~Voloshin,
Phys.\ Rev.\ C {\bf 58} (1998) 1671.
%
\bibitem{Pratt:su}
S.~Pratt,
Phys.\ Rev.\ Lett.\  {\bf 53} (1984) 1219.
%
\bibitem{Pratt:ev}
S.~Pratt,
Phys.\ Rev.\ D {\bf 33} (1986) 72.
%
\bibitem{Pratt:cc}
S.~Pratt,
Phys.\ Rev.\ D {\bf 33} (1986) 1314.
%
\bibitem{Pratt:zq}
S.~Pratt, T.~Cs\"org\H o and J.~Zim\'any,
Phys.\ Rev.\ C {\bf 42} (1990) 2646.
%
\bibitem{crab}
see S.~Pratt's web page: \\
\verb+http://www.nscl.msu.edu/~pratt/freecodes/crab/home.html+.
%
\bibitem{Ray:qj}
R.~L.~Ray,
Phys.\ Rev.\ C {\bf 57} (1998) 2523.
%
\bibitem{Rischke:1995cm}
D.~H.~Rischke and M.~Gyulassy,
Nucl.\ Phys.\ A {\bf 597} (1996) 701.
%
\bibitem{Rischke:1996em}
D.~H.~Rischke and M.~Gyulassy,
Nucl.\ Phys.\ A {\bf 608} (1996) 479.
%
\bibitem{Schlei:1992jj}
B.~R.~Schlei, U.~Ornik, M.~Plumer and R.~M.~Weiner,
Phys.\ Lett.\ B {\bf 293} (1992) 275.
\bibitem{Schlei:1996md}
B.~R.~Schlei,
Phys.\ Rev.\ C {\bf 55} (1997) 954.
%
\bibitem{Schlei:1998zy}
B.~R.~Schlei, D.~Strottman, J.~P.~Sullivan and H.~W.~van Hecke,
Eur.\ Phys.\ J.\ C {\bf 10} (1999) 483.
%
\bibitem{Shuryak:1972kq}
E.~V.~Shuryak,
Phys.\ Lett.\ B {\bf 44}, 387 (1973).
%
\bibitem{Sinyukov:1998fc}
Y.~Sinyukov, R.~Lednick\'y, S.~V.~Akkelin, J.~Pluta and B.~Erazmus,
Phys.\ Lett.\ B {\bf 432} (1998) 248.
%
\bibitem{Sinyukov:2002if}
Y.~M.~Sinyukov, S.~V.~Akkelin and Y.~Hama,
Phys.~Rev.~Lett.~{\bf 89} (2002) 052301.
%
\bibitem{Soff:2000eh}
S.~Soff, S.~A.~Bass and A.~Dumitru,
Phys.\ Rev.\ Lett.\  {\bf 86} (2001) 3981.
%
\bibitem{Soff:2001hc}
S.~Soff, S.~A.~Bass, D.~H.~Hardtke and S.~Y.~Panitkin,
Phys.\ Rev.\ Lett.\  {\bf 88} (2002) 072301
%
\bibitem{Soff:2002qw}
S.~Soff,
arXiv:hep-ph/0202240.
%
\bibitem{Sorge:vt}
H.~Sorge, H.~St\"ocker and W.~Greiner,
Nucl.\ Phys.\ A {\bf 498} (1989) 567C.
%
\bibitem{Sorge:dy}
H.~Sorge, H.~St\"ocker and W.~Greiner,
Annals Phys.\  {\bf 192} (1989) 266.
%
\bibitem{Ster:1999ib}
A.~Ster, T.~Cs\"org\H o and B.~L\"orstad,
Nucl.\ Phys.\ A {\bf 661} (1999) 419.
%
\bibitem{Sullivan:wb}
J.~P.~Sullivan {\it et al.},
Phys.\ Rev.\ Lett.\  {\bf 70} (1993) 3000.
%
\bibitem{Tomasik:1997eq}
B.~Tom\'a\v sik and U.~W.~Heinz,
Eur.\ Phys.\ J.\ C {\bf 4} (1998) 327.
%
\bibitem{Tomasik:1999kz}
B.~Tom\'a\v sik and U.~W.~Heinz,
Acta Phys.\ Slov.\  {\bf 49} (1999) 251.
%
\bibitem{Tomasik:1999cq}
B.~Tom\'a\v sik, U.~A.~Wiedemann and U.~W.~Heinz,
arXiv:nucl-th/9907096.
%
\bibitem{Tomasik:2001uz}
B.~Tom\'a\v sik and U.~W.~Heinz,
Phys.\ Rev.\ C {\bf 65} (2002) 031902.
%
\bibitem{Tomasik:2002qt}
B.~Tom\'a\v sik and U.~A.~Wiedemann,
arXiv:nucl-th/0207074.
%
\bibitem{Voloshin:1995mc}
S.~A.~Voloshin and W.~E.~Cleland,
Phys.\ Rev.\ C {\bf 53} (1996) 896.
%
\bibitem{Voloshin:1996ch}
S.~A.~Voloshin and W.~E.~Cleland,
Phys.\ Rev.\ C {\bf 54} (1996) 3212.
%
\bibitem{Voloshin:1997jh}
S.~Voloshin, R.~Lednick\'y, S.~Panitkin and N.~Xu,
Phys.\ Rev.\ Lett.\  {\bf 79} (1997) 4766.
%
\bibitem{Wiedemann:1995au}
U.~A.~Wiedemann, P.~Scotto and U.~W.~Heinz,
Phys.\ Rev.\ C {\bf 53} (1996) 918.
%
\bibitem{Wiedemann:1996ig}
U.~A.~Wiedemann and U.~W.~Heinz,
Phys.\ Rev.\ C {\bf 56} (1997) 3265.
%
\bibitem{Wiedemann:1996ej}
U.~A.~Wiedemann and U.~W.~Heinz,
Phys.\ Rev.\ C {\bf 56} (1997) 610.
%
\bibitem{Wiedemann:1997cr}
U.~A.~Wiedemann,
Phys.\ Rev.\ C {\bf 57} (1998) 266.
%
\bibitem{Wiedemann:1998ng}
U.~A.~Wiedemann, D.~Ferenc and U.~W.~Heinz,
Phys.\ Lett.\ B {\bf 449} (1999) 347.
%
\bibitem{Wiedemann:1999qn}
U.~A.~Wiedemann and U.~W.~Heinz,
Phys.\ Rept.\  {\bf 319} (1999) 145.
%
\bibitem{Wu:1996wk}
Y.~F.~Wu, U.~W.~Heinz, B.~Tom\'a\v sik and U.~A.~Wiedemann,
Eur.\ Phys.\ J.\ C {\bf 1} (1998) 599.
%
\bibitem{Yano:gk}
F.~B.~Yano and S.~E.~Koonin,
Phys.\ Lett.\ B {\bf 78} (1978) 556.
%
\bibitem{Zajc:vb}
W.~A.~Zajc {\it et al.},
Phys.\ Rev.\ C {\bf 29} (1984) 2173.
%
\bibitem{Zajc:1986sq}
W.~A.~Zajc,
Phys.\ Rev.\ D {\bf 35} (1987) 3396.
%
\bibitem{Zschiesche:2001dx}
D.~Zschiesche, S.~Schramm, H.~St\"ocker and W.~Greiner,
Phys.\ Rev.\ C {\bf 65} (2002) 064902.
%

\centerline{\underline{Experimental papers:}}

\vspace{0.5cm}
%
\bibitem{e802-qm95}
V.~Cianciolo  [E802 Collaboration],
Nucl.\ Phys.\ A {\bf 590} (1995) 459C.
%
\bibitem{e802-arxiv}
L.~Ahle  [E802 Collaboration],
Phys.\ Rev.\ C {\bf 66} (2002) 054906.
%
\bibitem{e814-plb1994}
J.~Barrette {\it et al.}  [E814 Collaboration.],
Phys.\ Lett.\ B {\bf 333} (1994) 33.
%
\bibitem{e877-qm96hbt}
D.~Mi\'skowiec {\it et al.}  [E877 Collaboration],
Nucl.\ Phys.\ A {\bf 610} (1996) 227C.
%
\bibitem{Barrette:1997fj}
J.~Barrette {\it et al.}  [E877 Collaboration],
Phys.\ Rev.\ Lett.\  {\bf 78} (1997) 2916.
%
\bibitem{e877-qm97}
J.P.~Wessels {\it et al.}  [E877 Collaboration],
Nucl.\ Phys.\ A {\bf 638} (1998) 69.
%
\bibitem{EOS-TPC}
G.~Rai {\it et al.}, 
IEEE Trans.\ Nucl.\ Sci.\  {\bf 37} (1990) 56.
%
\bibitem{Lisa:2000xj}
M.~A.~Lisa {\it et al.}  [E895 Collaboration],
Phys.\ Lett.\ B {\bf 496} (2000) 1.
%
\bibitem{e895-edep}
M.A.~Lisa {\em et al.} [E895 Collaboration], Phys. Rev. Lett. {\bf 84}
(2000) 2798.
%
\bibitem{Panitkin:2001qb}
S.~Y.~Panitkin {\it et al.}  [E895 Collaboration],
Phys.\ Rev.\ Lett.\  {\bf 87} (2001) 112304.
%
\bibitem{e895-qm01}
M.~A.~Lisa {\it et al.}  [E895 Collaboration],
Nucl.\ Phys.\ A {\bf 698} (2002) 185.
%
\bibitem{e917-atlanta}
B.~Holzmann for the E917 Colaboration, 
  {\it Heavy Ion Physics from Bevalac to RHIC}, ed. by R.~Seto, 
  (World Scientific, Singapore, 1999).
%
\bibitem{na35-O}
A.~Bamberger {\it et al.}  [NA35 Collaboration],
Phys.\ Lett.\ B {\bf 203} (1988) 320.
%
\bibitem{na35-hum}
T.~J.~Humanic {\it et al.}  [NA35 Collaboration],
Z.\ Phys.\ C {\bf 38} (1988) 79.
%
\bibitem{na35-S}
T.~Alber {\it et al.}  [NA35 Collaboration],
Z.\ Phys.\ C {\bf 66} (1995) 77.
%
\bibitem{na35-kt}
T.~Alber {\it et al.}  [NA35 Collaboration.],
Phys.\ Rev.\ Lett.\  {\bf 74} (1995) 1303.
%
\bibitem{na35-qm95}
T.~Alber  [NA35 and NA49 Collaborations],
Nucl.\ Phys.\ A {\bf 590} (1995) 453C.
%
\bibitem{na44-93}
H.~Boggild {\it et al.} [NA44 Collaboration],
Phys.\ Lett.\ B {\bf 302} (1993) 510
[Erratum-ibid.\ B {\bf 306} (1993) 418].
%
\bibitem{na44-kaon94}
H.~Beker {\it et al.} [NA44 Collaboration],
Z.\ Phys.\ C {\bf 64} (1994) 209.
%
\bibitem{na44-3d}
H.~Boggild {\it et al.}  [NA44 Collaboration],
Phys.\ Lett.\ B {\bf 349} (1995) 386.
%
\bibitem{NA44-mperp}
H.~Beker {\it et al.} [NA44 Collaboration],
Phys.\ Rev.\ Lett.\  {\bf 74} (1995) 3340.
%
\bibitem{na44-mult}
K.~Kaimi {\it et al.} [NA44 Collaboration],
Z.\ Phys.\ C {\bf 75} (1997) 619.
%
\bibitem{na44-98}
I.~G.~Bearden {\it et al.} [NA44 Collaboration],
Phys.\ Rev.\ C {\bf 58} (1998) 1656.
%
\bibitem{na44-3part}
H.~Boggild {\it et al.}  [NA44 Collaboration],
Phys.\ Lett.\ B {\bf 455} (1999) 77.
%
\bibitem{murray:psd}
M.~Murray and B.~Holzer, Phys. Rev. {\bf C} 63 (2001) 054901.
%
\bibitem{na44-01}
I.~G.~Bearden {\it et al.}  [NA44 Collaboration],
Phys.\ Rev.\ Lett.\  {\bf 87} (2001) 112301.
%
\bibitem{Adamova:2002wi}
D.~Adamov\'a {\it et al.}  [NA45 CERES Collaboration],
arXiv:nucl-ex/0207005.
%
\bibitem{Adamova:2002ff}
D.~Adamov\'a {\it et al.}  [NA45 CERES Collaboration],
arXiv:nucl-ex/0207008.
%
\bibitem{NA49-app}
H.~Appelsh\"auser {\it et al.}  [NA49 Collaboration],
Eur.\ Phys.\ J.\ C {\bf 2} (1998) 661.
%
\bibitem{NA49-ganz}
R.~Ganz {\it et al.}  [NA49 Collaboration],
arXiv:nucl-ex/9808006.
%
\bibitem{na49-qm99}
R.~Ganz  [NA49 Collaboration],
Nucl.\ Phys.\ A {\bf 661} (1999) 448.
%
%
\bibitem{Seyboth:2002wu}
P.~Seyboth {\it et al.} [NA49 Collaboration],
arXiv:hep-ex/0206046.
%
\bibitem{na49QM2002}
C.~Blume for the NA49 Collaboration, arXiv:nucl-ex/0208020,
(Proceedings of Quark Matter 2002,
to be published in Nucl.~Phys.~A),
see also transparencies at 
\verb+http://alice-france.in2p3.fr/qm2002/Transparencies/+\\
\verb+19Plenary/Blume.pdf+.
%
\bibitem{Afanasiev:2002fv}
S.~V.~Afanasiev {\it et al.} [NA49 Collaboration],
arXiv:nucl-ex/0210018.

\bibitem{wa80-1992}
R.~Albrecht {\it et al.}  [WA80 Collaboration],
Z.\ Phys.\ C {\bf 53} (1992) 225.
%
\bibitem{wa80-inter}
R.~Albrecht {\it et al.}  [WA80 Collaboration],
Phys.\ Rev.\ C {\bf 50} (1994) 1048.
%
\bibitem{wa80-1995}
T.~C.~Awes {\it et al.}  [WA80 Collaboration],
Z.\ Phys.\ C {\bf 69} (1995) 67.
%
\bibitem{Antinori:2001yi}
F.~Antinori {\it et al.}  [WA97 Collaboration],
J.\ Phys.\ G {\bf 27} (2001) 2325.
%
\bibitem{Aggarwal:2000zs}
M.~M.~Aggarwal {\it et al.}  [WA98 Collaboration],
Eur.\ Phys.\ J.\ C {\bf 16} (2000) 445.
%
%
\bibitem{Aggarwal:2000ex}
M.~M.~Aggarwal {\it et al.}  [WA98 Collaboration],
Phys.\ Rev.\ Lett.\  {\bf 85} (2000) 2895.
%
\bibitem{wa98inprep}
M.~M.~Aggarwal  [WA98 Collaboration],
arXiv:nucl-ex/0210002.
%
\bibitem{Adcox:2002uc}
K.~Adcox {\it et al.}  [PHENIX Collaboration],
Phys.\ Rev.\ Lett.\  {\bf 88} (2002) 192302.
%
\bibitem{phenixQM2002}
A.~Enokizono for the PHENIX Collaboration, arXiv:nucl-ex/0209026,
(Proceedings of Quark Matter 2002, to be published in Nucl.~Phys.~A).
%
\bibitem{phobosHBT}
S.~Manly for the PHOBOS Collaboration, arXiv:nucl-ex/0210036,
(Proceedings of Quark Matter 2002, to be published in Nucl.~Phys.~A).
%
\bibitem{Adler:2001zd}
C.~Adler {\it et al.}  [STAR Collaboration],
Phys.\ Rev.\ Lett.\  {\bf 87} (2001) 082301.
%
\bibitem{Retiere:2001ed}
F.~Reti\`ere  [STAR Collaboration],
arXiv:nucl-ex/0111013.
%
\bibitem{STARQM02}
L.~Ray for the STAR Collaboration, arXiv:nucl-ex/0211030
(Proceedings of Quark Matter 2002, to be published in Nucl.~Phys.~A).
%
\bibitem{star02-HBT}
M.~L\'opez Noriega for the STAR Collaboration,
arXiv:nucl-ex/0210031 
(Proceedings of Quark Matter 2002, to be published in Nucl.~Phys.~A).
%
\bibitem{Willson-3pi}
R.~Willson for the STAR Collaboration, arXiv:nucl-ex/0211026
(Proceedings of Quark Matter 2002, to be published in Nucl.~Phys.~A).









\end{thebibliography}
\end{document}